\documentclass[structabstract]{aa}  
\usepackage{graphicx}

\usepackage{natbib}
\usepackage{verbatim}
\usepackage{pdfpages}
\usepackage{caption}
\usepackage{lscape} 
\bibpunct{(}{)}{;}{a}{}{,}

\usepackage{txfonts}
\begin{document}

   \title{Infrared study of transitional disks in Ophiuchus with \textit{Herschel} \thanks{ \textit{Herschel} is an ESA space observatory with science instruments provided by European-led Principal Investigator consortia and with important participation from NASA.} \thanks{Table 4 and Table 1 of the appendix are only available in electronic format the CDS via anonymous ftp to cdsarc.u-strasbg.fr (130.79.128.5)or via http://cdsweb.u-strasbg.fr/cgi-bin/qcat?J/A+A/}}
   \authorrunning{Isabel Rebollido et al.}
   \titlerunning{Infrared study of disks in the centre of Ophiuchus with \textit{Herschel} }
     
   \author{Isabel Rebollido \inst{1}
   \and Bruno Mer\'in \inst{1}
   \and \'Alvaro Ribas \inst{1,2,3}
   \and Ignacio Bustamante \inst{1,2,3}
   \and Herv\'e Bouy \inst{2}
   \and Pablo Riviere-Marichalar \inst{1}
   \and Timo Prusti \inst{4}
   \and G\"oran L. Pilbratt \inst{4}
   \and Philippe Andr\'e \inst{5}
   \and P\'eter \'Abrah\'am \inst{6}
               }

   \institute{European Space Astronomy Centre (ESA), P.O. Box, 78, 28691 Villanueva de la Ca\~{n}ada, Madrid, Spain
   \and Centro de Astrobiolog\'ia, INTA-CSIC, P.O. Box - Apdo. de correos 78, Villanueva de la Ca\~{n}ada Madrid 28691, Spain
   \and ISDEFE - ESAC, P.O. Box, 78, 28691 Villanueva de la Ca\~{n}ada, Madrid, Spain
   \and ESA, Scientific Support Office, Directorate of Science and Robotic Exploration, European Space Research and Technology Centre (ESTEC/SRE-S), Keplerlaan 1, 2201 AZ Noordwijk, The Netherlands
   \and Laboratoire AIM Paris, Saclay, CEA/DSM, CNRS, Universit\'e Paris Diderot, IRFU, Service d'Astrophysique, Centre d'Etudes de Saclay, Orme des Merisiers, 91191 Gif-sur-Yvette, France
      \and Konkoly Observatory, Research Centre for Astronomy and Earth Sciences, Hungarian Academy of Sciences, PO Box 67, 1525 Budapest, Hungary
   }

      \date{Received xxx; accepted xxx}

 
  \abstract
   {Observations of nearby star-forming regions with the \textit{Herschel} Space Observatory  complement our view of the protoplantary disks in Ophiuchus with information about the outer disks. } 
   {The main goal of this project is to provide new far-infrared fluxes for the known disks in the core region of Ophiuchus and to identify potential transitional disks using data from \textit{Herschel}. }
   {We obtained PACS and SPIRE photometry of previously spectroscopically confirmed young stellar objects (YSO) in the region and analysed their spectral energy distributions.}
  {From an initial sample of 261 objects with spectral types in Ophiuchus, we detect 49 disks in at least one \textit{Herschel} band. We provide new far-infrared fluxes for these objects. One of them is clearly a new transitional disk candidate. }
{The data from \textit{Herschel} Space Observatory provides fluxes that complement previous infrared data and that we use to identify a new transitional disk candidate.}

   
 \maketitle
%

\section{Introduction}

Protoplanetary disks around young stars are objects of major interest as they lead us to a better understanding of star and planet formation. Transitional disks are key in this study since they appear to have an unusual radial structure. They have been proposed as the environment for planet formation \citep{Marsh1992} and have  other proposed formation mechanisms, such as photo-evaporation by ultraviolet light emitted by the central star \citep{clarke2001}, grain growth \citep{dullemond2005}, and gravitational instabilities \citep[see][for a recent review on transitional disks]{Espaillat2014}. Recently, \cite{kim2013} studied accretion towards transitional disks in Orion A and concluded that planet formation was the most likely explanation for their observations. What characterizes a transitional disk is a lack of excess in the near- or mid-IR region (usually around 8 or 10 $\mu$m) and typical Class-II excesses in mid- to far-IR. This lack of near- and mid-IR excess denotes an inner disk opacity hole, which is related to the dust distribution in the surroundings of the star and reveals inner holes. These objects are thought to be an intermediate stage between Class II objects (optically thick disks) and Class III objects (smaller amount of material in the disk).

The first transitional disks were reported by \cite{Strom1989}. Since that time, the known population of these objects has grown substantially  thanks to data from \textit{Spitzer} Space Telescope \citep{SPITZER}. The study of these objects has improved with new and more powerful telescopes, such as the \textit{Herschel} Space Obsevatory \citep{Pilbratt2010}, which provides a wider range of wavelengths. \textit{Herschel}  also represents an improvement \textit{Spitzer}'s sensitivity and spatial resolution at long wavelengths, which allows for a reduction in the noise level of measurements at these wavelengths. Studies in the  millimetric and sub-millimetric range allowed for the imaging and direct measurement of hole sizes, such as those made by \cite{Andrews2011} with the Sub-Millimeter Array \citep{SMA}. Recent work with ALMA \citep{ALMA} has achieved new results in the field as seen in works by \cite{vanderMarel2013} and the recent study on the transitional disk HL Tauri \citep{HLTau2015}.

The aim of this work is to study the young stellar objects (YSOs) in the Ophiuchus star-forming region. \textit{Herschel} data provide us with accurate fluxes of the detected objects, and  enables the construction of the spectral energy distributions (SEDs) along with other multi-wavelength photometric data, collected by \cite{Ribas2014}. The study of the SEDs also allows us to classify the transitional disks in the region  by identifying the previously mentioned lack of excess in near and/or mid-IR region.


The structure of this work is as follows: Sect. 2 describes the \textit{Herschel} observations and data reduction, explaining the source detection and photometry extraction processes. In Sect. 3 we explain the results obtained from the data reduction, both with the method described in \cite{Ribas2013} and with the analysis of the SED of each individual object. In Sect. 4 we discuss the detection statistics and compare them with other similar studies. In Sect. 5 we present the conclusions of this work.

\section{Observations and data reduction}
\label{observations}

The Ophiuchus cloud complex was observed by \textit{Herschel} as part of the \textit{Herschel} Gould Belt Survey \citep{Andre2010} and by a deeper PACS survey \citep{AlvesdeOliveira2013}. It is one of the closest star-forming regions, located at an estimated distance of 130 pc and with an age between 2 and 5 Myr, although it has been suggested that it is younger \cite[see][for a review on the region]{Wilking2008}. Apart from its proxmity,  most significant characteristic of this cloud is its dense core, which is the focus of our study. The cloud itself appears as a large scale structure with complex filaments, heated by B-type stars, and is strongly emitting at infrared wavelengths. We might consider the cloud itself as a possible source of contamination. Figure 1 shows the region with the objects in the sample overplotted and marked differently according to their state of non-detected, detected, or transitional.

\begin{figure*}
  \centering
     \includegraphics[width=1.\textwidth]{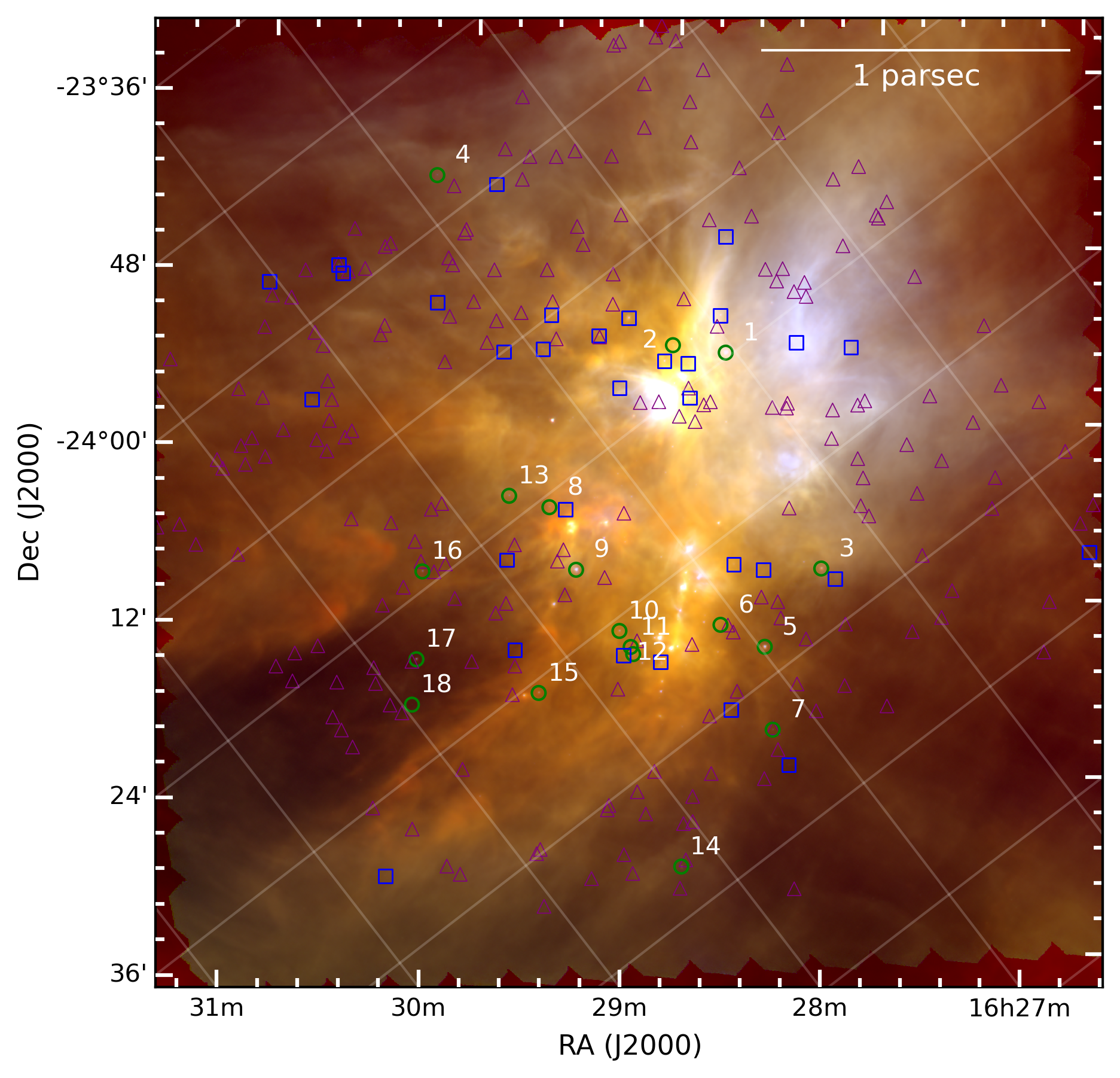} 
   \caption{RGB image (PACS 70 in blue, PACS 100 in green, and PACS 160 in red) of the observed Ophiuchus region with the marked objects. Transitional disks are marked with green circles. The blue squares mark the detected objects, and the purple triangles mark the rest of the sample of known young stars in the region.  }
 \label{fig:transitional}
\end{figure*}

The maps used to examine this region were obtained from two sets of observations, using the parallel mode of both PACS \citep{Poglitsch2010} and SPIRE \citep{SPIRE} on board the \textit{Herschel} Space Observatory. For SPIRE maps (250 $\mu$m, 350 $\mu$m and 500 $\mu$m) the parallel mode was used, with a speed of 60\arcsec/s (program $KPGT\_pandre\_1$), but for the PACS maps (70 $\mu$m and 160 $\mu$m) the scan mapping mode was used, with a cross-scan speed of 20\arcsec/s (program $OT1\_pabraham\_3$). We used this later PACS data because it goes deeper, as shown by the sensitivity limits given in the SPIRE/PACS Observers' Manual and in Table 1 in this work. In both cases (PACS and SPIRE), a single pair of scan and cross-scan was obtained. 
The obsids for the SPIRE maps are (1342-) 205093 \& 205094, and for PACS maps (1342-) 238816 \& 238817.

\begin{table*}
\centering
\caption{Sensitivity of the \textit{Herschel} observations used in this study}  
\begin{tabular}[width=1\textwidth]{lc c c c cl}

\hline
\hline
  & F$_{70} \,[mJy] $   & F$_{160}\,[mJy]$  & F$_{250}\,[mJy]$  & F$_{350}\,[mJy]$  & F$_{500}\,[mJy] $ \\
\hline
Sensitivity per scan & 12.2     &       14.3    &       12.6    &       10.5    &       15.0\\ 
Effective Sensitivity & 2       &       2.4     &       6.3     &       5.2     &       7.5\\
Minimum Flux Detected & 20      &       50      &       70      &       10      &       30\\ 
\hline
\end{tabular}

\end{table*}

The maps were produced using Scanamorphos \citep{Scanamorphos}, an IDL software designed to process \textit{Herschel} maps. Version 24.0 was used for PACS maps, with calibration file version 65, and version 22.0 for SPIRE maps, with calibration files version 12.3. 

\subsection{Sample selection and point source photometry}
\label{fluxes}

The objective of this work is to determine the nature of the sources detected in the maps produced with PACS and SPIRE observations and to obtain fluxes in the mid- to far-IR regions when detected. The YSO candidates were selected from the work of \cite{Ribas2014}  \citep[which for Ophiuchus contains objects from][]{Natta2002, Wilking2005, AlvesDeOliveira2010,Erickson2011}, for having known spectral type from spectroscopy, making a list of 258 sources, all of them located in the core of the cloud. For a more complete study, the objects from \cite{cieza2010} contained in the region of interest were also considered. The final sample consists of 261 objects. We used photometry from the 2MASS K-band \citep{2MASS} and WISE4 \citep{WISEINST} to classify the disks according to the method described by \cite{Lada1987}. The only object that does not fulfill the criteria is IRS 48, which has a positive slope that is almost flat. This object cannot be classified as Class\ I and remains Class II in the final classificaiton because of the lack of strong emission in mid- and far-IR. 
  
To determine whether our candidates are transitional disks or not, the first step was to detect sources in our maps and extract the photometry from them. We used the algorithm \textit{Sussextractor} in the \textit{Herschel Interactive Processing Environment} (HIPE), version 12.1, with a threshold of S/N $>$ 3. Additionally, visual inspection was applied to all sources. Only sources clearly separated from filaments and distinguishable from the background were selected as valid detections. 

We extracted aperture photometry for these sources using the sourceExtractorDaophot task in HIPE \citep{HIPE}. The fluxes were aperture corrected to account for the shape of the PSF and they are listed in Table 4. Those corrections were obtained from \cite{balog2014} for PACS and from the SPIRE Observers' Manual (now named SPIRE Handbook) version 2.5. In Table 4, objects marked with an asterisk  suffered from high nearby background emission and, in those cases, we used a special sky estimate  from a rectangular area identified after visual inspection. 
The set of apertures and the corrections applied to the photometry extracted are shown in Table 2. We tested several point-source extraction algorithms, including Sussextractor, Daophot, AnnularSkyAperturePhotometry, and Hyper \citep{Traficante2014}, and different apertures. The combination above provides the best fit to the MIPS70  \citep{Rieke2004} fluxes of clean selected sources in the field. 

\begin{table*}
\centering
\caption{Aperture photometry parameters}        
\begin{tabular}[width=1\textwidth]{lc c c c cl}

\hline
\hline
Band & FWHM (") & Radius (") & Inner Annulus (") & Outer Annulus (")    &       Aperture Correction Factors \\
\hline
PACS-70 & 5.4   &       6       &       25      &       35      &       1.5711\\ 
PACS-160 & 10.5 &       12      &       25      &       35      &       1.4850\\ 
SPIRE-250 & 18  &       22      &       60      &       90      &       1.2584\\
SPIRE-350 & 24  &       30      &       60      &       90      &       1.2242\\
SPIRE-500 & 36  &       42      &       60      &       90      &       1.1975\\
\hline
\end{tabular}

\end{table*}

At the end of the process, we had 49 successfully detected sources in at least one PACS band, and 19 in a SPIRE band; PACS also detected all of these. Images of the detected objects can be found in Figures 3 to 9 of the online appendix, for visual inspection.

The calibration errors for PACS and SPIRE are 5\% and 7\% (as stated in the PACS Observer's Manual, version 2.3 and in the SPIRE Observer's Manual version 2.5), however, as a more conservative estimation, we used 25\% for both. These measurement uncertainties account for the high variations in the results of the different extraction methods and apertures listed above.

In Table 3 we show the parameters of the detected sources, as given in the literature, and for an easier reference we list the identification numbers (I.D.) used in this work as seen in Figure 1. From now on, the I.D. number is referenced in parenthesis for its respective source. Table 4 gives the fluxes measured with \textit{Herschel} for each source.

\begin{table*}

\centering
\caption{Parameters of detected sources as extracted from the literature. Also, we give the I.D. number used to identify the transitional disks.}
\begin{tabular}[width=1\textwidth]{lc c c c  c c cl}
\hline
\hline
I.D. & Name & R.A. (deg) & Dec (deg) & SpT & $A_v*$ & References \\
\hline
-    &   2MASSJ16245974-2456008 & $16:24:59.63$ & $-24:55:59.32$ & M3.5 & $0.10$ & 1 \\
-    &   V*V852Oph & $16:25:24.38$ & $-24:29:43.77$ & M4.5 & $0.21$ & 2 \\
-    &   2MASSJ16253673-2415424 & $16:25:36.73$ & $-24:15:42.32$ & K4 & $0.10$ & 1  \\
-    &   2MASSJ16253958-2426349 & $16:25:39.58$ & $-24:26:34.27$ & M2 & $0.13$ & 2 \\
-    &   V*V2058Oph & $16:25:56.18$ & $-24:20:47.77$ & K4.5 & $0.61$ & 1 \\
$1$  &   Haro1-6 & $16:26:03.02$ & $-24:23:36.04$ & K1 & $5.70$ & 4  \\
-    &   2MASSJ16261684-2422231 & $16:26:16.83$ & $-24:22:23.32$ & K6 & $0.11$ & 1 \\
$2$  &   DoAr24 & $16:26:17.08$ & $-24:20:21.47$ & K4.5 & $0.13$ & 2  \\
-    &   2MASSJ16262189-2444397 & $16:26:21.88$ & $-24:44:39.67$ & M8 & $0.10$ & 2 \\
-    &   DoAr24E & $16:26:23.28$ & $-24:20:59.37$ & G6 & $0.10$ & 2 \\
$3$  &   DoAr25 & $16:26:23.68$ & $-24:43:13.57$ & K5 & $0.21$ & 2 \\
-    &   GSS32 & $16:26:24.03$ & $-24:24:48.32$ & K5 & $0.63$ & 1 \\
-    &   2MASSJ16262407-2416134 & $16:26:24.08$ & $-24:16:13.27$ & K5.5 & $0.19$ & 2 \\
-    &   2MASSJ16263297-2400168 & $16:26:32.98$ & $-24:00:16.77$ & M4.5 & $0.09$ & 2 \\
-    &   2MASSJ16263682-2415518 & $16:26:36.93$ & $-24:15:52.32$ & M0 & $0.55$ & 1 \\
-    &   $[GY92]93$ & $16:26:41.28$ & $-24:40:17.87$ & M5 & $0.09$ & 2 \\
-    &   2MASSJ16264285-2420299 & $16:26:42.83$ & $-24:20:30.32$ & M1 & $0.11$ & 1 \\
-    &   2MASSJ16264643-2412000 & $16:26:46.48$ & $-24:11:59.97$ & G3.5 & $0.10$ & 2 \\
$4$  &   WSB40 & $16:26:48.58$ & $-23:56:34.57$ & K5.5 & $0.09$ & 2 \\
-    &   WL18 & $16:26:48.98$ & $-24:38:25.07$ & K6.5 & $0.59$ & 2 \\
-    &   2MASSJ16265677-2413515 & $16:26:56.68$ & $-24:13:51.47$ & K7 & $0.11$ & 2\\
$5$  &   SR24S & $16:26:58.48$ & $-24:45:36.67$ & K1 & $0.18$ & 2\\
$6$  &   2MASSJ16270659-2441488 & $16:27:06.68$ & $-24:41:49.07$ & M5.5 & $0.09$ & 2 \\
-    &   2MASSJ16270907-2412007 & $16:27:09.03$ & $-24:12:01.32$ & M2.5 & $0.11$ & 1\\
$7$  &   WSB46 & $16:27:15.08$ & $-24:51:38.77$ & M2 & $0.09$ & 2\\
-    &   $[WMR2005]4-10$ & $16:27:17.48$ & $-24:05:13.67$ & M3.5 & $0.10$ & 2 \\
-    &   2MASSJ16271836-2454537 & $16:27:18.38$ & $-24:54:52.77$ & M3.75 & $0.16$ & 2\\
-    &   WSB49 & $16:27:22.98$ & $-24:48:07.07$ & M4.25 & $0.09$ & 2 \\
-    &   2MASSJ16272658-2425543 & $16:27:26.58$ & $-24:25:54.47$ & M8 & $0.14$ & 2 \\
$8$  &   2MASSJ16273084-2424560 & $16:27:30.88$ & $-24:24:56.37$ & M3.25 & $0.12$ & 2 \\
-    &   2MASSJ16273311-2441152 & $16:27:33.13$ & $-24:41:14.32$ & K6 & $0.50$ & 1\\
$9$  &   IRS48 & $16:27:37.23$ & $-24:30:34.32$ & A0 & $0.76$ & 1 \\
$10$ &   IRS49 & $16:27:38.28$ & $-24:36:58.67$ & K5.5 & $0.15$ & 2 \\
-    &   2MASSJ16273832-2357324 & $16:27:38.28$ & $-23:57:32.97$ & K6 & $0.14$ & 2\\
$11$ &   2MASSJ16273863-2438391 & $16:27:38.60$ & $-24:38:39.00$ & M6 & $0.24$ & 3 \\
-    &   2MASSJ16273901-2358187 & $16:27:38.98$ & $-23:58:19.17$ & K5.5 & $0.15$ & 2 \\
$12$ &   WSB52 & $16:27:39.48$ & $-24:39:15.87$ & K5 & $0.09$ & 2 \\
$13$ &   SR9 & $16:27:40.28$ & $-24:22:04.37$ & K5 & $0.17$ & 2 \\
-    &   2MASSJ16274270-2438506 & $16:27:42.68$ & $-24:38:51.27$ & M2 & $0.11$ & 2\\
-    &   V*V2059Oph & $16:27:55.58$ & $-24:26:18.27$ & M2 & $0.11$ & 2 \\
-    &   2MASSJ16280256-2355035 & $16:28:02.58$ & $-23:55:03.61$ & M3 & $4.30$ & 4 \\
-    &   2MASSJ16281379-2432494 & $16:28:13.83$ & $-24:32:49.32$ & M4 & $0.10$ & 1 \\
$14$ &   2MASSJ16281385-2456113 & $16:28:13.83$ & $-24:56:10.32$ & M0 & $0.10$ & 1 \\
$15$ &   WSB60 & $16:28:16.58$ & $-24:36:58.57$ & M4.5 & $0.11$ & 2 \\
-    &   2MASSJ16281673-2405142 & $16:28:16.83$ & $-24:05:15.32$ & K6 & $0.08$ & 1 \\
$16$ &   SR20W & $16:28:23.38$ & $-24:22:40.87$ & K5 & $0.50$ & 2 \\
$17$ &   SR13 & $16:28:45.28$ & $-24:28:19.27$ & M3.75 & $0.20$ & 2 \\
$18$ &   2MASSJ16285694-2431096 & $16:28:57.03$ & $-24:31:09.32$ & M5.5 & $0.13$ & 1 \\
-    &   2MASSJ16294427-2441218 & $16:29:44.28$ & $-24:41:21.80$ & M4 & $0.80$ & 4 \\
\hline
\end{tabular}
\tablefoot{
 References: 1) \cite{Erickson2011}, 2) \cite{Wilking2005}, 3) \cite{Natta2002}, 4) \cite{cieza2010}  \\** All $A_v$ are calculated according to \cite{Ribas2014}
}
\end{table*}

\begin{table*}

\centering
\caption{Point source fluxes of each of the 49 sources detected}
\begin{tabular}[width=1\textwidth]{lc c c c c c c cl}
\hline
\hline
I.D. & Name &  F$_{70} \,[Jy] $   & F$_{160}\,[Jy]$  & F$_{250}\,[Jy]$  & F$_{350}\,[Jy]$  & F$_{500}\,[Jy] $ \\
\hline
-    &   2MASSJ16245974-2456008 & $0.05 \pm 0.01$ & - & - & - & - \\
-    &   V*V852Oph & $0.73 \pm 0.18$ & - & - & - & - \\
-    &   2MASSJ16253673-2415424 & $0.74 \pm 0.18$ & - & - & - & - \\
-    &   2MASSJ16253958-2426349 & $0.92 \pm 0.23$ & - & - & - & - \\
-    &   V*V2058Oph & $3.86 \pm 0.97$ & $3.77 \pm 0.94$ & $12.56 \pm 3.14$ & - & - \\
$1$  &   Haro1-6 & $10.70 \pm 2.68$ & $7.48 \pm 1.87$ & - & - & - \\
-    &   2MASSJ16261684-2422231 & $0.20 \pm 0.05$ & - & - & - & - \\
$2$  &   DoAr24 & $0.50 \pm 0.12$ & - & - & - & - \\
-    &   2MASSJ16262189-2444397 & $0.10 \pm 0.02$ & - & - & - & - \\
-    &   DoAr24E & $4.17 \pm 1.04$ & $3.91 \pm 0.98$ & - & - & - \\
$3$  &   DoAr25 & $1.39 \pm 0.35$ & $3.52 \pm 0.88$ & $4.17 \pm 1.04$ & $5.30 \pm 1.33$ & $2.05 \pm 0.51$ \\
-    &   GSS32 & $3.70 \pm 0.93$ & - & - & - & - \\
-    &   2MASSJ16262407-2416134 & $3.04 \pm 0.76$ & $6.10 \pm 1.53$ & $4.73 \pm 1.18$ & $2.66 \pm 0.66$ & $1.12 \pm 0.28$ \\
-    &   2MASSJ16263297-2400168 & $0.08 \pm 0.02$ & $0.13 \pm 0.03$ & - & $0.01 \pm 0.00$ & - \\
-    &   2MASSJ16263682-2415518* & $1.37 \pm 0.34$ & $0.72 \pm 0.18$ & $0.44 \pm 0.11$ & - & - \\
-    &   $[GY92]93$ & $0.02 \pm 0.00$ & - & - & - & - \\
-    &   2MASSJ16264285-2420299 & $0.65 \pm 0.16$ & - & - & - & - \\
-    &   2MASSJ16264643-2412000 & $0.57 \pm 0.14$ & - & - & - & - \\
$4$  &   WSB40 & $0.42 \pm 0.10$ & $0.29 \pm 0.07$ & - & - & - \\
-    &   WL18 & $0.42 \pm 0.10$ & - & - & - & - \\
-    &   2MASSJ16265677-2413515 & $0.21 \pm 0.05$ & $0.79 \pm 0.20$ & - & - & - \\
$5$  &   SR42S & $9.70 \pm 2.42$ & $7.88 \pm 1.97$ & $5.06 \pm 1.26$ & $3.02 \pm 0.76$ & $1.55 \pm 0.39$ \\
$6$  &   2MASSJ16270659-2441488 & $0.10 \pm 0.02$ & - & - & - & - \\
-    &   2MASSJ16270907-2412007 & $0.07 \pm 0.02$ & $0.23 \pm 0.06$ & - & $0.11 \pm 0.03$ & - \\
$7$  &   WSB46 & $0.25 \pm 0.06$ & $0.15 \pm 0.04$ & - & - & - \\
-    &   $[WMR2005]4-10$ & $0.18 \pm 0.04$ & $0.32 \pm 0.08$ & $0.51 \pm 0.13$ & $0.37 \pm 0.09$ & - \\
-    &   2MASSJ16271836-2454537* & $0.07 \pm 0.02$ & $0.05 \pm 0.01$ & $0.07 \pm 0.02$ & - & - \\
-    &   WSB49 & $0.04 \pm 0.01$ & - & - & - & - \\
-    &   2MASSJ16272658-2425543 & $0.04 \pm 0.01$ & - & - & - & - \\
$8$  &   2MASSJ16273084-2424560 & $0.50 \pm 0.13$ & - & - & - & - \\
-    &   2MASSJ16273311-2441152 & $1.27 \pm 0.32$ & $0.08 \pm 0.02$ & - & - & - \\
$9$  &   IRS48 & $37.57 \pm 9.39$ & $12.85 \pm 3.21$ & $6.48 \pm 1.62$ & $2.51 \pm 0.63$ & - \\
$10$ &   IRS49* & $1.29 \pm 0.32$ & $0.95 \pm 0.24$ & $0.07 \pm 0.02$ & $0.67 \pm 0.17$ & - \\
-    &   2MASSJ16273832-2357324* & $0.63 \pm 0.16$ & $0.25 \pm 0.06$ & $0.52 \pm 0.13$ & $0.30 \pm 0.07$ & - \\
$11$ &   2MASSJ16273863-2438391 & $0.09 \pm 0.02$ & - & - & - & - \\
-    &   2MASSJ16273901-2358187 & $0.44 \pm 0.11$ & $0.52 \pm 0.13$ & $0.10 \pm 0.02$ & $0.09 \pm 0.02$ & $0.03 \pm 0.01$ \\
$12$ &   WSB52* & $2.37 \pm 0.59$ & $2.23 \pm 0.56$ & $3.43 \pm 0.86$ & $0.76 \pm 0.19$ & $0.28 \pm 0.07$ \\
$13$ &   SR9 & $0.86 \pm 0.21$ & $0.22 \pm 0.05$ & - & - & - \\
-    &   2MASSJ16274270-2438506 & $0.02 \pm 0.01$ & - & - & - & - \\
-    &   V*V2059Oph & $0.06 \pm 0.01$ & - & - & - & - \\
-    &   2MASSJ16280256-2355035 & $0.02 \pm 0.01$ & - & - & - & - \\
-    &   2MASSJ16281379-2432494 & $0.05 \pm 0.01$ & - & - & - & - \\
$14$ &   2MASSJ16281385-2456113 & $0.74 \pm 0.18$ & $0.68 \pm 0.17$ & $0.29 \pm 0.07$ & $0.19 \pm 0.05$ & $0.14 \pm 0.03$ \\
$15$ &   WSB60 & $0.87 \pm 0.22$ & $1.19 \pm 0.30$ & $1.26 \pm 0.32$ & $1.79 \pm 0.45$ & - \\
-    &   2MASSJ16281673-2405142 & $0.12 \pm 0.03$ & - & - & - & - \\
$16$ &   SR20W & $0.89 \pm 0.22$ & $0.60 \pm 0.15$ & $0.57 \pm 0.14$ & $0.46 \pm 0.12$ & $0.37 \pm 0.09$ \\
$17$ &   SR13 & $1.36 \pm 0.34$ & $1.32 \pm 0.33$ & $0.79 \pm 0.20$ & $0.51 \pm 0.13$ & $0.20 \pm 0.05$ \\
$18$ &   2MASSJ16285694-2431096 & $0.02 \pm 0.01$ & - & - & - & - \\
-    &   2MASSJ16294427-2441218 & $0.05 \pm 0.01$ & - & - & - & - \\
\hline
\end{tabular}
\tablefoot{
\tablefoottext{*}{Sources near bright background emission; sky measurement was done in clean regions.} \\
}
\end{table*}

\subsection{Sensitivity and non-detections}

The high background present in the region increases the sensitivity limit of these observations, and affects the number of detections. In Table 1 we present the sensitivities given in the SPIRE/PACS Parallel Mode Observers' Manual v2.1 (section 2.3), and the minimum flux detected for each band. Also, the effective sensitivity is calculated taking  the number of scans made per map into account; this includes six for the PACS maps and two for the SPIRE maps.
We also report a lower value of fluxes in the cleanest areas of the map, which agrees with the fact that the extended emission from the molecular cloud reduces the effective sensitivity.

Another study took  into account the sources detected by \textit{Spitzer} and catalogued by c2d \citep{Evans2009} in each of the apertures defined per band and per detected source  to estimate possible contamination due to \textit{Herschel} resolution. The flux in MIPS-24 band of each of this c2d sources was used to extrapolate via the median SED of the Ophiuchus region (see Table 5) the expected fluxes at 70 $\mu$m. In Table 1 of the online appendix, we list a sum of all the expected fluxes for this contaminating sources, where it is possible to check whether the expected contamination is lower than the flux error. 
 Notice that this contamination flux is only an estimation, as only the reported object has been detected in the aperture.

\section{Results}
\label{results}

The fluxes of all the detected sources per band are given in Table 4, as mentioned above. As expected, the higher flux corresponds to the most known object in the region, IRS 48 (\#9). All the objects proposed as transitional disk candidates present relatively high fluxes, being 2MASS J16285694-2431096 (\#18) the the candidate with the lower flux at 70 $\mu$m (20 mJy). The detection on more than one band for these objects seems to be arbitrary. Less than half of the candidates have been detected either in one band or in all of them. This is probably related to the physical properties of each of the disks rather than to instrumental issues.

\subsection{Identification of transitional disks}
Once we measured the fluxes, we use the criterion by \cite{Ribas2013} to classify transitional disks with \textit{Herschel} photometry. As previously noted, transitional disks display little or no excess at near- to mid-IR, and large excess at long wavelengths, which translates into a change of slope in the SED of the object around 12 $\mu$m, i.e. from a negative to  positive slope. To proceed with the identification, we defined two indexes , following: 
\begin{equation}
\centering
$$\alpha$ $_{\lambda_{1}-\lambda_{2}}$ = $\frac{log(\lambda_{1}F_{\lambda_{1}})-log(\lambda_{2}F_{\lambda_{2}})}{log(\lambda_{1})-log(\lambda_{2})}$$
,\end{equation}
where $\lambda$ is measured in $\mu$m and $F_{\lambda}$ in  erg $\cdot$ s$^{-1}$ $\cdot$ cm$^{-2}$. The first index is defined for the band K acquired from 2MASS and 12 $\mu$m acquired from WISE. The second index is defined for the 12 $\mu$m band and the 70 $\mu$m band acquired from \textit{Herschel}. 
The criterion to determine whether an object is a transitional disk candidate is  $\alpha$ $_{12-70}$ ${>}$ 0, since we define transitional disks as objects with a deficit of excess flux at near to  mid-IR fluxes, but standard excesses at longer wavelengths. 
For DoAr 24(\#2), the 12 $\mu$m-band was not available and, therefore, we used 8 $\mu$m band from \textit{Spitzer/IRAC} \citep{IRAC}  instead.

To understand the result of this process, we represent in Fig. 2 a slope-slope diagram, which shows the value of the two different slopes in the two axes. The figure shows that we obtained two candidates with a positive slope between 12 $\mu$m and 70 $\mu$m, clearly separated from the Class II population. These two candidates, 2MASS J16281385-2456113 (\#14) and Haro1-6 (\#1), clearly fulfil the criteria of having a lack of mid-IR excess since Haro1-6 (\#1) is an object previously classified as debris disk in \cite{cieza2010}. The candidate 2MASS J16281385-2456113 (\#14), has never been reported before as a transitional disk. The other objects above the threshold do not have this kind of  clear 12 $\mu$m flux deficit, which is indicative of the presence of a flatter slope in the mid- to far- infrared wavelengths. If we construct the SEDs of those objects (see Fig. 4 ), we see that in general they do not present the characteristic gap expected from a transitional disk. However, as they fulfil the criteria, we  classified them here as tentative candidates. Some of these candidates were also classified as transitional disks in previous works. Object  WSB 60 (\#15) was imaged in \cite{Andrews2009}, detecting a small inner hole in dust continuum observations done with the Sub-Millimeter Array \citep{SMA}. In the work by \cite{cieza2010} DoAr 25 (\#3) was already suggested as a candidate.
We consider objects SR 42 S (\#5), WSB 46 (\#7), and SR 20 W (\#16) as tentative candidates despite they do not fulfill the criteria for their nominal values, but we are considering a high error in PACS photometry. In particular, SR 42 S (\#5) has already been classified as transitional disk, and appears in \cite{Espaillat2014} as so. For WSB 46 (\#7) and SR 20 W (\#16), further study is needed to determine their nature.

\begin{figure}
  \centering
     \includegraphics[width=0.5\textwidth]{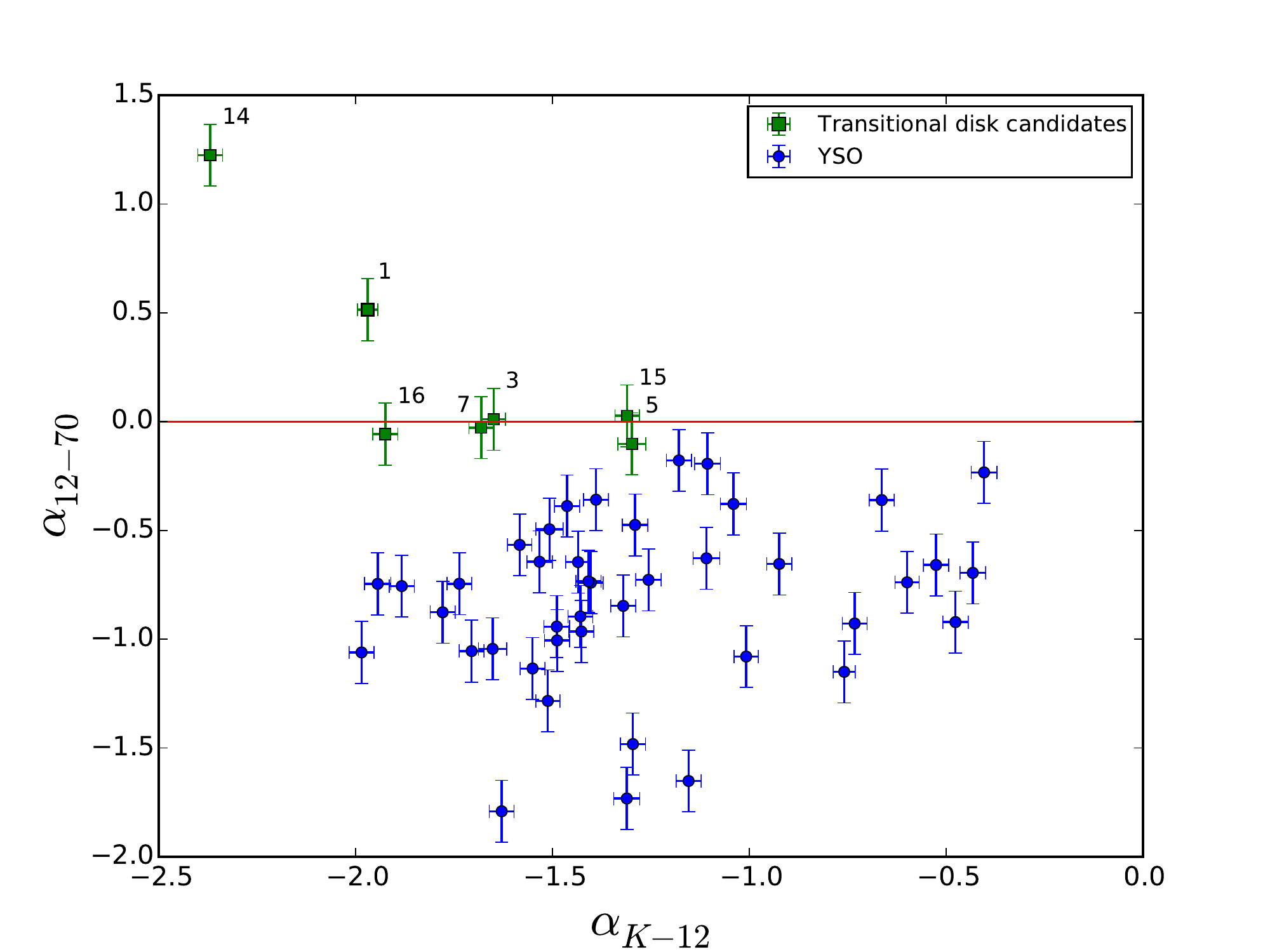} 
   \caption{SED slope between 12 and 70 $\mu$m as a function of the SED slope between the K-band and 12 $\mu$m. Transitional disks are marked with green squares.}
 \label{fig:transitional}
\end{figure}

\subsection{Complementary identification with spectral energy distributions}
\label{seds}

        To better analyse the nature of these objects, we built their SEDs using data from optical to mid-infrared from both ground-based and space telescopes (all references for the photometry can be found in section 2.1). We completed these SEDs  with the photometry shown in Table 2 and extracted from \textit{Herschel} data, therefore, we cover a range between 0.35 to 500 $\mu$m. Fig. 4 and Fig. 1 and 2 of the online appendix show the obtained SEDs, plus the NextGen atmosphere model for each object \citep{Nextgen}, which are the best approximation of how a naked photosphere would emit as a function of its spectral type. 
    
    We also built the median SED of all the objects detected in the region for comparison (see Table 5), which is plotted along each object's SEDs. Because of the lack of detections for fluxes under the sensitivity limits given in Table 4, the median SED might be slightly overestimated, but we assume the effect in our result is negligible. 

\begin{table}
\small
\centering
\caption{Median SED of detected disks in Ophiuchus}
\begin{tabular}[width=1\textwidth]{lc c c c c c cl}
\hline\hline
Band & Median & First Quartile & Third Quartile & Detections \\
\hline
J & $1.0000$ & $1.0000$ & $1.0000$ & $49$ \\
H & $0.6109$ & $0.5664$ & $0.6577$ & $49$ \\
K & $0.3305$ & $0.3018$ & $0.3662$ & $49$ \\
IRAC-3.6 & $0.0743$ & $0.0628$ & $0.0958$ & $45$ \\
IRAC-4.5 & $0.0380$ & $0.0305$ & $0.0537$ & $47$ \\
IRAC-5.8 & $0.0212$ & $0.0162$ & $0.0293$ & $46$ \\
IRAC-8.0 & $0.0121$ & $0.0088$ & $0.0190$ & $48$ \\
MIPS-24 & $0.0018$ & $0.0013$ & $0.0030$ & $48$ \\
PACS-70 & $0.0003$ & $0.0001$ & $0.0008$ & $49$ \\
PACS-160 & $8.53\cdot10^{-5}$ & $5.54\cdot10^{-5}$ & $0.0002$ & $26$ \\
SPIRE-250 & $3.48\cdot10^{-5}$ & $2.32\cdot10^{-5}$ & $0.0001$ & $17$ \\
SPIRE-350 & $9.60\cdot10^{-6}$ & $5.88\cdot10^{-6}$ & $1.44\cdot10^{-5}$ & $15$ \\
SPIRE-500 & $2.62\cdot10^{-6}$ & $2.12\cdot10^{-6}$ & $4.18\cdot10^{-6}$ & $8$ \\
\hline

\end{tabular}

\end{table}

    To determine the interstellar extinction, we used the procedure in \cite{Ribas2014} and the extinction law from \cite{Weingartner2001},  which uses a model of grains to estimate the interstellar extinction, scattering, and infrared emission. In each plot of Fig. 4 and 1 and 2 of the online appendix, the observed fluxes are also shown as empty circles. The de-reddened fluxes, according to the associated interstellar extinction,  are shown as filled circles.

        When visually inspecting the SEDs, we detect objects with a lack of mid-IR excess that were not classified as transitional disks, according to the criteria of \cite{Ribas2013} described in section 3.1. Some of these objects are well known in the literature, such as IRS 48 (\#9). To make a more reliable study of the region, we add in this subsection a complementary criterion based on Spitzer data to identify the other transitional disks or transitional disks candidates detected by \textit{Herschel} with a change of slope between 12 $\mu$m and 24 $\mu$m. Figure 3 shows the new slope-slope diagram.

\begin{figure}
  \centering
    \includegraphics[width=0.5\textwidth]{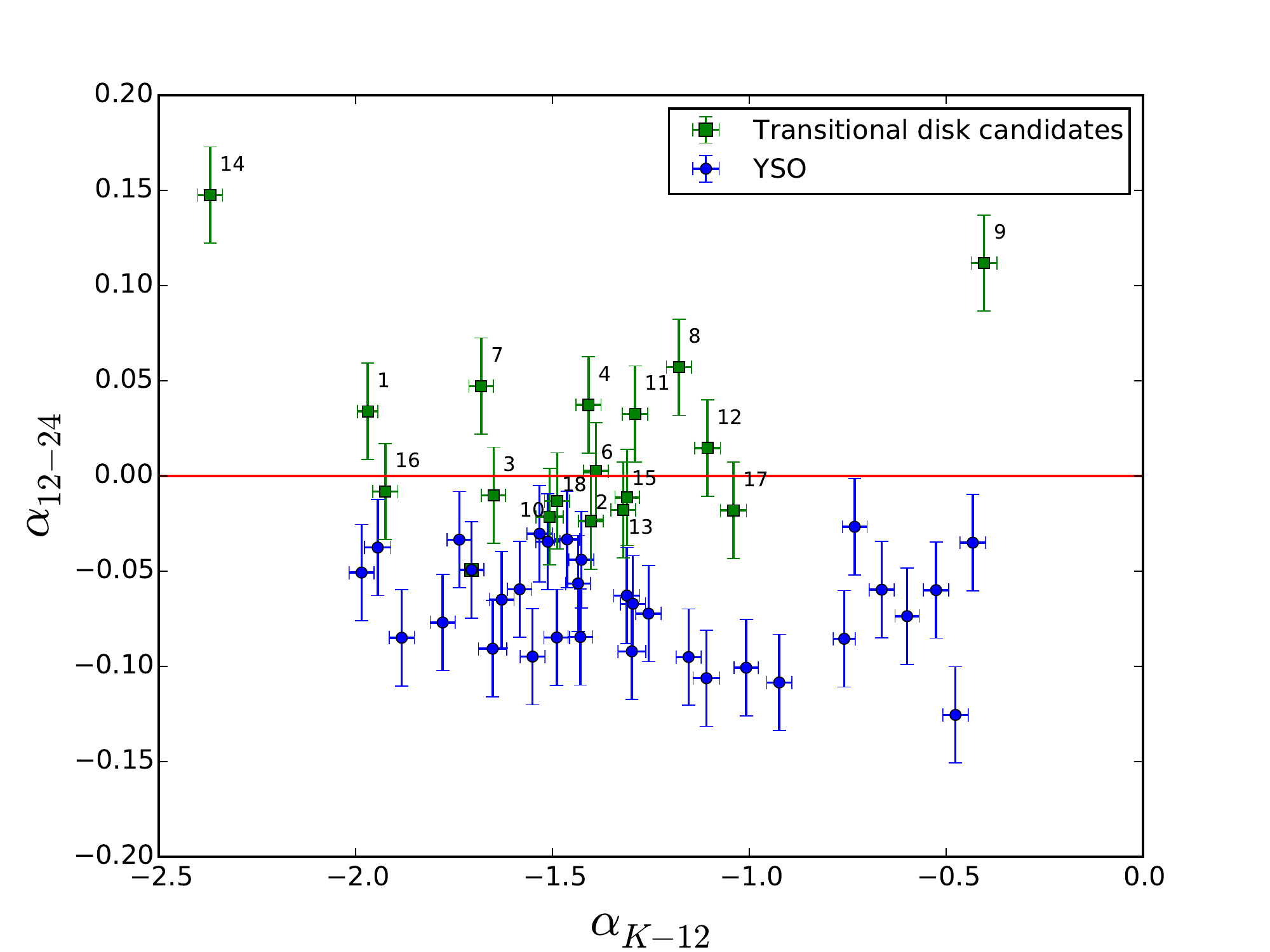} 
  \caption{SED slope between 12 and 24 $\mu$m as a function of the SED slope between the K-band and 12 $\mu$m. Transitional disks are marked with green squares.}
 \label{fig:transitional}
\end{figure}

We detected  17 transitional disks candidates with this criteria. Six of the objects detected with the previous criteria appear now as well, with the exception of SR 24 S (\#5), which is a confirmed transitional disk.

Figure 4 shows the SEDs for the 18 transitional disks candidates, where the change of slope is not equally noticeable for all of them. In the case of 2MASS J16281385-2456113 (\#14), the change of slope is evident, in agreement with that expected from the slope-slope diagram. In cases such as WSB 60 (\#15) or IRS 48 (\#9), which are well-known transitional disks, the slope is flatter, and located in a different position in the wavelength axis. For SR 24 S (\#5), we see an unexpected behaviour, where we get a discrepancy between MIPS-24 and WISE4 fluxes, but we still see an increase in the far-IR emission with respect to the near-IR. Another object classified as a transitional disk in \cite{cieza2010} is SR 9 (\#13), although it has a continuous decreasing slope, as the criterion shows, and looks more like a Class II SED. 
The rest of the objects in the sample have been widely studied, but not previously considered  transitional disks. 
The variation in the outputs of both methods illustrates the complexity in defining a selection criteria.
\begin{figure*}[SEDtransitional]
 \centering
 
    \includegraphics[width=0.3\textwidth,page=6]{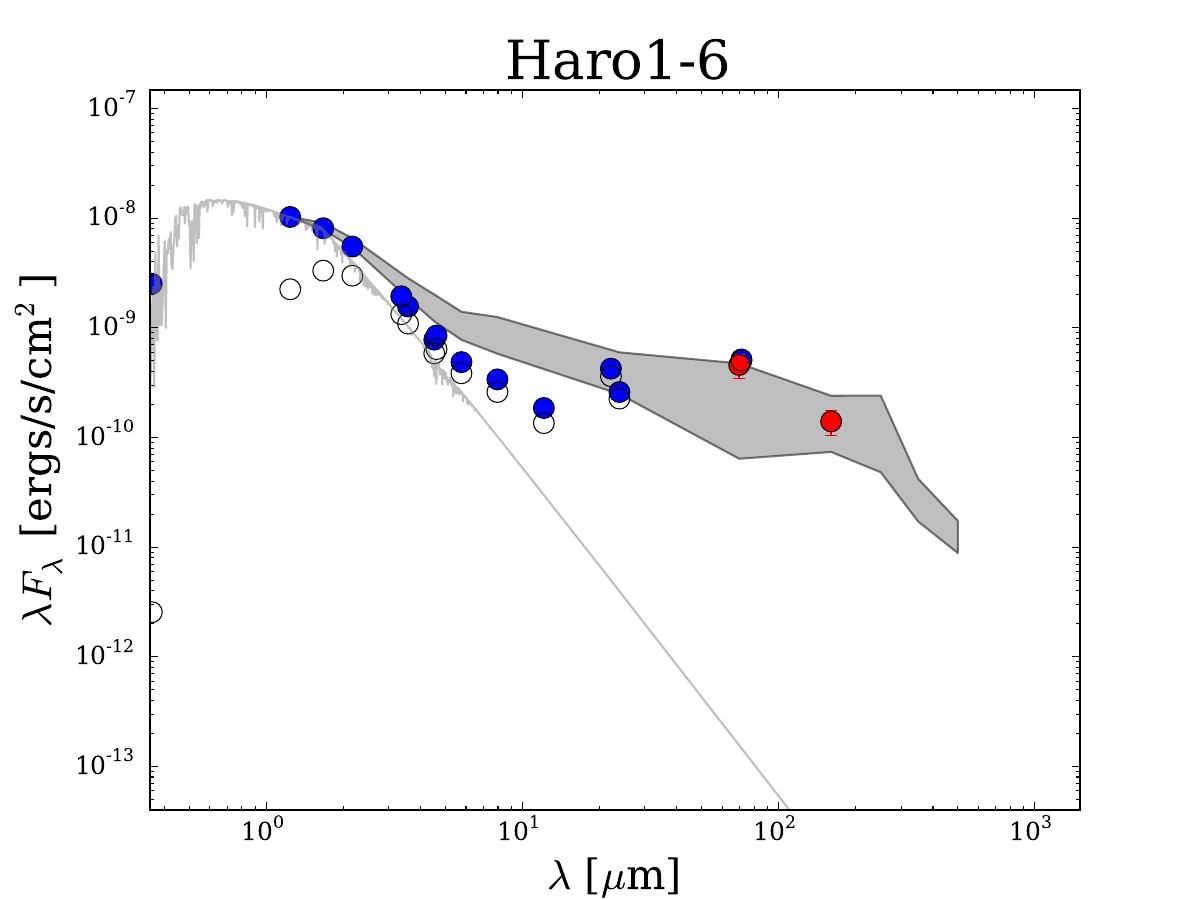}
    \includegraphics[width=0.3\textwidth,page=8]{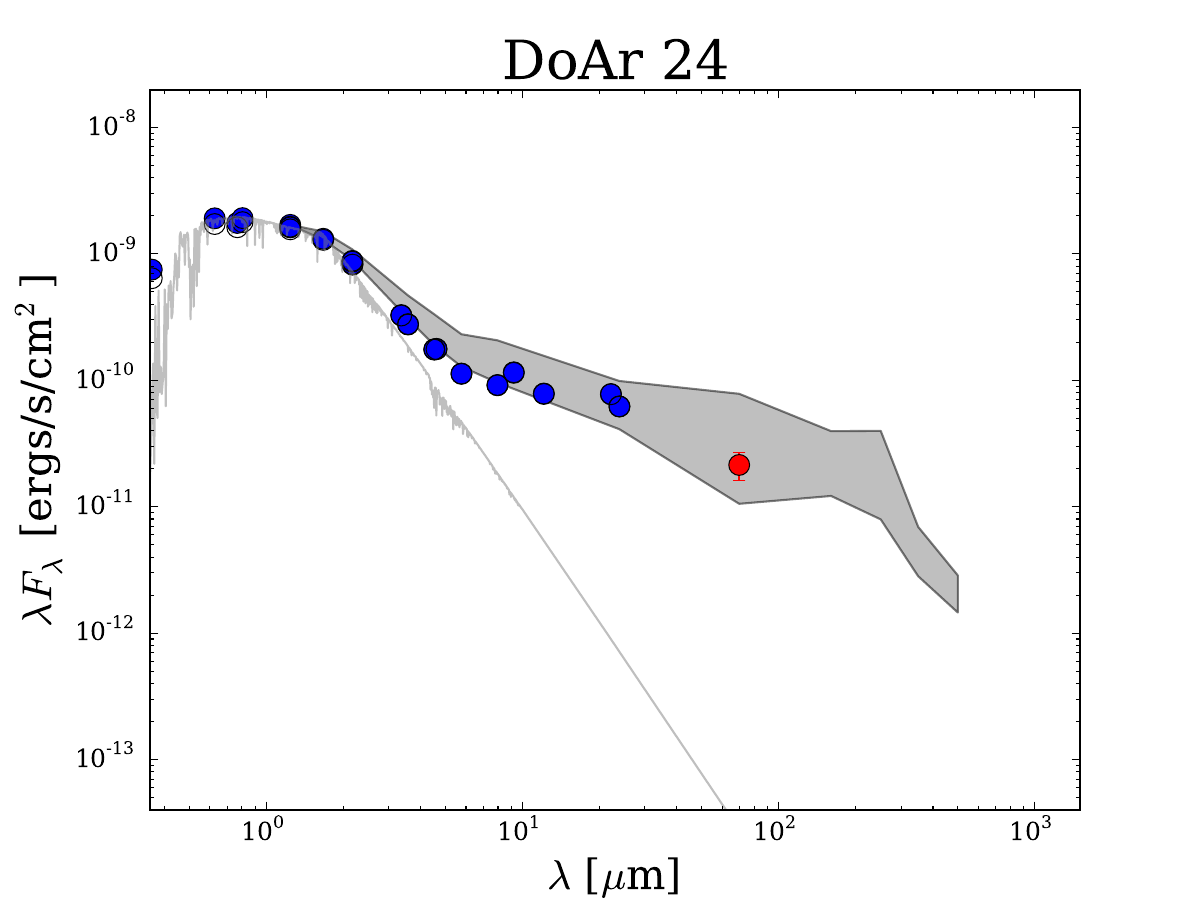}
    \includegraphics[width=0.3\textwidth,page=11]{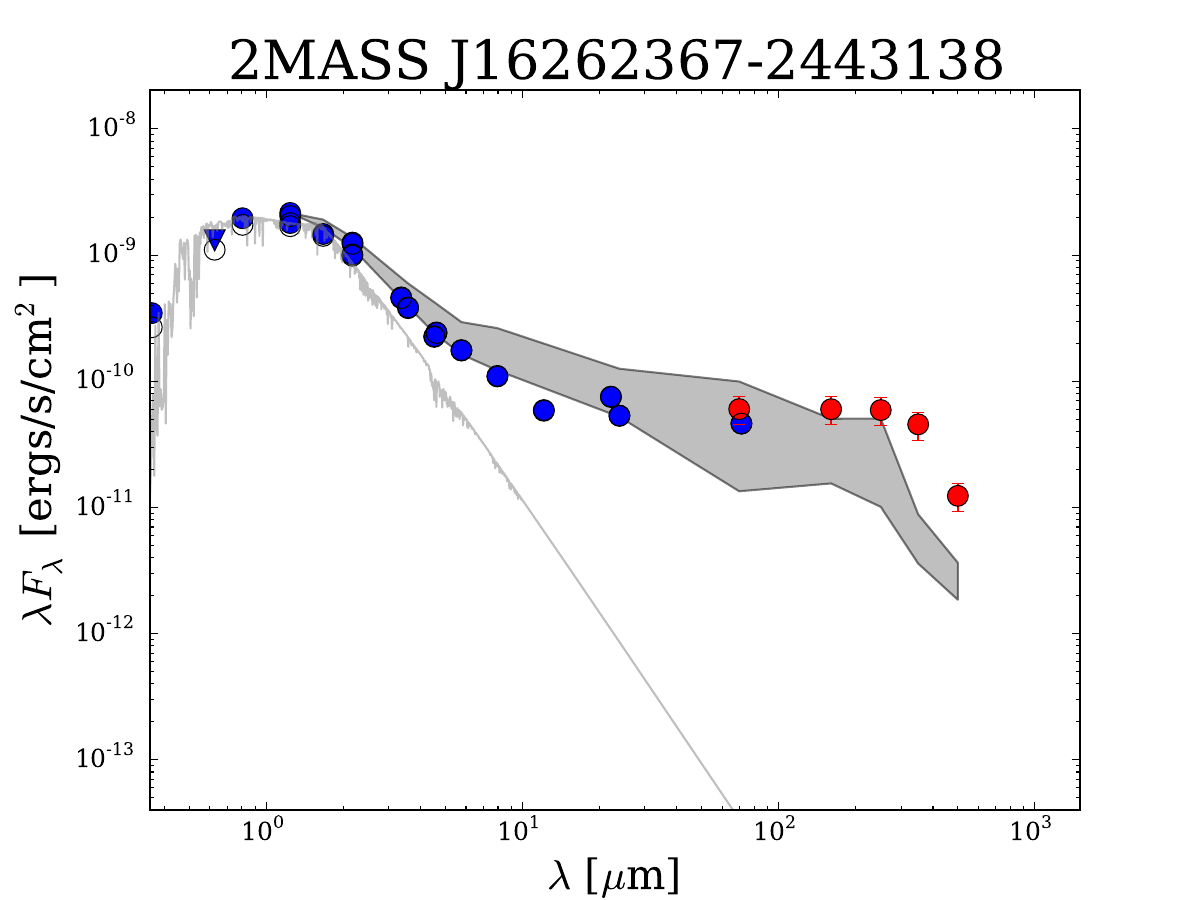}
    \includegraphics[width=0.3\textwidth,page=19]{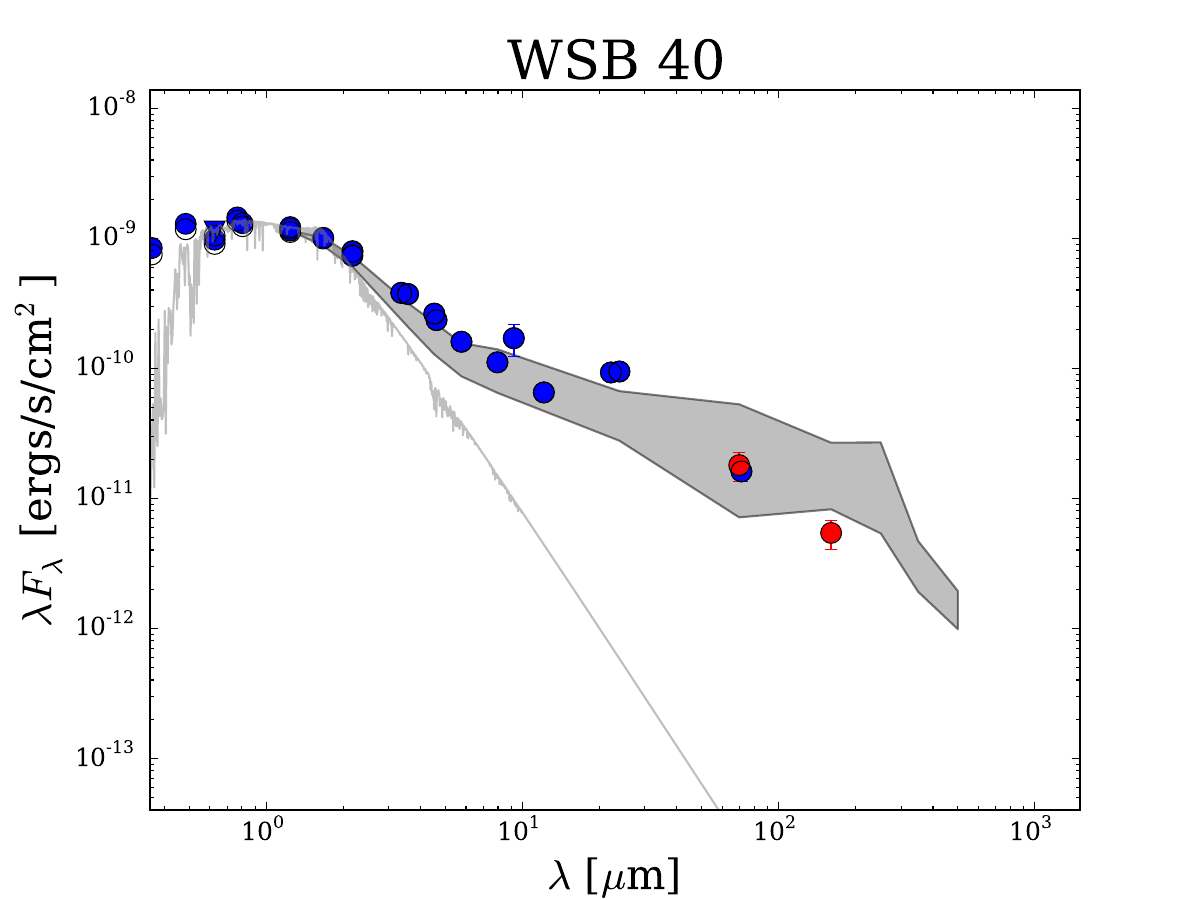}
    \includegraphics[width=0.3\textwidth,page=22]{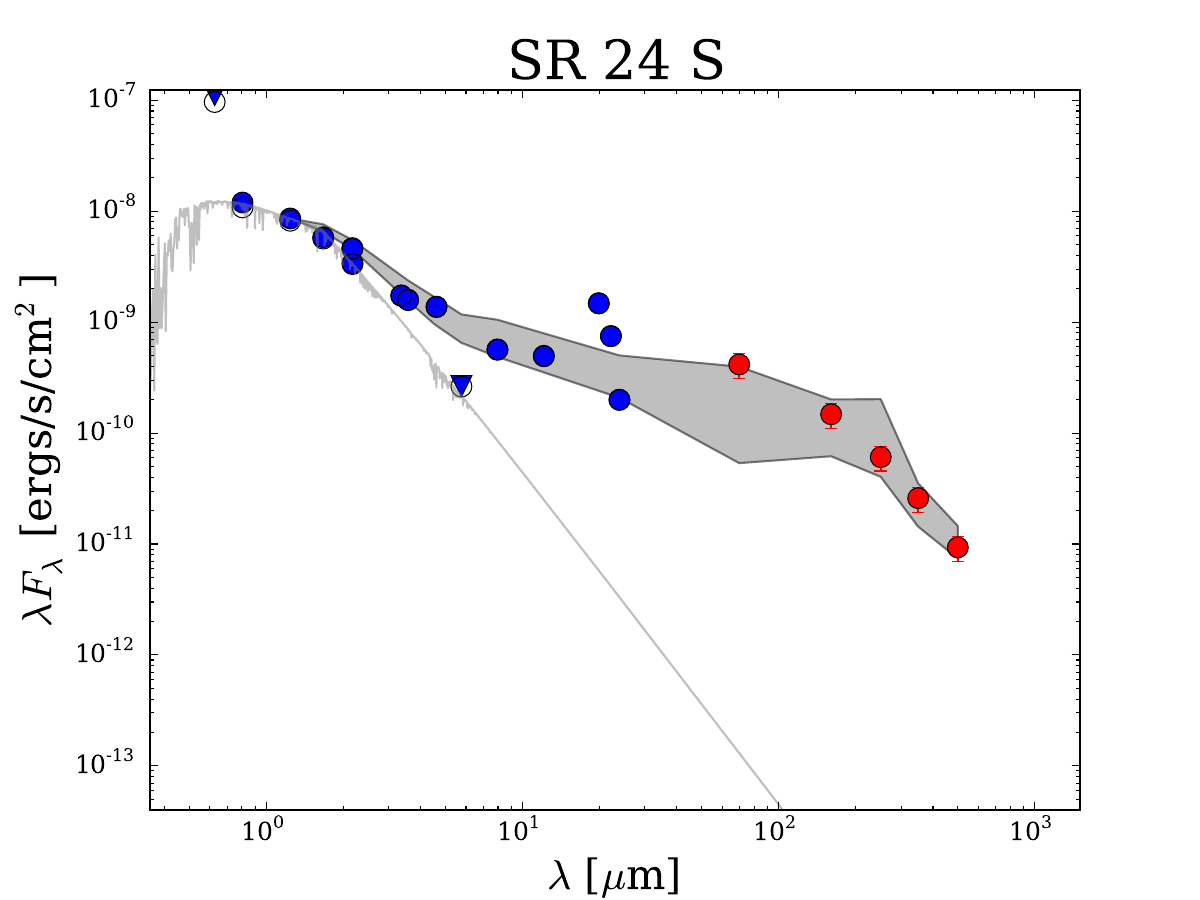}
    \includegraphics[width=0.3\textwidth,page=23]{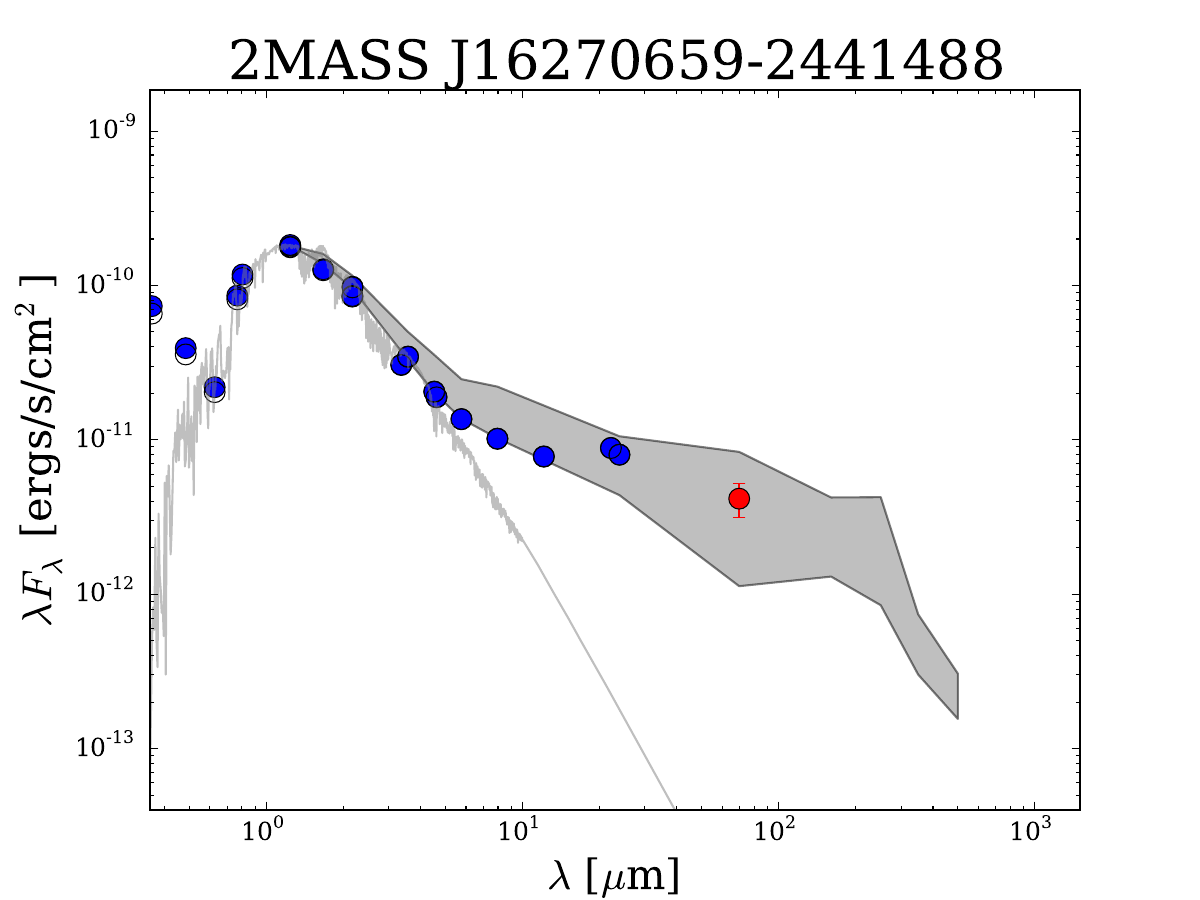}
    \includegraphics[width=0.3\textwidth,page=25]{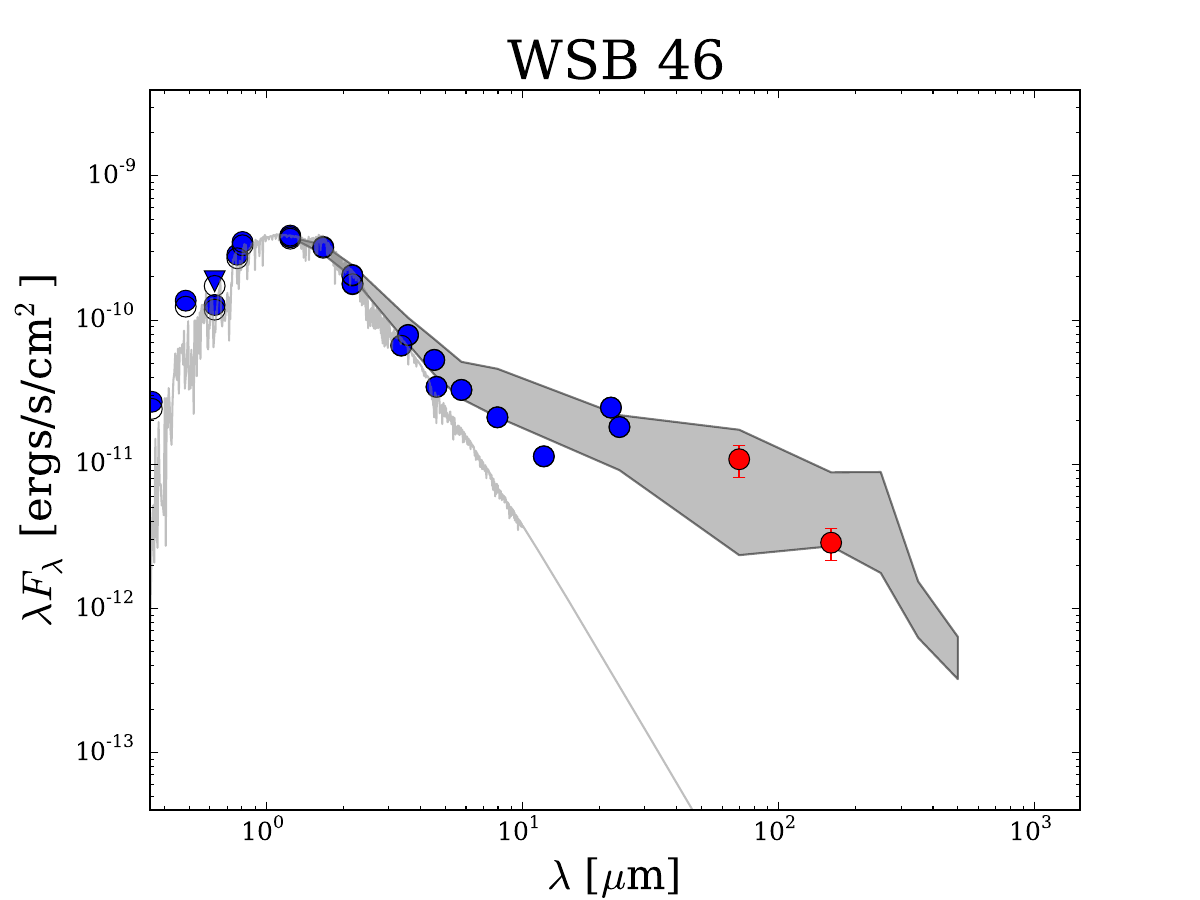}
    \includegraphics[width=0.3\textwidth,page=30]{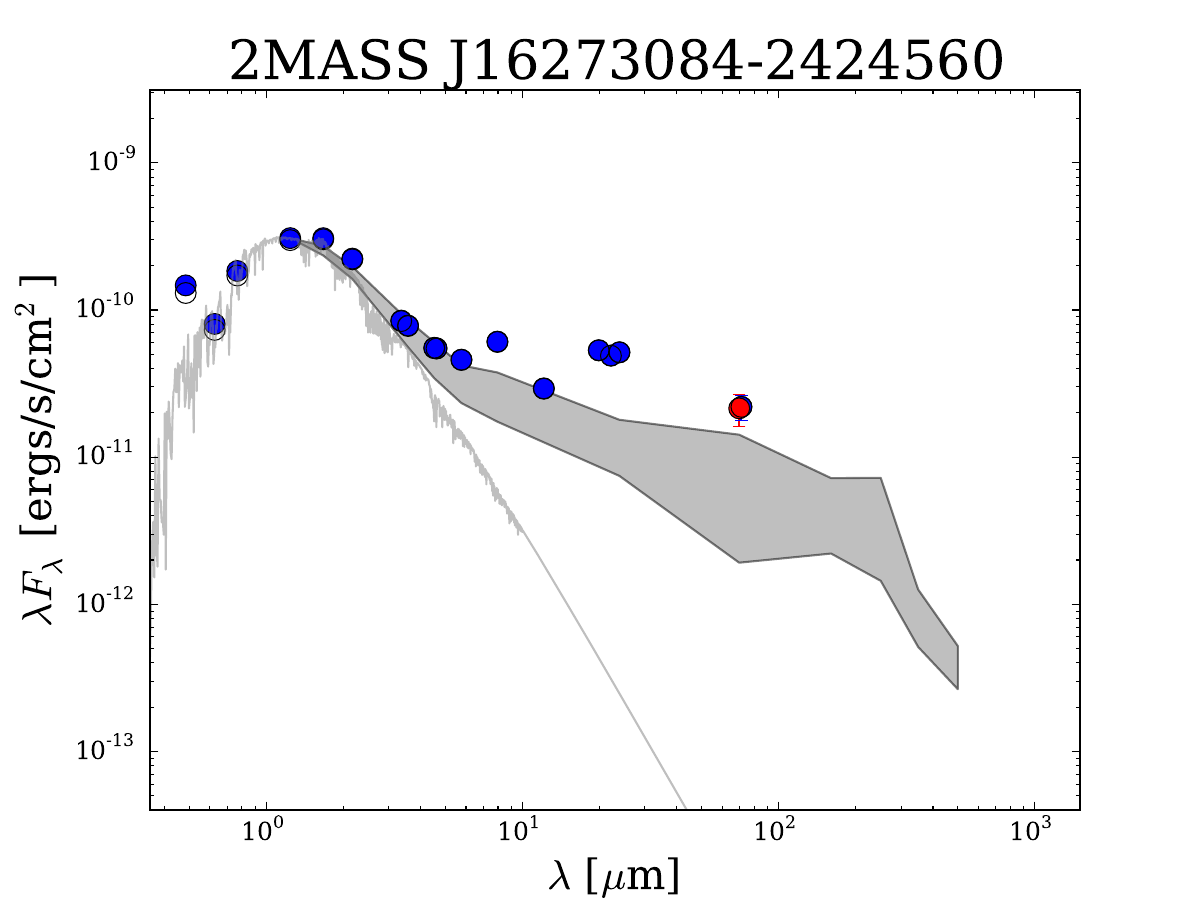}
    \includegraphics[width=0.3\textwidth,page=32]{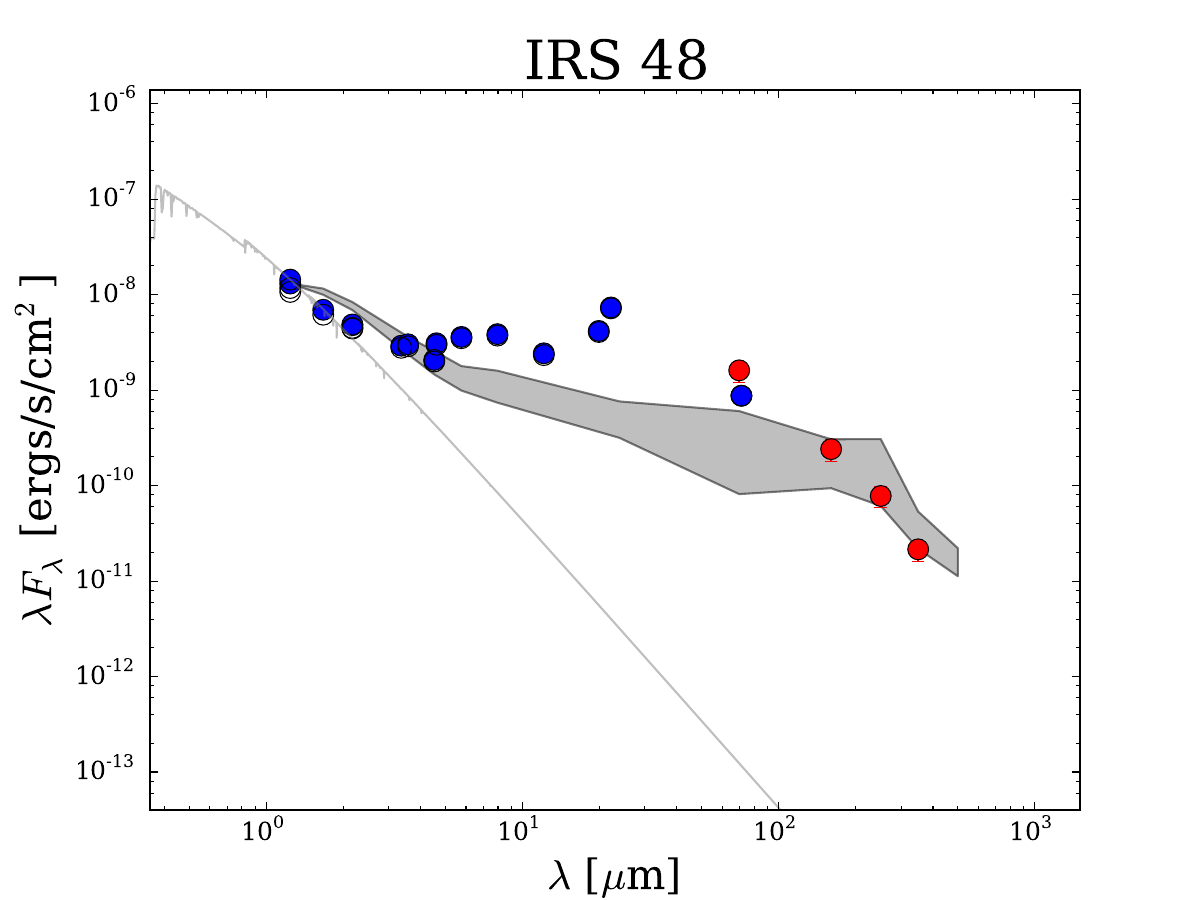}
    \includegraphics[width=0.3\textwidth,page=33]{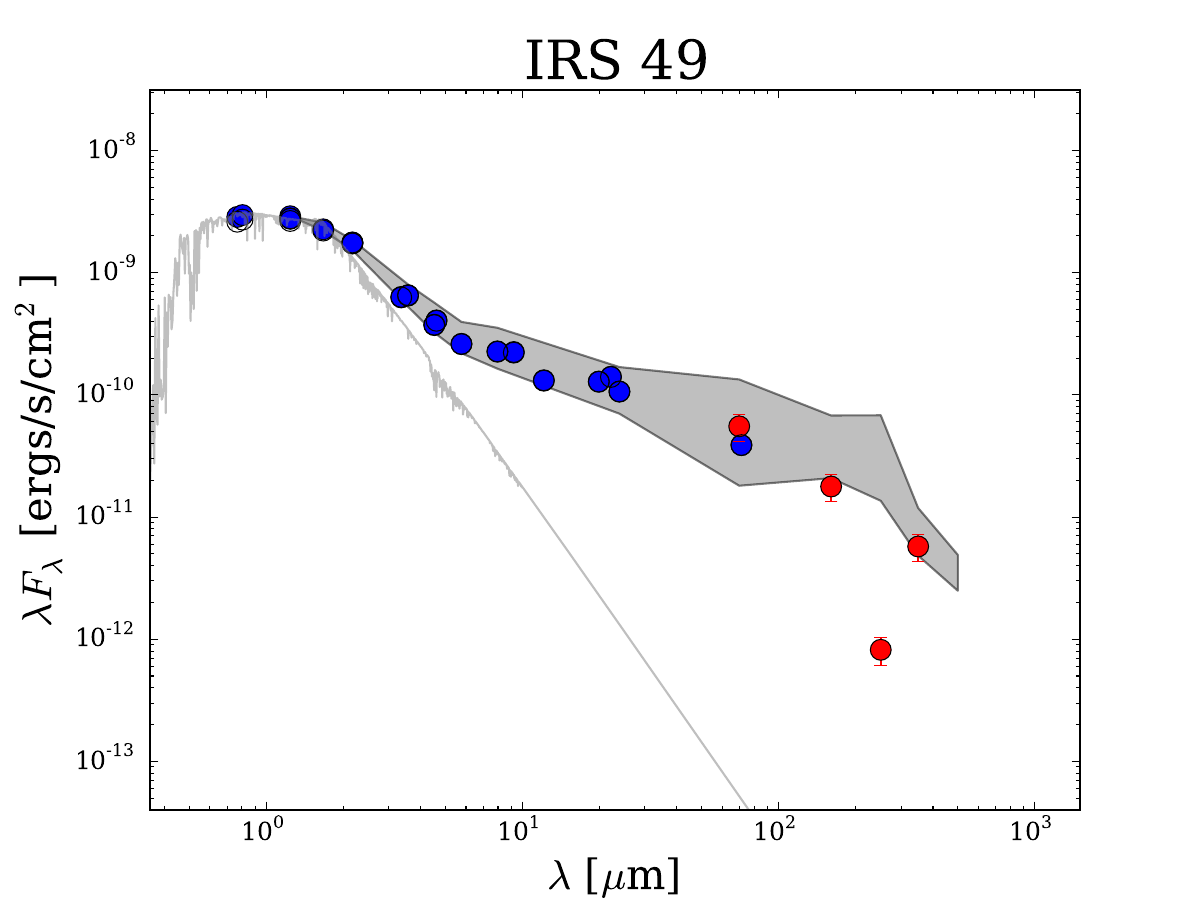}
    \includegraphics[width=0.3\textwidth,page=35]{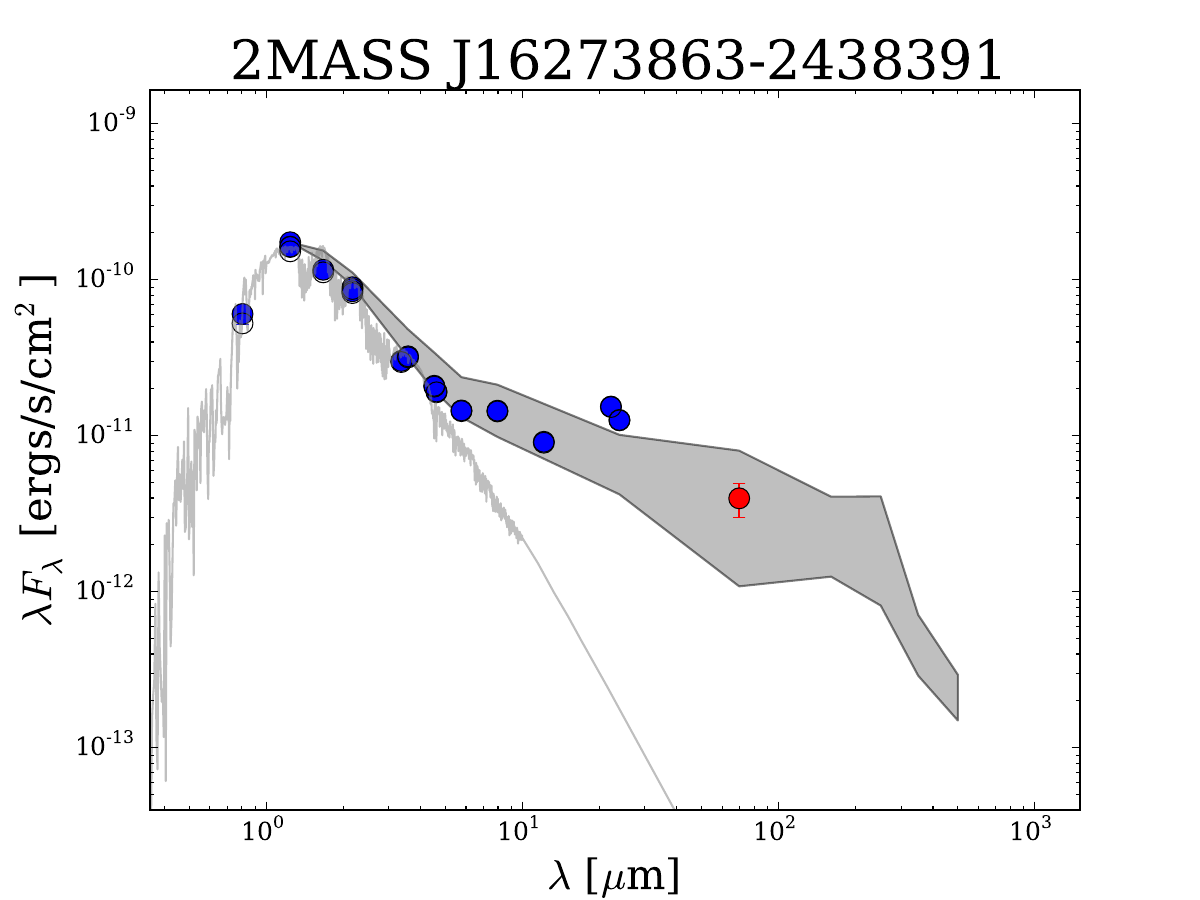}
    \includegraphics[width=0.3\textwidth,page=37]{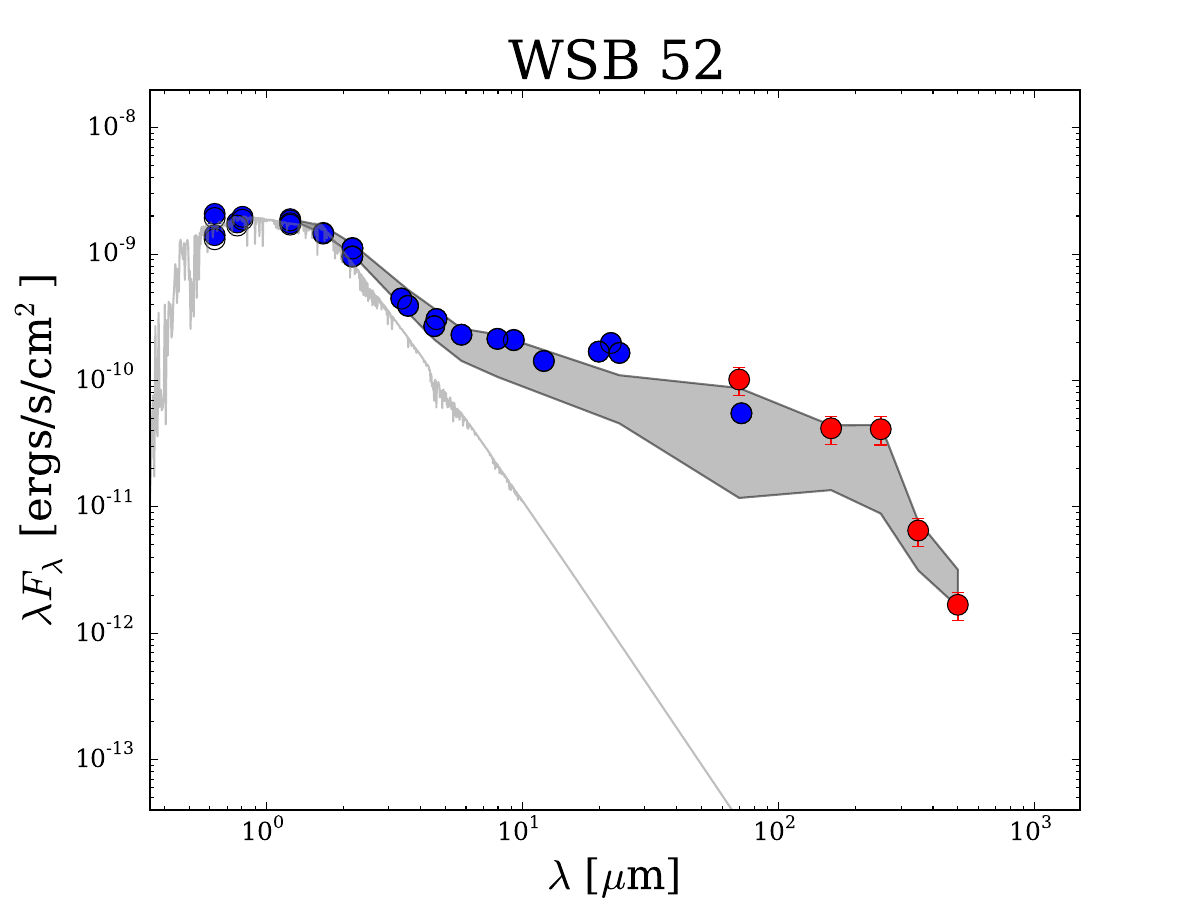}
    \includegraphics[width=0.3\textwidth,page=38]{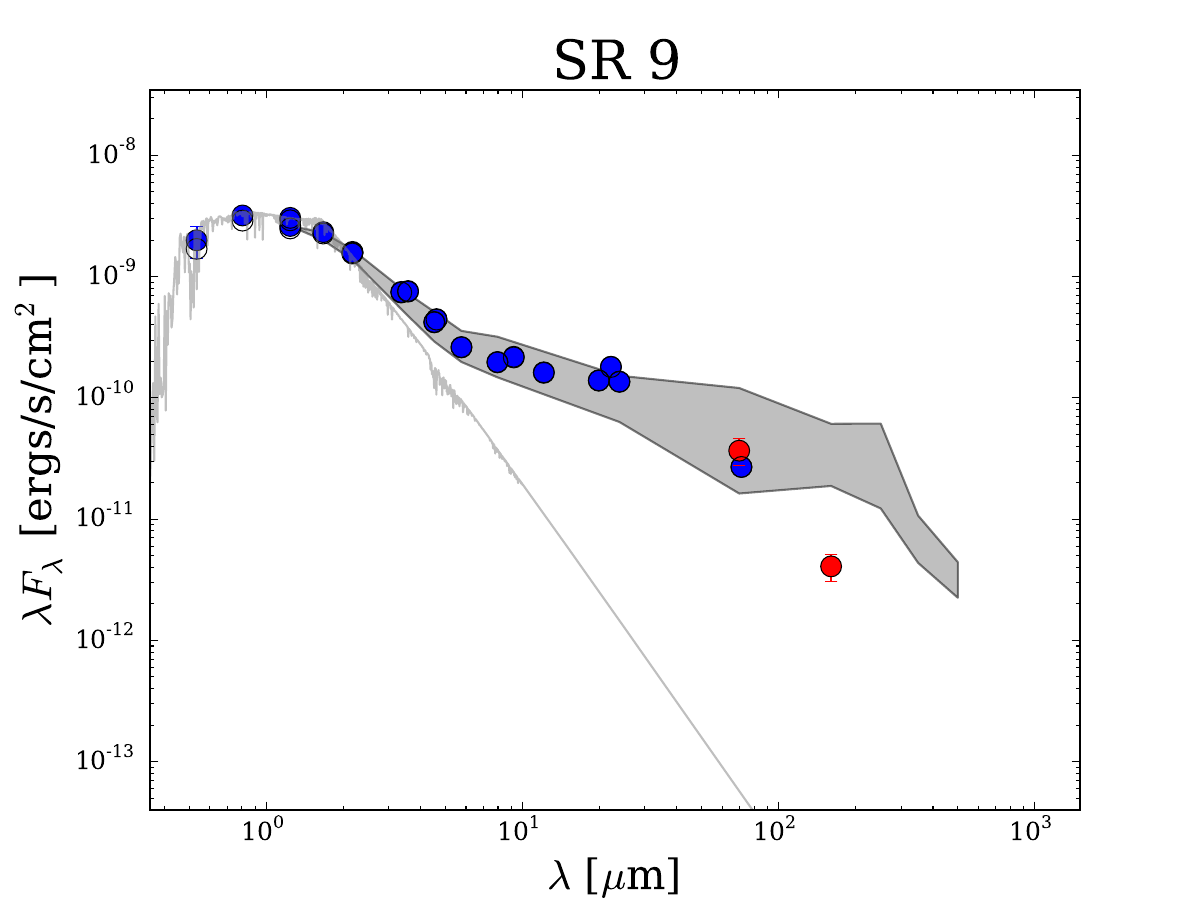}
    \includegraphics[width=0.3\textwidth,page=43]{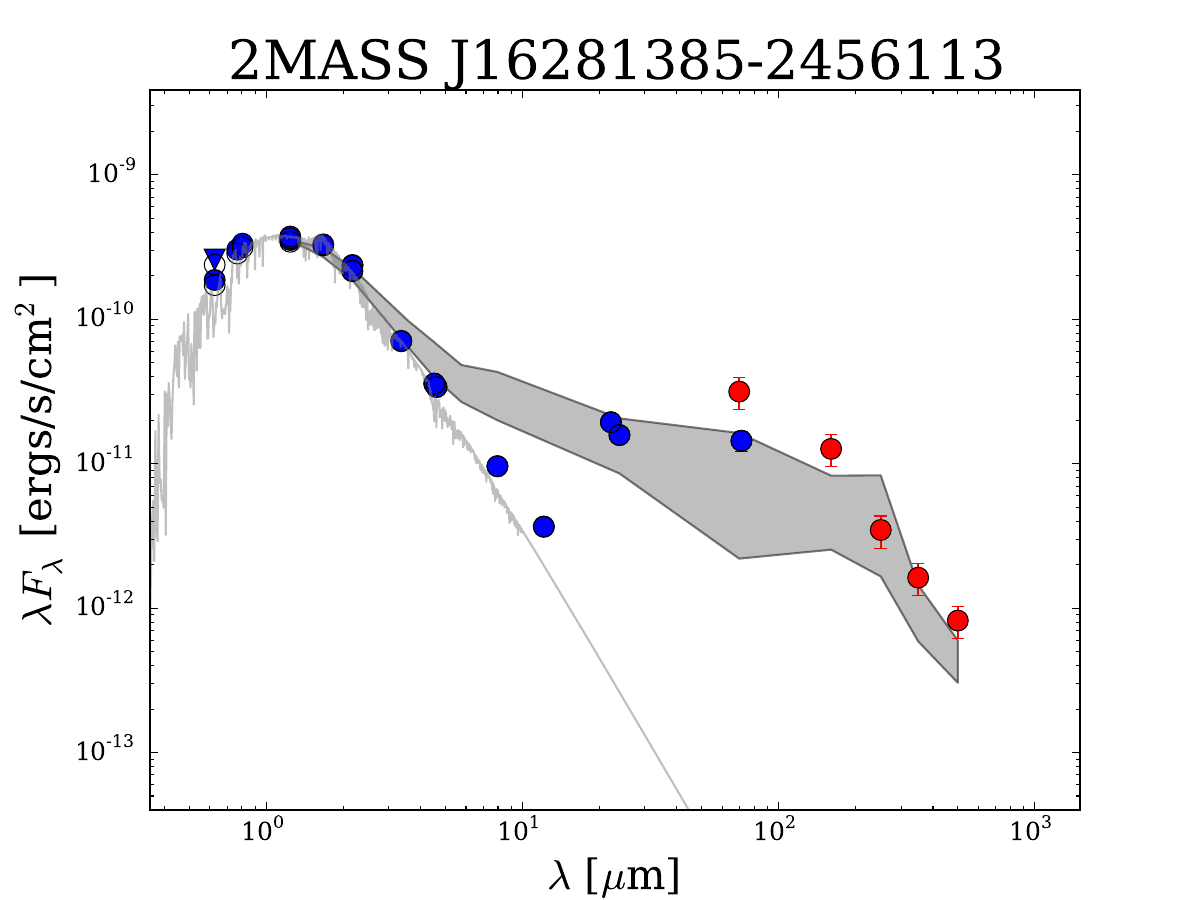}
    \includegraphics[width=0.3\textwidth,page=44]{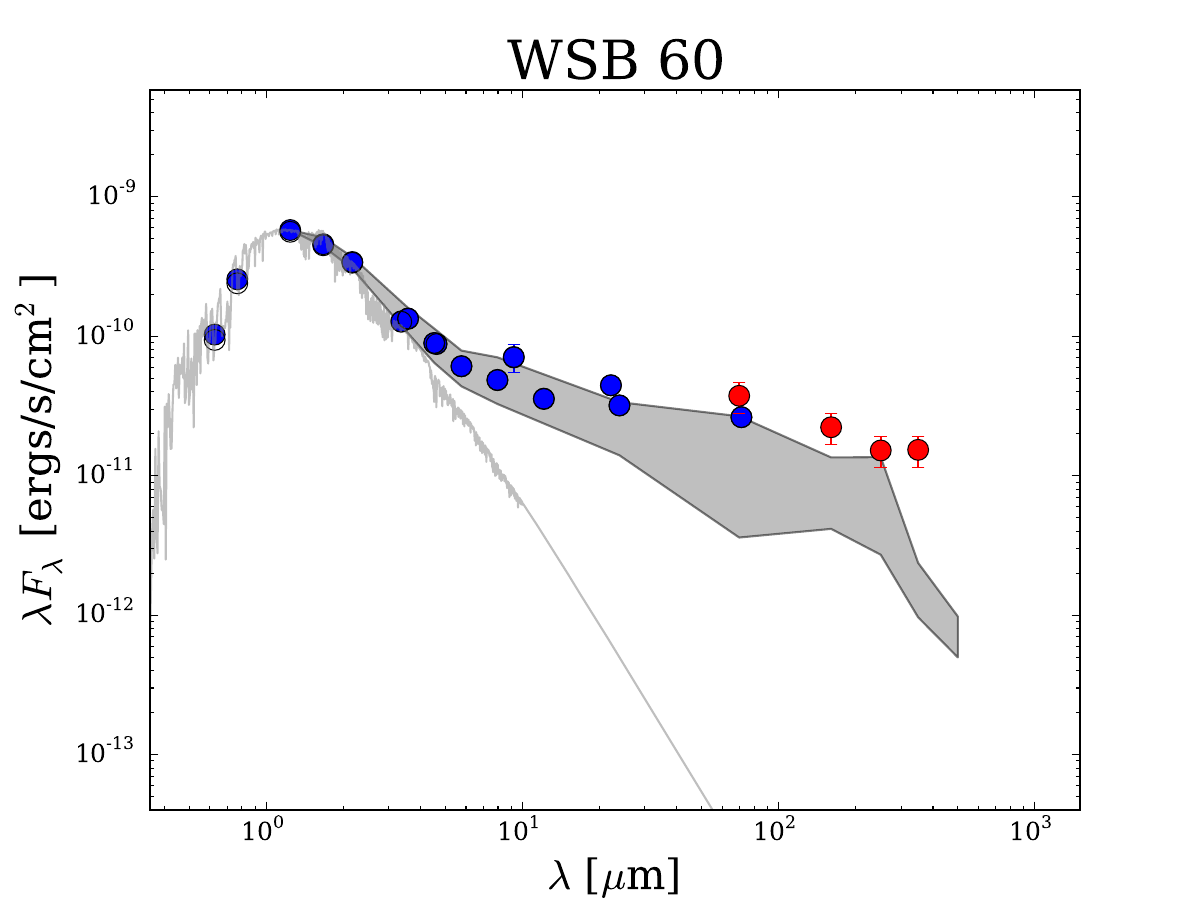}
    \includegraphics[width=0.3\textwidth,page=46]{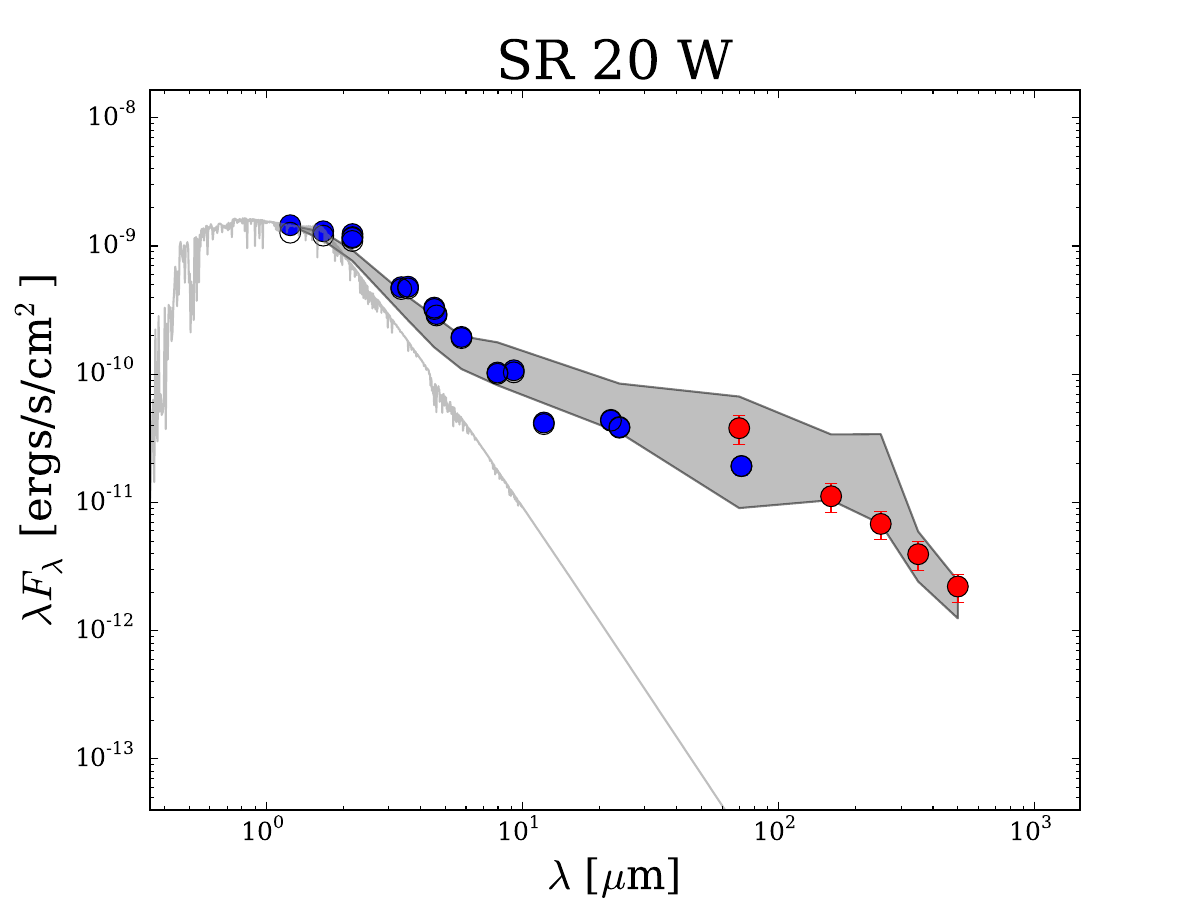}
    \includegraphics[width=0.3\textwidth,page=47]{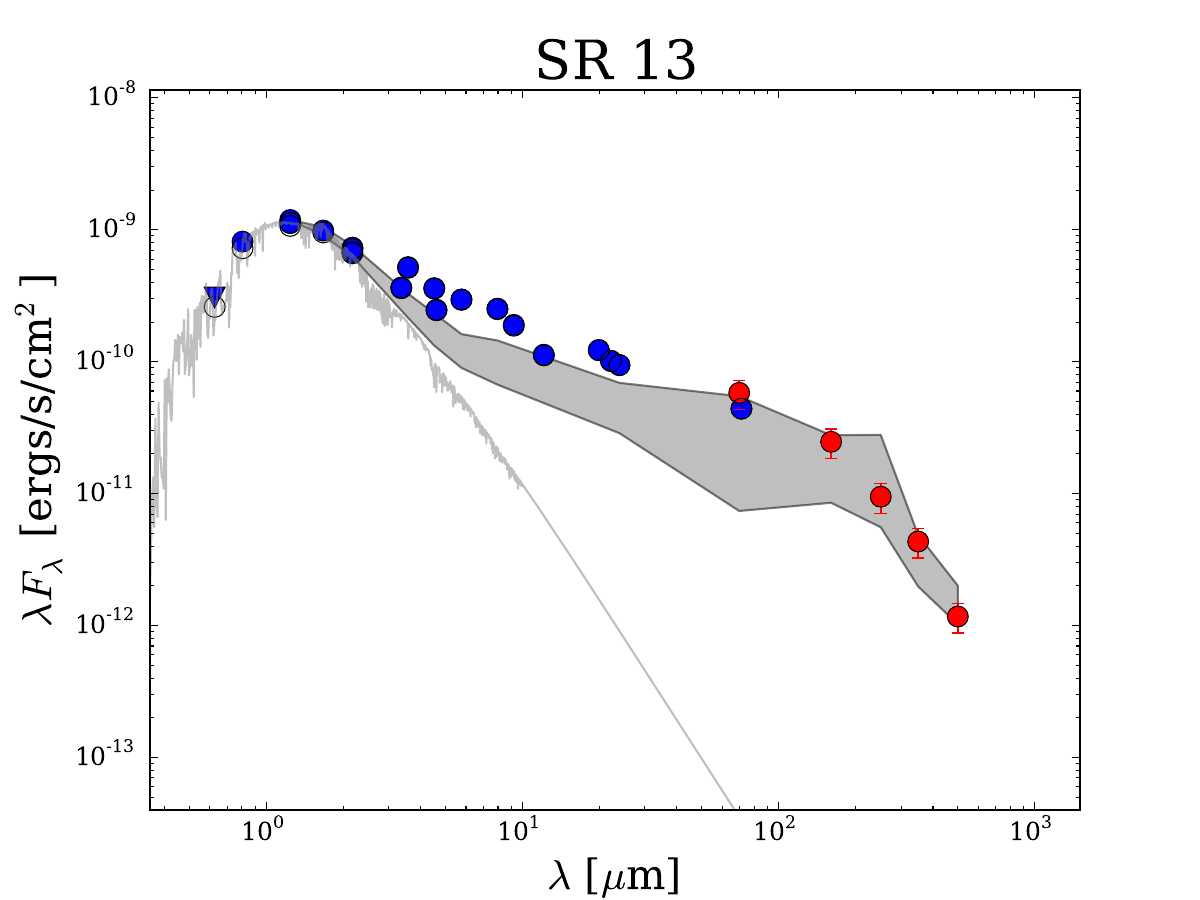}
    \includegraphics[width=0.3\textwidth,page=48]{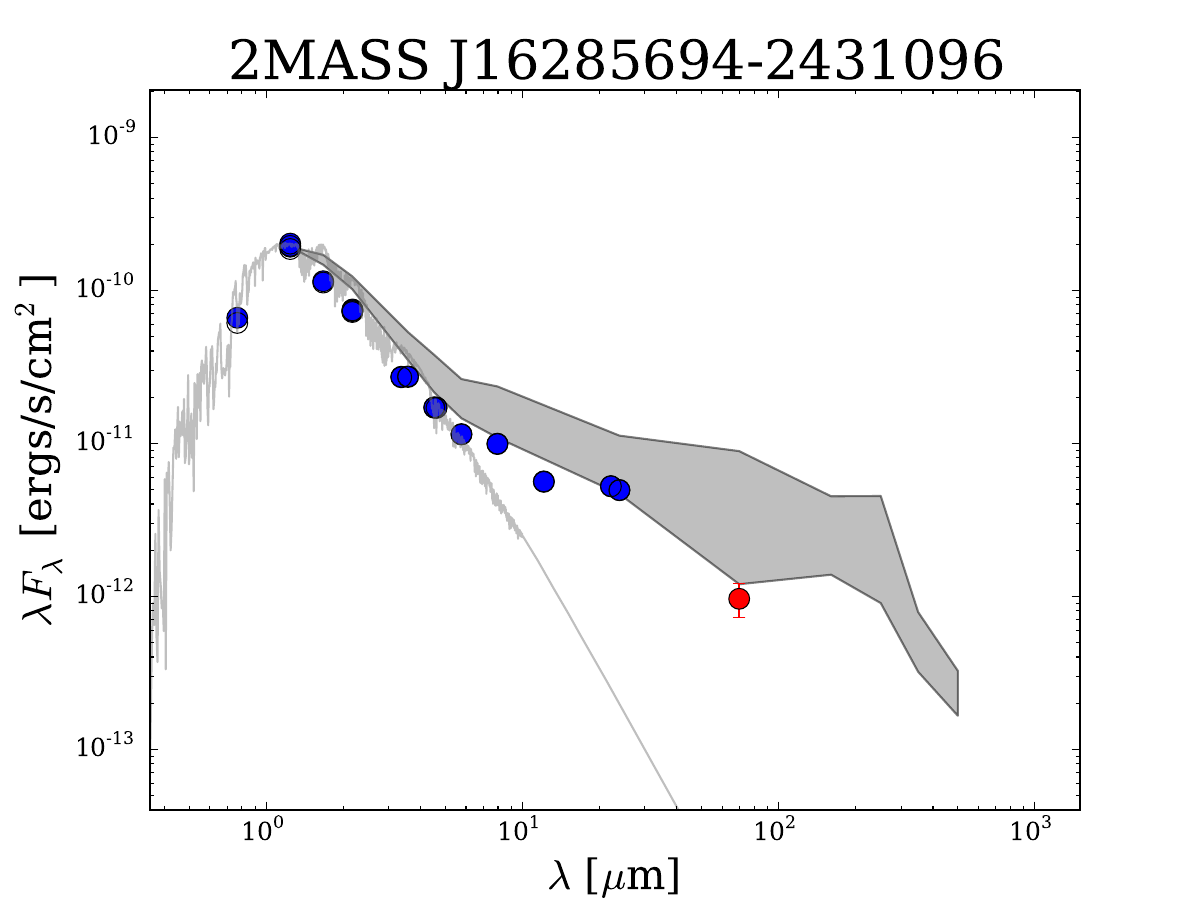}    
  \caption{Spectral energy distribution (SED) of the sources classified as transitional disks candidates. Blue dots show data acquired from the literature, red dots are photometric fluxes obtained from \textit{Herschel} data. Grey dashed line is the photosphere model according to the spectral type, and the grey shaded area is the filled area between the first and third quartile of all the disk fluxes. Observed fluxes are shown with empty circles and $A_v$ values used are in Table 3.}
  \label{fig:transitionalsed}
\end{figure*}

\section{Discussion}
\label{discussion}

\subsection{Detection statistics}

The initial sample was composed of 261 YSO objects in the centre of Ophiuchus, with known spectral type from optical spectroscopy and all classified as Class II objects. Our sample is different from that in \cite{Evans2009}, since they had photometrically selected objects, including many objects with other classes, while we have only spectroscopically confirmed Class II YSOs. All of these YSOs fell within the coverage of the maps used, and 49 were detected in at least one \textit{Herschel} band, 49 in PACS, and 19 in SPIRE. This leads to a \textit{Herschel} detection rate of $18.77\% \pm 2.6\%$, which is much smaller than the percentage of detections obtained in similar studies in Chamaeleon \citep{Ribas2013} and Lupus \citep{Bustamante2015} of around 30 \%. Given that Ophiuchus is closer than those regions (150$\sim$200 pc), the low detection rate is probably due to the higher background, which is emitting at mid- and long-IR wavelengths, and precludes the detection of faint objects. 

\subsection{Incidence of transitional disks in the centre of Ophiuchus}
We report here the detection of 18 transitional disk candidates in the cloud complex in the centre of Ophiuchus based on new \textit{Herschel} and previous known data of spectroscopically confirmed YSO sample. Despite the fact that all of these objects fulfil either one or both of the criteria exposed previously, only a few of them have evident changes of slopes in the SEDs. The candidate with the biggest change in slope, 2MASS J16281385-2456113 (\#14), is new to the literature.

The fraction of transitional disk candidates observed in Ophiuchus based on our \textit{Herschel} sample, is $37\% ^{+7}_{-6}$. We have considered two classification criteria, depending on the change of slope, that is, on the wavelength position of the lack of mid-IR excess. If we only consider  our main criteria, the fraction is lower, at $14\%^{+6}_{-4}$, but  it is compatible with the fractions measured  in other regions with similar ages in previous works \citep{Espaillat2014}. Even though we have identified  18 transitional disks candidates in total, some of them present a relatively flat slope and, in the slope-slope diagram, are represented  very close to, or even below,  the threshold . Having objects with nearly a Class-II slope explains the  high fraction of transitional disks in the region, as many of them might be fulfilling the criteria due to the large error in their PACS fluxes. These objects would need further study for their safe classification. One of the objects close to the threshold, however, namely YLW 58 or WSB 60  (\#15), had been imaged with SMA and shows a small but conspicuous inner hole \citep{Andrews2011}.
Despite this, most of the objects in the first criterion were also detected by the complementary criterion;  SR 42 S (\#5) was not, even though it is a confirmed transitional disk \citep{Andrews2011}. The fact that \textit{Herchel} data is including transitional candidates to the sample, confirms the improvement that \textit{Herschel} represents regarding reaching to further regions in disks. The longer wavelengths now accessible allow us to detect wider and larger cavities, which we could not have identified solely with  data from \textit{Spitzer} at mid-infrared ranges. 

\subsection{Other interesting objects}

Even though  we have combined two criteria to identify new candidates, and that these criteria select even very small changes of slope, there might be objects in the sample with the characteristics of a transitional disk, which are, as previously noted,  a lack of excess around 10 $\mu$m and normal excesses at mid- to far-IR. If we inspect Figs. 1 and 2 of the online appendix, we observe that several objects present these features.
There are two clear cases of objects previously classified as transitional disk candidates: 2MASS J16280256-2355035 and 2MASS J16294427-2441218. Both of these candidates present a lack of excess around 8 $\mu$m and excesses in longer wavelengths, but because of the restrictions in the criteria, neither appear as transitional candidates.
The objects  V* V852 Oph, 2MASS J16253958-2426349 and 2MASS J16262189-2444397 are similar cases that were never classified as transitional disks, but have a gap in their SED, which could be an indication of a hole in their disks. The object
2MASS J16272658-2425543 also shows a lack of excess around 10 $\mu$m with a larger excess in longer wavelengths. This object  could also be considered  a transitional candidate, despite the fact that it does not fulfil any of the criteria.
All of these objects are shown in Table 6 along with the previous objects classified as transitional disk candidates.

\subsection{Study of 70 micron fluxes in transitional disks}

For a comparison of the 70 $\mu$m flux of both the transitional disk candidates and tentative candidates, we constructed the median SED of all the detected objects in the sample, using the photometric data in Table 5. This would show if, apart from a lack of excess in the near mid-IR, transitional disks also show another remarkable characteristic that could be used in classification or, possibly, in disk modelling.
The median SED lacks the contribution from the faintest objects, which might remain undetected by \textit{Herschel} or some of the other surveys, and hence is just an upper limit to the true median SED of Ophiuchus.

When we overplot the median SED to the  detected fluxes (as a grey shaded area in Figs. 4 and 1 and 2 of the online appendix), we only find the  70 $\mu$m flux to be higher  for some of the objects classified as transitional disks candidates. Hence,  we cannot conclude that we have detected a trend in the transitional disk population, such as as has been observed in previous investigations \citep[][\cite{Bustamante2015}]{Ribas2013}. We also find the case of 2MASS J16285694-2431096 (\#18), where the 70 $\mu$m flux is not only lower than the median SED, but lower than the third quartile. Because of the lack of large excess in mid- to far-IR and the fact that this object presents a slightly smaller flux in the 70 $\mu$m band than the rest of transitional disks candidates, it needs further study to clarify its nature. 

\section{Conclusions}
\label{conclusions}

We have detected 49 objects in the central region of Ophiuchus in at least one PACS band and 19 in at least one SPIRE band. We obtained accurate photometric fluxes for the detected objects by means of aperture photometry.

Seven of the detected objects were classified as transitional disk candidates by the criterion in \cite{Ribas2013}, and 11 more were classified according to the complementary criterion, generating a total sample of 18 transitional disk candidates. Some of these candidates were already imaged in previous works, and hence, confirmed. Six more objects are added to the final classification of candidates that have transitional features in their SEDs, rather than fulfilling any of the criterion. All of the transitional disk candidates are shown in Table 6 along with their classification criteria. This large difference between the identification methods can be due to the different nature or evolutionary stages of the disks, creating different geometries and leading to a large diversity of SEDs.

Several of the objects classified as transitional disks candidates have not been considered candidates before, but 2MASSJ16281385-2456113 (\#14) appears to be an attractive object for follow-up because of its prominent change of slope when compared to the rest of the sample, including previously imaged disks, such as IRS 48 (\#9). 

So far, \textit{Herschel} data has proved to be very useful to improve the characterization of the outer regions of protoplanetary systems because of its long-wavelength coverage, unattainable until now, and its improved sensitivity and spatial resolution compared with previous IR missions. A study of the SED population of the disk sample detected with \textit{Herschel} should give us more information on the true nature of these disks, but this study is outside the scope of this work.

\begin{table}
\small
\centering
\caption{Summary of all transitional disk candidates in Ophiuchus}
\begin{tabular}[width=1\textwidth]{lc c cl}
\hline\hline
I.D. & Name & Classification Criteria \\
\hline

$1$  &   Haro1-6 & $12-70$*\\
$2$  &   DoAr24 & $12-24$ \\
$3$  &   DoAr25 & $12-70$*\\
$4$  &   WSB40 & $12-24$ \\
$5$  &   SR42S & $12-70$ \\
$6$  &   2MASSJ16270659-2441488 & $12-24$ \\
$7$  &   WSB46 & $12-70$*\\
$8$  &   2MASSJ16273084-2424560 & $12-24$ \\
$9$  &   IRS48 & $12-24$ \\
$10$ &   IRS49 & $12-24$ \\
$11$ &   2MASSJ16273863-2438391 & $12-24$ \\
$12$ &   WSB52 & $12-24$ \\
$13$ &   SR9 & $12-24$ \\
$14$ &   2MASSJ16281385-2456113 $\dagger$ & $12-70$* \\
$15$ &   WSB60 & $12-70$* \\
$16$ &   SR20W & $12-70$* \\
$17$ &   SR13 & $12-24$ \\
$18$ &   2MASSJ16285694-2431096 & $12-24$ \\
\hline
-   &   V*V852Oph & SED \\
-   &   2MASSJ16253958-2426349 & SED \\
-   &   2MASSJ16262189-2444397 & SED\\
-   &   2MASSJ16272658-2425543 & SED \\
-   &   2MASSJ16280256-2355035 & SED \\
-   &   2MASSJ16294427-2441218 & SED \\
\hline

\end{tabular}
\tablefoot{ *These objects have also been classified with the complementary criterion of 12-24 \\     $\dagger$ This transitional disk candidate is new to the literature.} \\
\end{table}

\begin{acknowledgements}
We thank the referee for his/her constructive comments. This work has been possible thanks to the ESAC Space Science Faculty for funding with code ESAC-321, ESAC Traineeship program and of the Herschel Science Centre. We thank G\'abor Marton for useful discussions on Herschel data evaluation. This work was partly supported by the Hungarina research grant OTKA 101393, and by the Momentum grant of the MTA CSFK Lendület Disk Research Group. PACS was developed by a consortium of institutes led by MPE (Germany),  including UVIE (Austria); KUL, CSL, IMEC (Belgium); CEA, OAMP (France); MPIA (Germany); IFSI, OAP/AOT, OAA/CAISMI, LENS, SISSA (Italy); IAC (Spain). This development has been supported by the funding agencies BMVIT (Austria), ESA-PRODEX (Belgium), CEA/CNES (France), DLR (Germany), ASI (Italy), and CICT/MCT (Spain). SPIRE was developed by a consortium of institutes led by Cardiff Univ. (UK), including Univ. Lethbridge (Canada); NAOC (China); CEA, LAM (France); IFSI, Univ. Padua (Italy); IAC (Spain); Stockholm Observatory (Sweden); Imperial College London, RAL, UCL-MSSL, UKATC, Univ. Sussex (UK); and Caltech, JPL, NHSC, Univ. Colorado (USA). This development has been supported by national funding agencies: CSA (Canada); NAOC (China); CEA, CNES, CNRS (France); ASI (Italy); MCINN (Spain); SNSB (Sweden); STFC (UK); and NASA (USA).This study also makes use of the data products from the Two Micron All Sky Survey (2MASS), a joint project of the University of Massachusetts and IPAC/Caltech, funded by NASA and the National Science Foundation; data products from the Wide-field Infrared Survey Explorer (WISE), a joint project of the University of California, Los Angeles, and the Jet Propulsion Laboratory (JPL)/California Institute of Technology (Caltech); data products from DENIS, a project partly funded by the SCIENCE and the HCM plans of the European Commission under grants CT920791 and CT940627; the NASA Infrared Processing and Analysis Center (IPAC) Science Archive; and the SIMBAD database.
\end{acknowledgements}

\bibliographystyle{aa}
\bibliography{mybiblio}

\clearpage

\begin{appendix}
\section{Appendix}

\begin{table*}

\centering
\caption{Estimation of contaminating flux contained in the aperture for each band, according to the median SED extrapolation.}
\begin{tabular}[width=1\textwidth]{lc c c c c c c c cl}
\hline\hline
I.D. & Name & F$_{70} \,[Jy] $   & F$_{160}\,[Jy]$  & F$_{250}\,[Jy]$  & F$_{350}\,[Jy]$  & F$_{500}\,[Jy] $ \\
    & & $(6")$& $(12")$& $(22")$& $(30")$& $(42")$& \\
\hline
-    &   2MASSJ16245974-2456008 & - & $2.66\cdot10^{-5}$ & $2.94\cdot10^{-4}$ & $9.00\cdot10^{-5}$ & $2.77\cdot10^{-5}$ \\
-    &   V*V852Oph & - & $1.44\cdot10^{-4}$ & $1.48\cdot10^{-4}$ & $4.96\cdot10^{-5}$ & $6.11\cdot10^{-5}$ \\
-    &   2MASSJ16253673-2415424 & - & - & $4.47\cdot10^{-5}$ & $1.23\cdot10^{-5}$ & $3.17\cdot10^{-5}$ \\
-    &   2MASSJ16253958-2426349 & - & - & $9.65\cdot10^{-4}$ & $4.60\cdot10^{-4}$ & $3.15\cdot10^{-4}$ \\
-    &   V*V2058Oph & $3.34\cdot10^{-2}$ & $1.12\cdot10^{-2}$ & $5.29\cdot10^{-3}$ & $1.46\cdot10^{-3}$ & $3.97\cdot10^{-4}$ \\
$1$  &   Haro1-6 & $5.85\cdot10^{-2}$ & $1.55\cdot10^{-2}$ & $7.64\cdot10^{-3}$ & $2.15\cdot10^{-3}$ & $6.09\cdot10^{-4}$ \\
-    &   2MASSJ16261684-2422231 & - & - & $1.01\cdot10^{-5}$ & $1.15\cdot10^{-5}$ & $1.48\cdot10^{-5}$ \\
$2$  &   DoAr24 & $4.51\cdot10^{-3}$ & $9.56\cdot10^{-4}$ & $4.50\cdot10^{-4}$ & $1.32\cdot10^{-4}$ & $4.03\cdot10^{-5}$ \\
-    &   2MASSJ16262189-2444397 & $5.90\cdot10^{-3}$ & $1.27\cdot10^{-3}$ & $5.98\cdot10^{-4}$ & $2.07\cdot10^{-4}$ & $7.49\cdot10^{-5}$ \\
-    &   DoAr24E & - & - & - & $1.41\cdot10^{-4}$ & $3.85\cdot10^{-5}$ \\
$3$  &   DoAr25 & - & $1.23\cdot10^{-4}$ & $8.50\cdot10^{-5}$ & $3.09\cdot10^{-5}$ & $2.91\cdot10^{-5}$ \\
-    &   GSS32 & - & $3.76\cdot10^{-2}$ & $1.77\cdot10^{-2}$ & $5.67\cdot10^{-3}$ & $1.61\cdot10^{-3}$ \\
-    &   2MASSJ16262407-2416134 & - & - & - & $6.89\cdot10^{-5}$ & $2.34\cdot10^{-5}$ \\
-    &   2MASSJ16263297-2400168 & - & - & $1.32\cdot10^{-5}$ & $1.86\cdot10^{-5}$ & $1.06\cdot10^{-5}$ \\
-    &   2MASSJ16263682-2415518 & - & $1.39\cdot10^{-3}$ & $6.68\cdot10^{-4}$ & $2.07\cdot10^{-4}$ & $6.18\cdot10^{-5}$ \\
-    &   $[GY92]93$ & - & - & $1.36\cdot10^{-5}$ & $1.12\cdot10^{-5}$ & $1.59\cdot10^{-5}$ \\
-    &   2MASSJ16264285-2420299 & - & - & - & $2.40\cdot10^{-4}$ & $1.60\cdot10^{-4}$ \\
-    &   2MASSJ16264643-2412000 & - & $3.32\cdot10^{-4}$ & $1.65\cdot10^{-4}$ & $4.94\cdot10^{-5}$ & $1.69\cdot10^{-5}$ \\
$4$  &   WSB40 & $2.09\cdot10^{-2}$ & $4.43\cdot10^{-3}$ & $2.09\cdot10^{-3}$ & $5.86\cdot10^{-4}$ & $1.65\cdot10^{-4}$ \\
-    &   WL18 & - & - & $4.75\cdot10^{-6}$ & $5.77\cdot10^{-6}$ & $5.37\cdot10^{-6}$ \\
-    &   2MASSJ16265677-2413515 & $1.59\cdot10^{-3}$ & $3.36\cdot10^{-4}$ & $1.58\cdot10^{-4}$ & $5.43\cdot10^{-5}$ & $2.13\cdot10^{-5}$ \\
$5$  &   SR24S & $3.92\cdot10^{-1}$ & $8.29\cdot10^{-2}$ & $3.91\cdot10^{-2}$ & $1.08\cdot10^{-2}$ & $2.95\cdot10^{-3}$ \\
$6$  &   2MASSJ16270659-2441488 & - & $2.93\cdot10^{-5}$ & $2.17\cdot10^{-5}$ & $1.28\cdot10^{-5}$ & $1.23\cdot10^{-5}$ \\
-    &   2MASSJ16270907-2412007 & - & - & $2.68\cdot10^{-5}$ & $1.86\cdot10^{-5}$ & $8.30\cdot10^{-6}$ \\
$7$  &   WSB46 & - & $4.76\cdot10^{-5}$ & $2.46\cdot10^{-5}$ & $1.65\cdot10^{-5}$ & $8.68\cdot10^{-6}$ \\
-    &   $[WMR2005]4-10$ & $5.65\cdot10^{-4}$ & $1.20\cdot10^{-4}$ & $7.63\cdot10^{-5}$ & $5.06\cdot10^{-5}$ & $2.52\cdot10^{-5}$ \\
-    &   2MASSJ16271836-2454537 & - & - & $7.27\cdot10^{-6}$ & $3.56\cdot10^{-6}$ & $3.40\cdot10^{-6}$ \\
-    &   WSB49 & - & - & $1.78\cdot10^{-5}$ & $6.58\cdot10^{-6}$ & $4.63\cdot10^{-6}$ \\
-    &   2MASSJ16272658-2425543 & - & - & $4.66\cdot10^{-5}$ & $3.35\cdot10^{-5}$ & $1.53\cdot10^{-5}$ \\
$8$  &   2MASSJ16273084-2424560 & - & - & $1.72\cdot10^{-5}$ & $1.30\cdot10^{-5}$ & $5.67\cdot10^{-6}$ \\
-    &   2MASSJ16273311-2441152 & - & - & - & $3.20\cdot10^{-6}$ & $5.10\cdot10^{-6}$ \\
$9$  &   IRS48 & - & - & - & $1.20\cdot10^{-4}$ & $4.42\cdot10^{-5}$ \\
$10$ &   IRS49 & - & - & $6.90\cdot10^{-6}$ & $7.97\cdot10^{-6}$ & $6.54\cdot10^{-6}$ \\
-    &   2MASSJ16273832-2357324 & - & - & $7.64\cdot10^{-4}$ & $2.19\cdot10^{-4}$ & $6.12\cdot10^{-5}$ \\
$11$ &   2MASSJ16273863-2438391 & $6.27\cdot10^{-3}$ & $1.62\cdot10^{-3}$ & $5.83\cdot10^{-5}$ & $3.59\cdot10^{-5}$ & $1.51\cdot10^{-3}$ \\
-    &   2MASSJ16273901-2358187 & - & - & $1.51\cdot10^{-4}$ & $5.32\cdot10^{-5}$ & $2.76\cdot10^{-5}$ \\
$12$ &   WSB52 & $1.45\cdot10^{-3}$ & $3.08\cdot10^{-4}$ & $1.56\cdot10^{-4}$ & $1.61\cdot10^{-5}$ & $1.25\cdot10^{-4}$ \\
$13$ &   SR9 & - & - & $1.70\cdot10^{-5}$ & $1.59\cdot10^{-5}$ & $1.02\cdot10^{-5}$ \\
-    &   2MASSJ16274270-2438506 & - & - & $5.44\cdot10^{-6}$ & $4.54\cdot10^{-6}$ & $6.99\cdot10^{-6}$ \\
-    &   V*V2059Oph & - & - & - & $8.23\cdot10^{-6}$ & $7.76\cdot10^{-6}$ \\
-    &   2MASSJ16280256-2355035 & - & - & $6.06\cdot10^{-8}$ & $5.08\cdot10^{-6}$ & $5.56\cdot10^{-6}$ \\
-    &   2MASSJ16281379-2432494 & - & - & $5.82\cdot10^{-6}$ & $4.25\cdot10^{-6}$ & $7.35\cdot10^{-6}$ \\
$14$ &   2MASSJ16281385-2456113 & - & - & $1.02\cdot10^{-4}$ & $2.94\cdot10^{-5}$ & $1.63\cdot10^{-5}$ \\
$15$ &   WSB60 & - & $2.13\cdot10^{-4}$ & $1.10\cdot10^{-5}$ & $1.00\cdot10^{-5}$ & $6.31\cdot10^{-6}$ \\
-    &   2MASSJ16281673-2405142 & - & - & - & $7.31\cdot10^{-6}$ & $8.51\cdot10^{-6}$ \\
$16$ &   SR20W & - & - & $4.36\cdot10^{-6}$ & $5.78\cdot10^{-6}$ & $6.48\cdot10^{-6}$ \\
$17$ &   SR13 & - & - & - & $4.01\cdot10^{-5}$ & $1.51\cdot10^{-5}$ \\
$18$ &   2MASSJ16285694-2431096 & - & - & $4.67\cdot10^{-6}$ & $1.83\cdot10^{-5}$ & $7.36\cdot10^{-6}$ \\
-    &   2MASSJ16294427-2441218 & - & - & $1.75\cdot10^{-5}$ & $1.35\cdot10^{-5}$ & $7.44\cdot10^{-6}$ \\
\hline
\end{tabular}
\tablefoot{
 The contaminating flux has been calculated extrapolating the MIPS-24 flux obtained from the Spitzer c2d catalogue \citep{Evans2009} to the median SED for each source contained in the different apertures for each band. }\\
\end{table*}
\clearpage

\begin{figure*}
 \centering
 
    \includegraphics[width=0.3\textwidth,page=1]{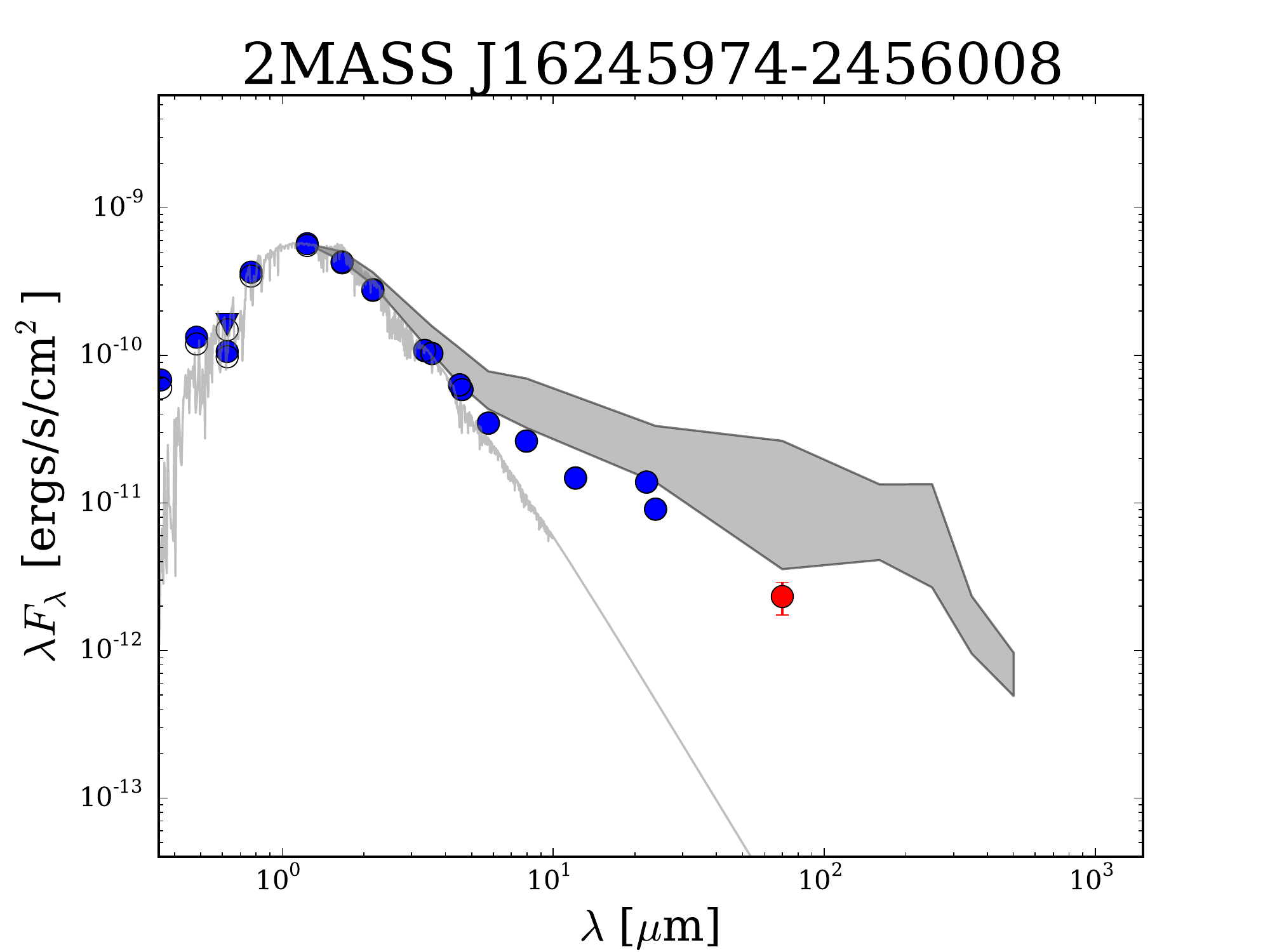}
    \includegraphics[width=0.3\textwidth,page=2]{Paper_SEDs.pdf}
    \includegraphics[width=0.3\textwidth,page=3]{Paper_SEDs.pdf}
    \includegraphics[width=0.3\textwidth,page=4]{Paper_SEDs.pdf}
    \includegraphics[width=0.3\textwidth,page=5]{Paper_SEDs.pdf}
    \includegraphics[width=0.3\textwidth,page=7]{Paper_SEDs.pdf}
    \includegraphics[width=0.3\textwidth,page=9]{Paper_SEDs.pdf}
    \includegraphics[width=0.3\textwidth,page=10]{Paper_SEDs.pdf}
    \includegraphics[width=0.3\textwidth,page=12]{Paper_SEDs.pdf}
    \includegraphics[width=0.3\textwidth,page=13]{Paper_SEDs.pdf}
    \includegraphics[width=0.3\textwidth,page=14]{Paper_SEDs.pdf}
    \includegraphics[width=0.3\textwidth,page=15]{Paper_SEDs.pdf}
    \includegraphics[width=0.3\textwidth,page=16]{Paper_SEDs.pdf}
    \includegraphics[width=0.3\textwidth,page=17]{Paper_SEDs.pdf}
    \includegraphics[width=0.3\textwidth,page=18]{Paper_SEDs.pdf}
    \includegraphics[width=0.3\textwidth,page=20]{Paper_SEDs.pdf}    
    \includegraphics[width=0.3\textwidth,page=21]{Paper_SEDs.pdf}    
    \includegraphics[width=0.3\textwidth,page=24]{Paper_SEDs.pdf}    
   \caption{Spectral energy distribution (SED's) of the sources detected in at least one band by \textit{Herschel} and classified as non-transitional. Blue dots show data acquired from the literature, red dots are photometric fluxes obtained from \textit{Herschel} data. Grey dashed line is the photosphere model according to the spectral type, and the grey shaded area is the filled area between the first and third quartile of all the disk fluxes. Observed fluxes are shown with empty circles and $A_v$ values used are in Table 3. }
  \label{fig:imagessed}
\end{figure*}
\begin{figure*}[HT]

\caption{Fig. \ref{fig:imagessed} continued.}
    \centering
        
    \includegraphics[width=0.3\textwidth,page=26]{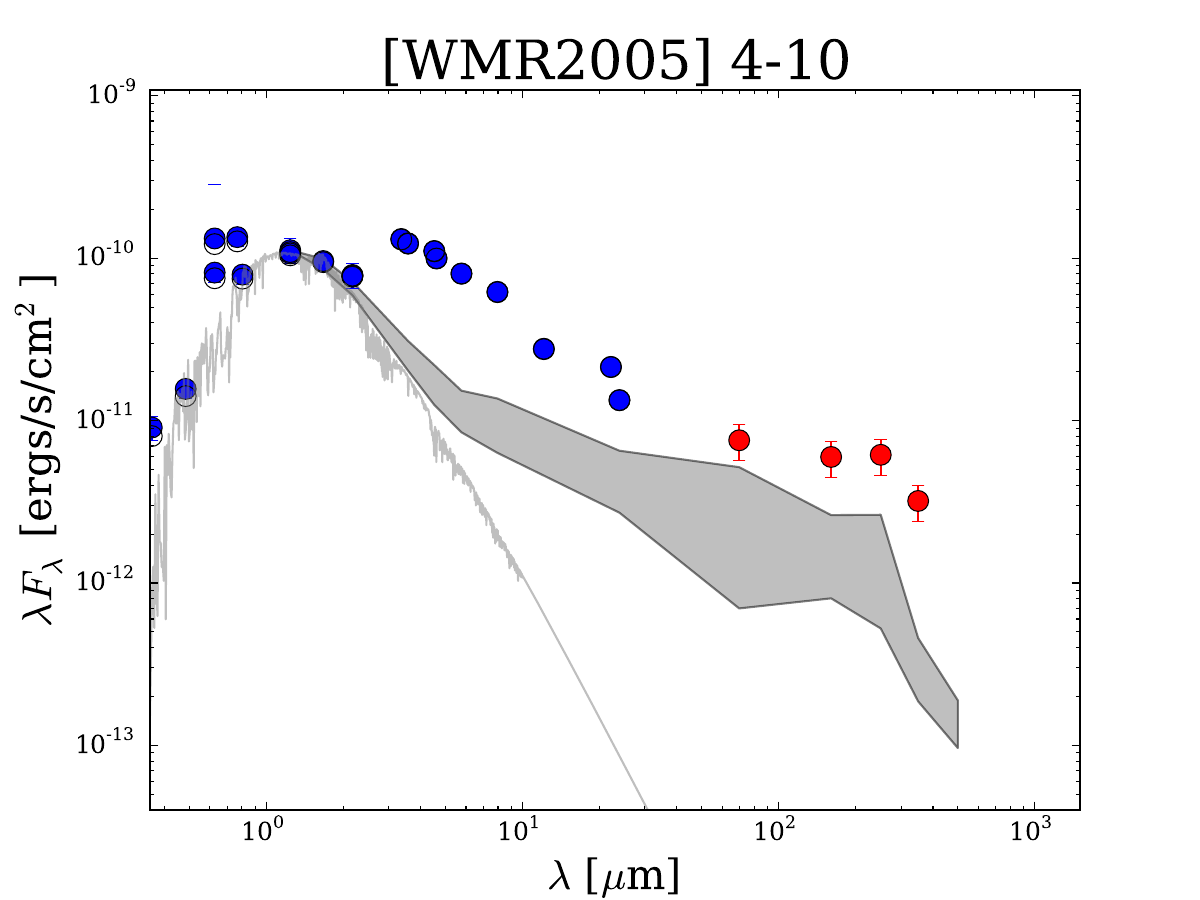}
        \includegraphics[width=0.3\textwidth,page=27]{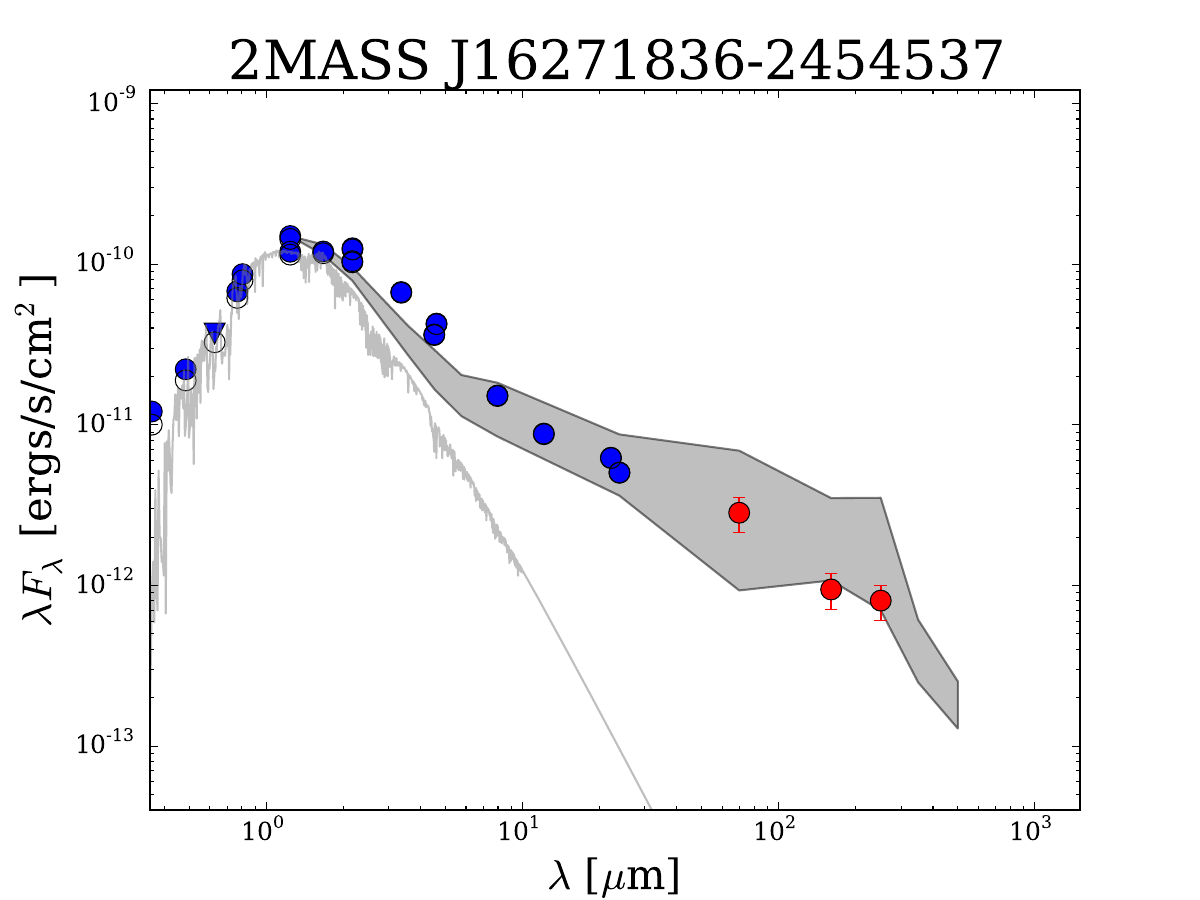}
        \includegraphics[width=0.3\textwidth,page=28]{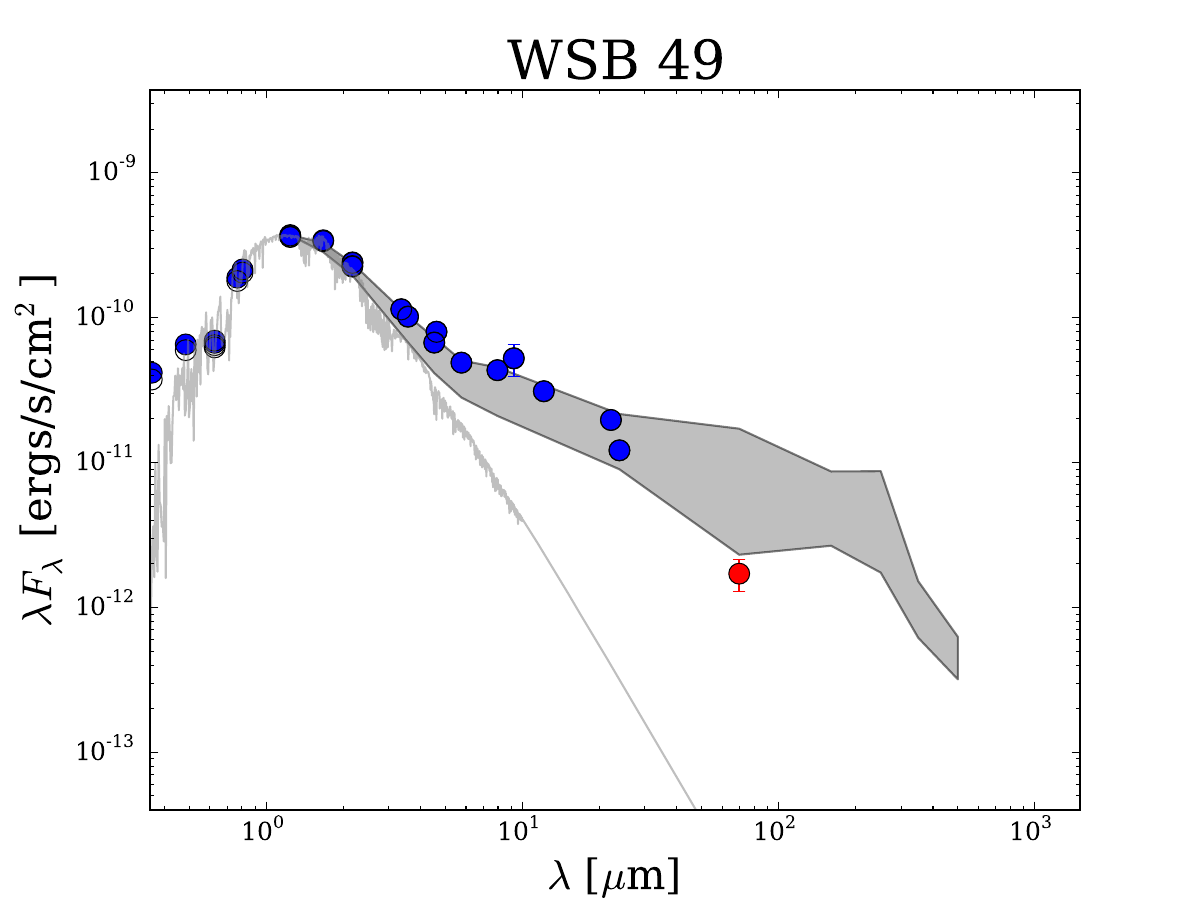}
    \includegraphics[width=0.3\textwidth,page=29]{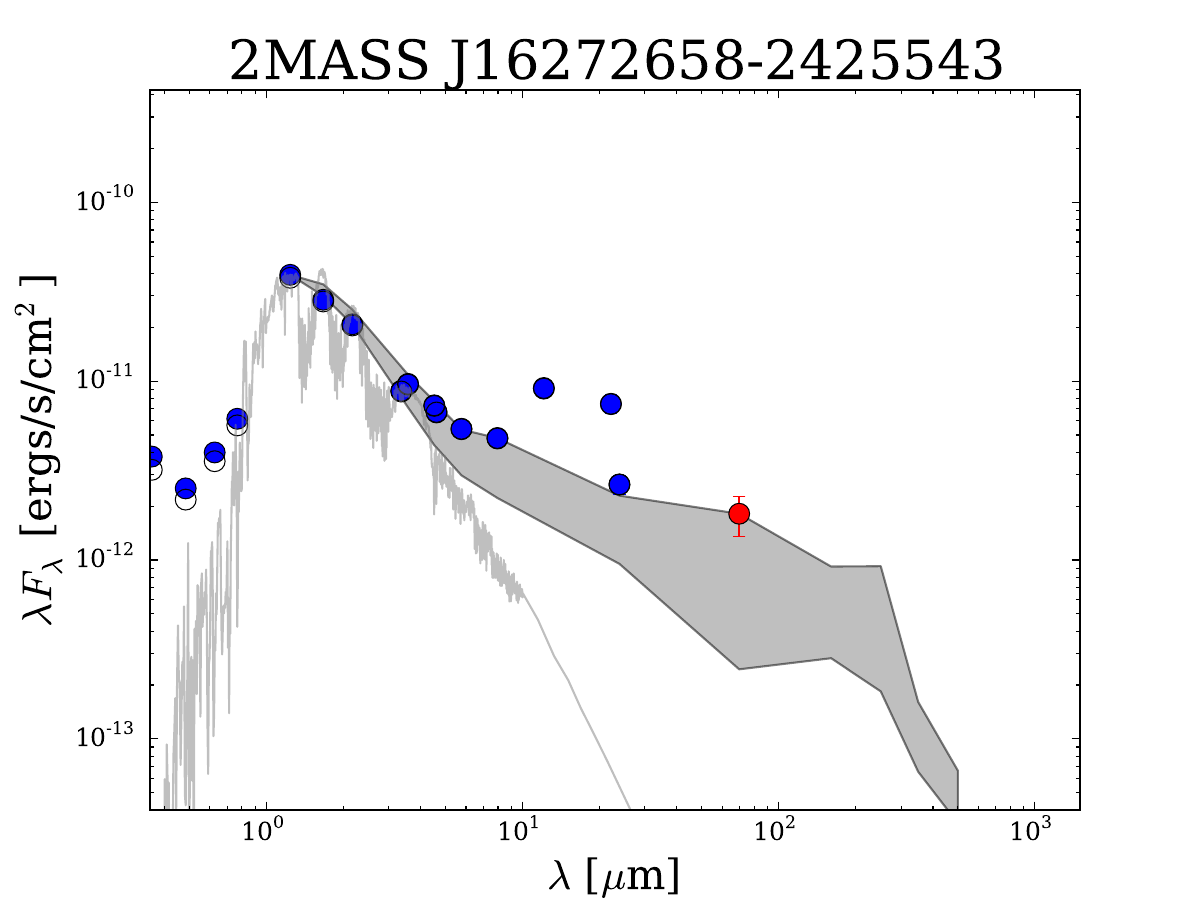}
    \includegraphics[width=0.3\textwidth,page=31]{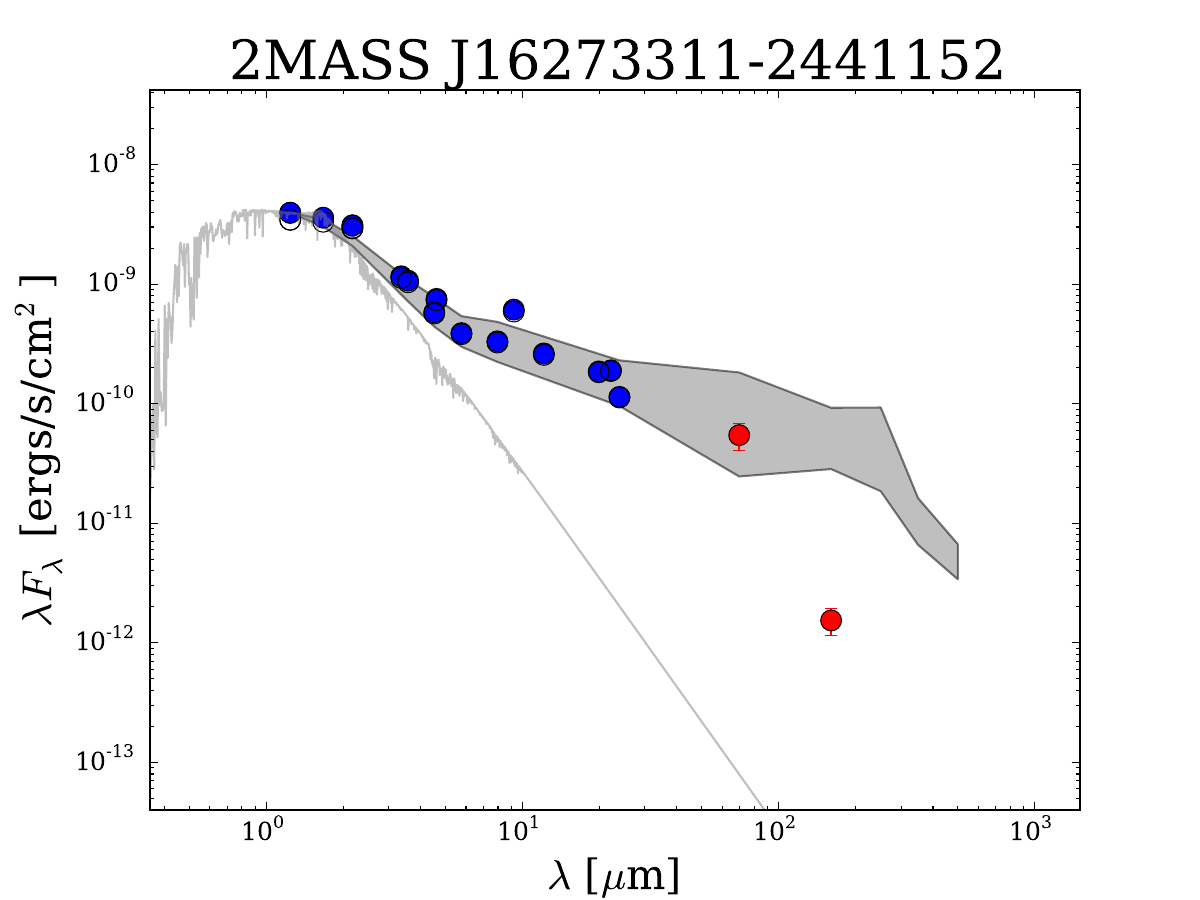}
    \includegraphics[width=0.3\textwidth,page=34]{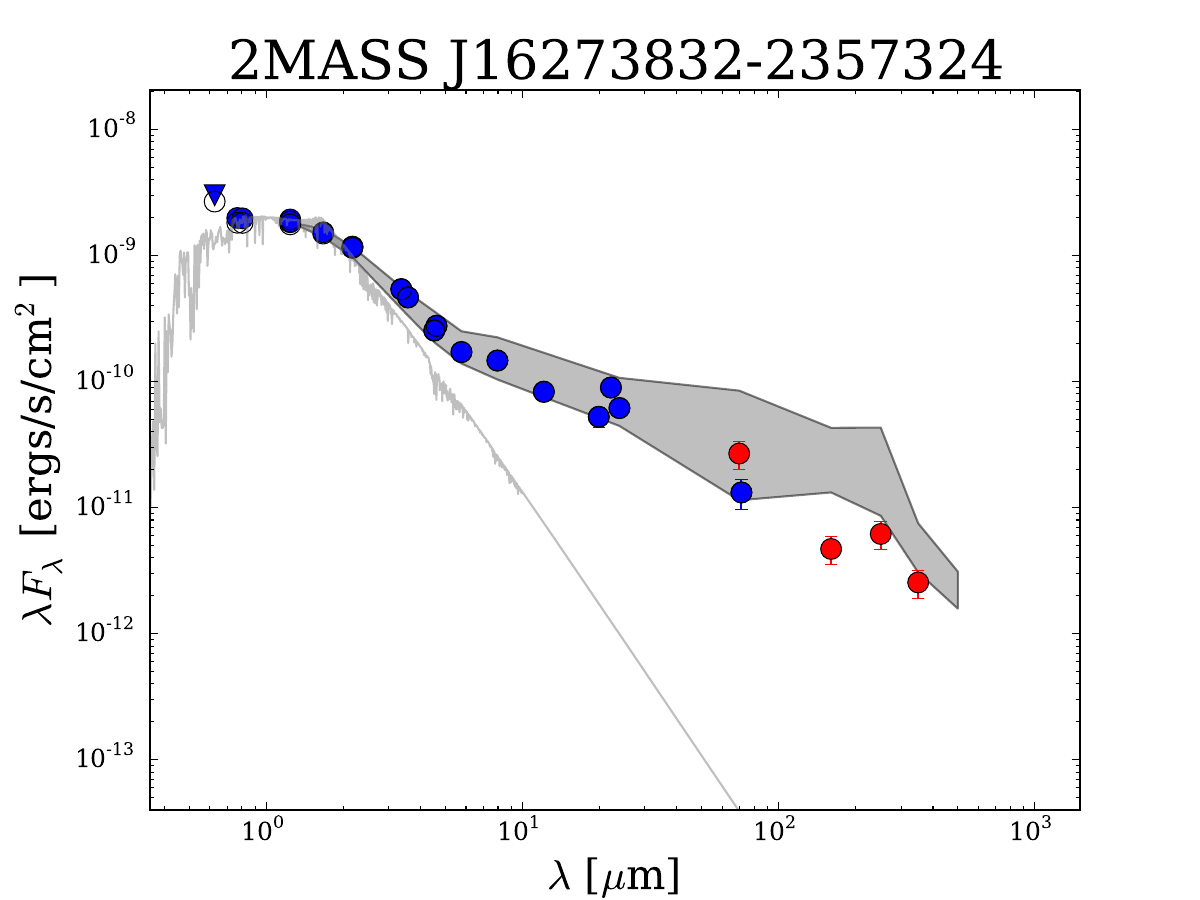}
    \includegraphics[width=0.3\textwidth,page=36]{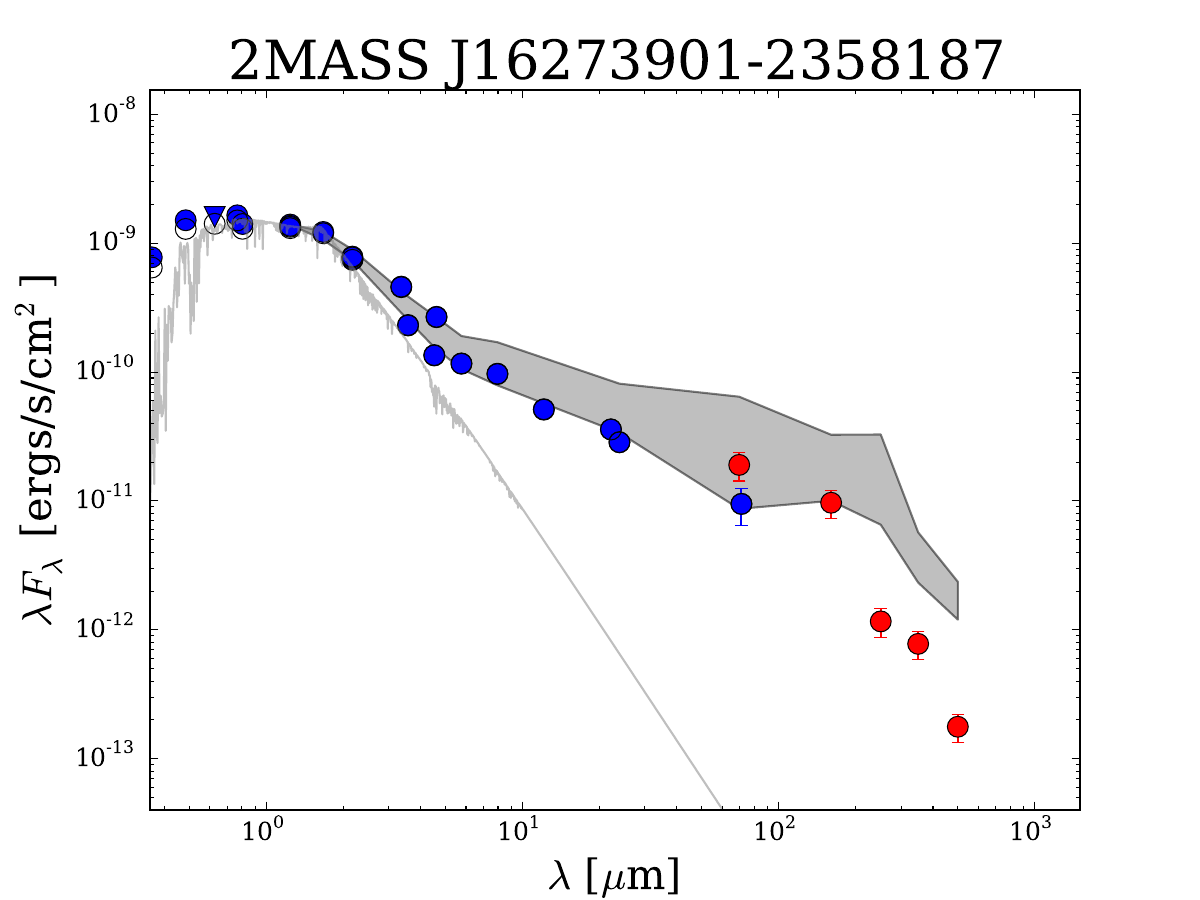}
    \includegraphics[width=0.3\textwidth,page=39]{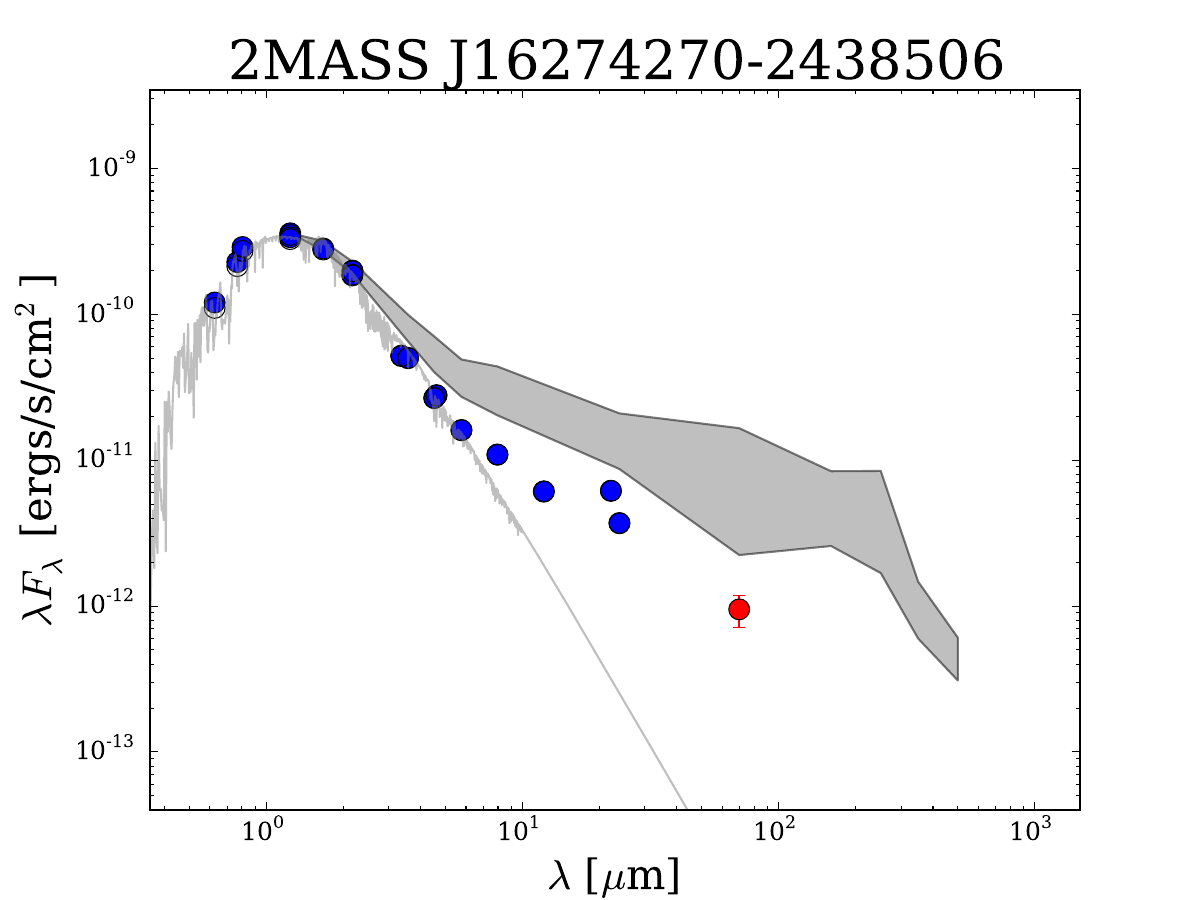}
    \includegraphics[width=0.3\textwidth,page=40]{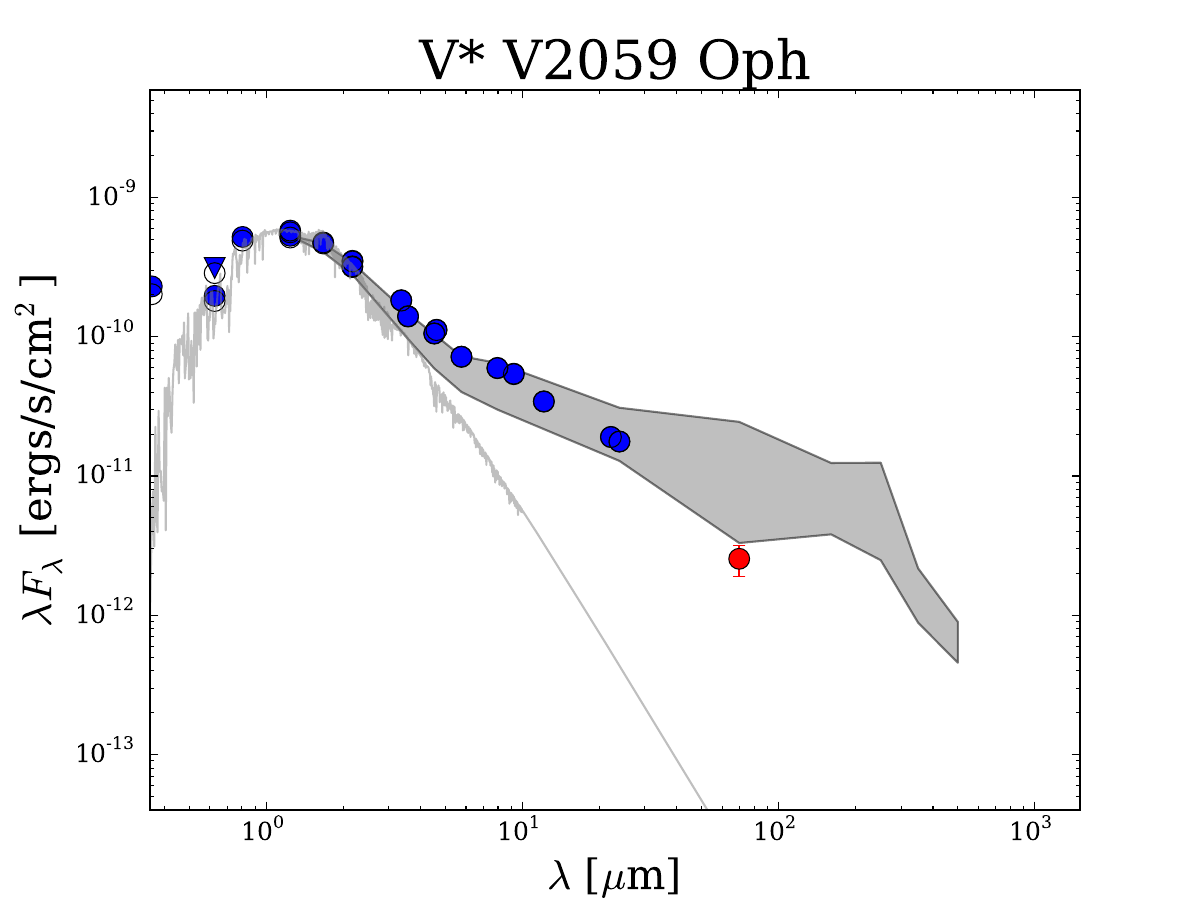}
    \includegraphics[width=0.3\textwidth,page=41]{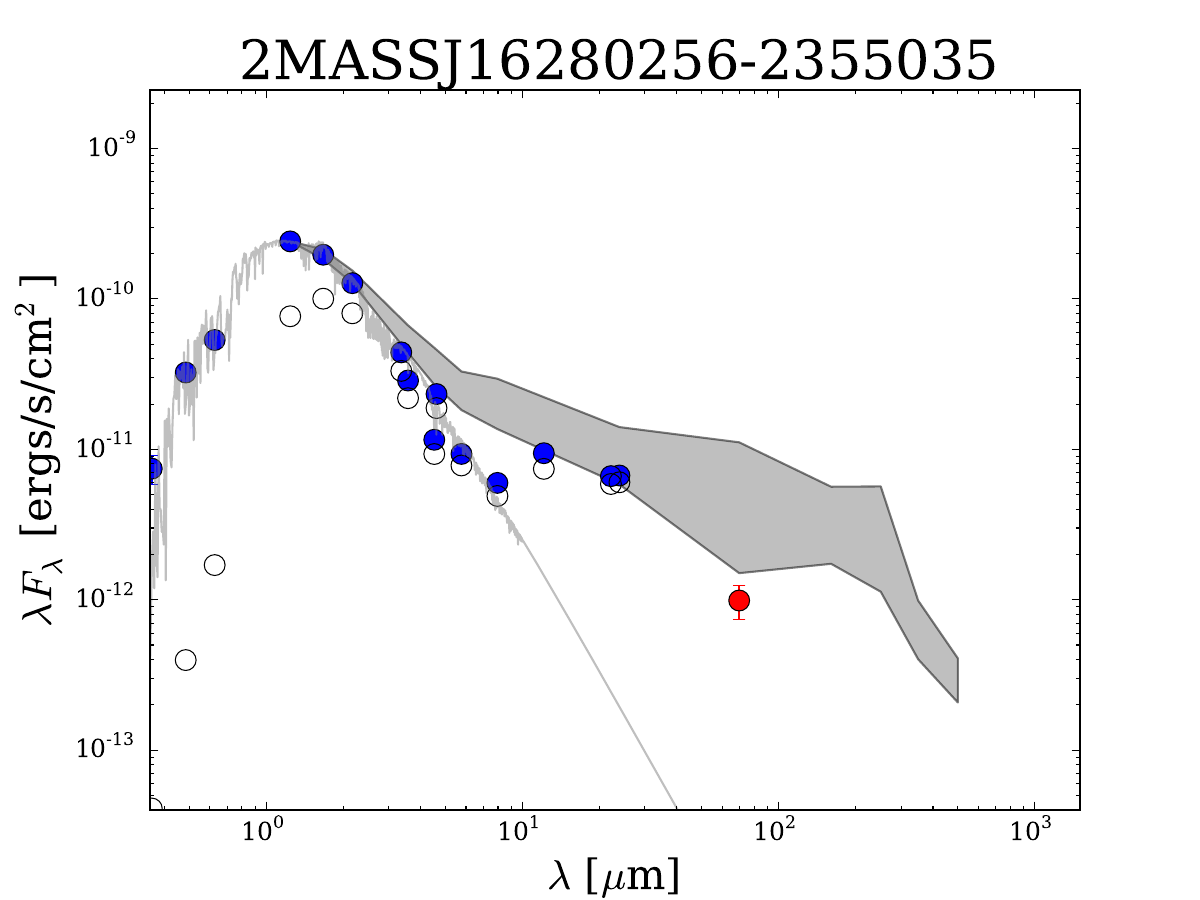}
    \includegraphics[width=0.3\textwidth,page=42]{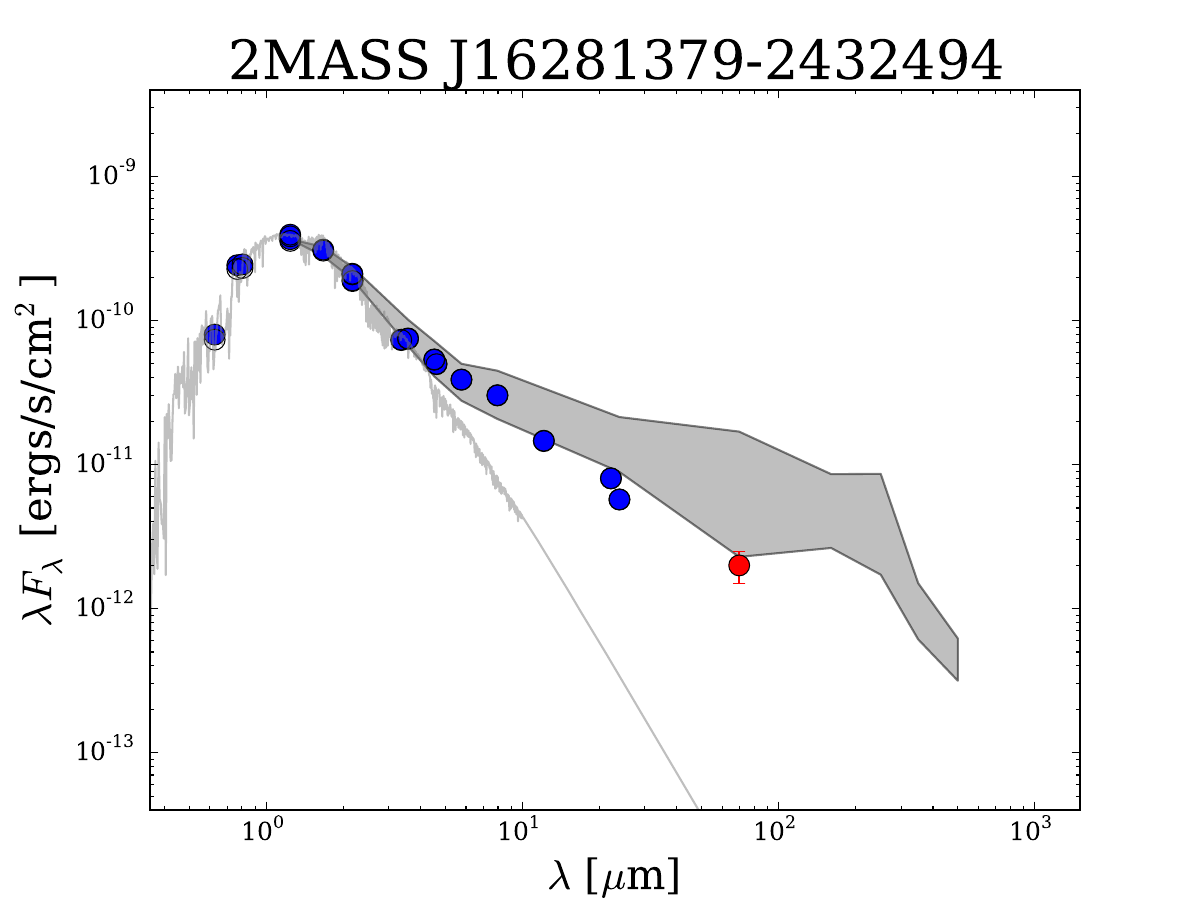}
    \includegraphics[width=0.3\textwidth,page=45]{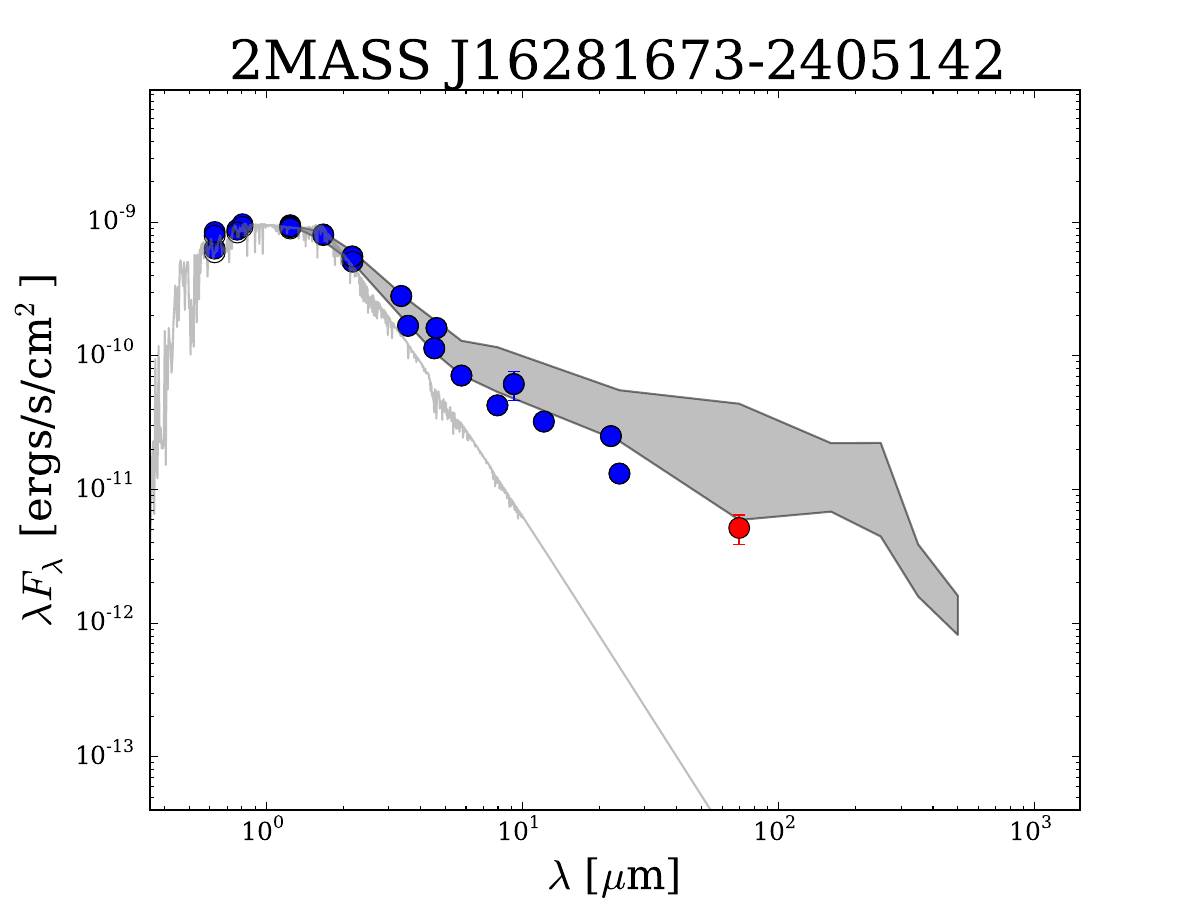}
    \includegraphics[width=0.3\textwidth,page=49]{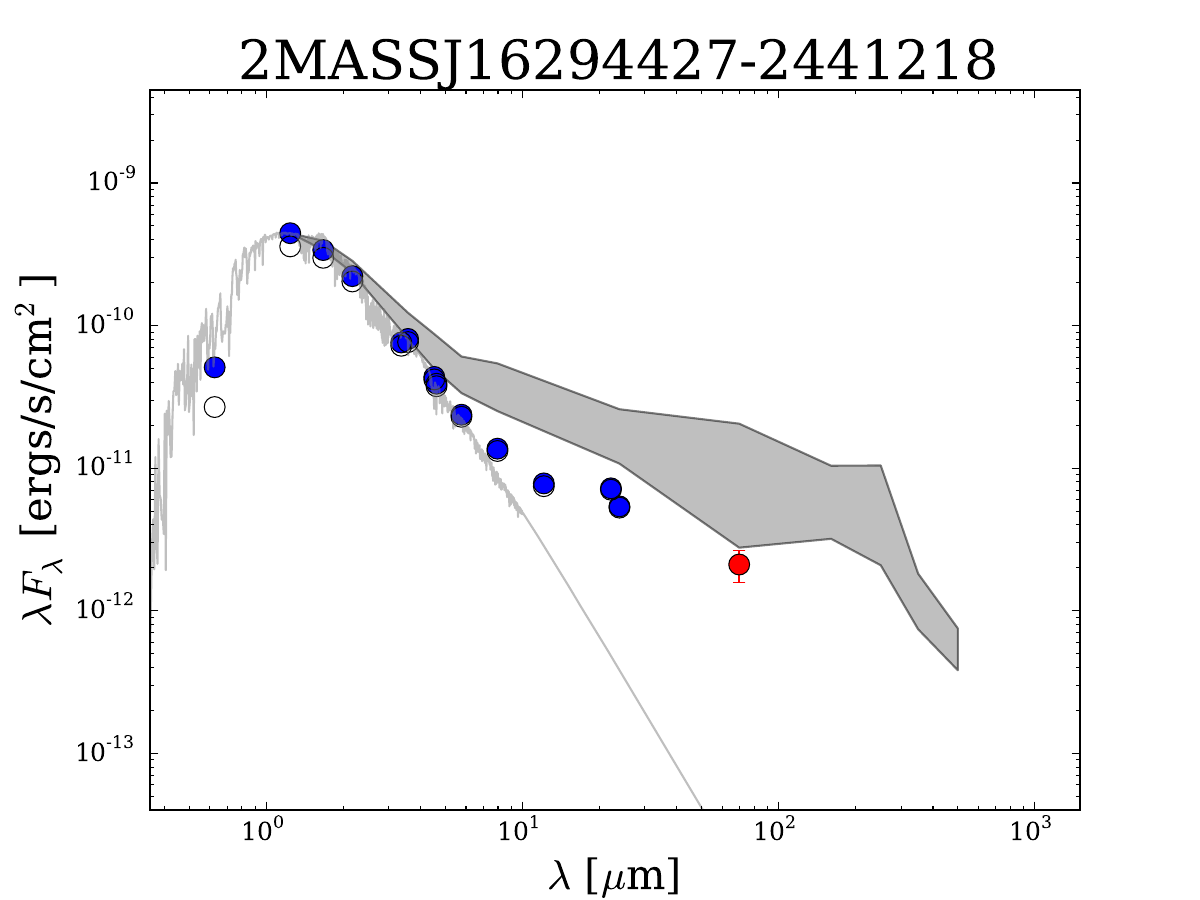}

  \label{fig:images2sed}
\end{figure*}

\begin{figure*}[HT]
  \centering
     \includegraphics[width=1\textwidth,page=1]{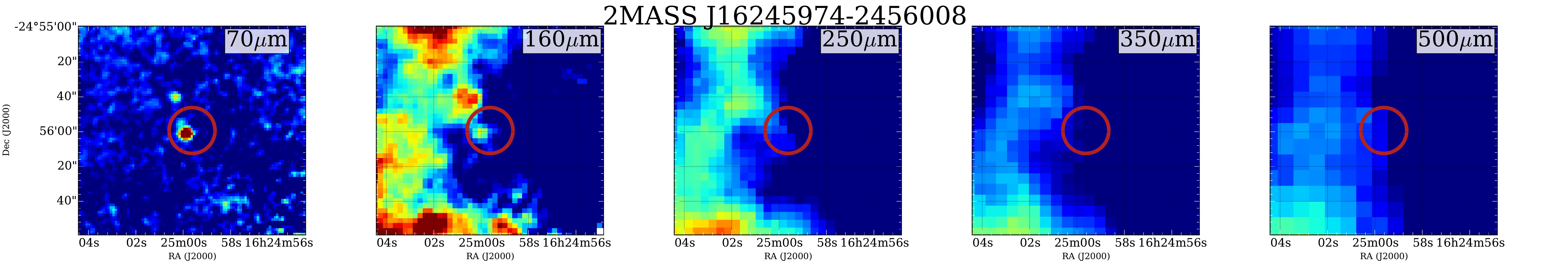}
    \includegraphics[width=1\textwidth,page=2]{Paper_Images.pdf}
    \includegraphics[width=1\textwidth,page=3]{Paper_Images.pdf}
    \includegraphics[width=1\textwidth,page=4]{Paper_Images.pdf}
    \includegraphics[width=1\textwidth,page=5]{Paper_Images.pdf}
    \includegraphics[width=1\textwidth,page=6]{Paper_Images.pdf}
    \includegraphics[width=1\textwidth,page=7]{Paper_Images.pdf}
  \caption{Thumbnail images of each of the 46 sources with at least one point source detected by \textit{Herschel}. All images are 60 \arcsec x 60 \arcsec with north up and east to the left. The cut levels are set to the RMS of the background for the minimum and three times that value for the maximum with linear stretch.}
  \label{fig:images}
\end{figure*}

\begin{figure*}[HT]
  \centering
\caption{Fig. \ref{fig:images} continued.}
    \includegraphics[width=1\textwidth,page=8]{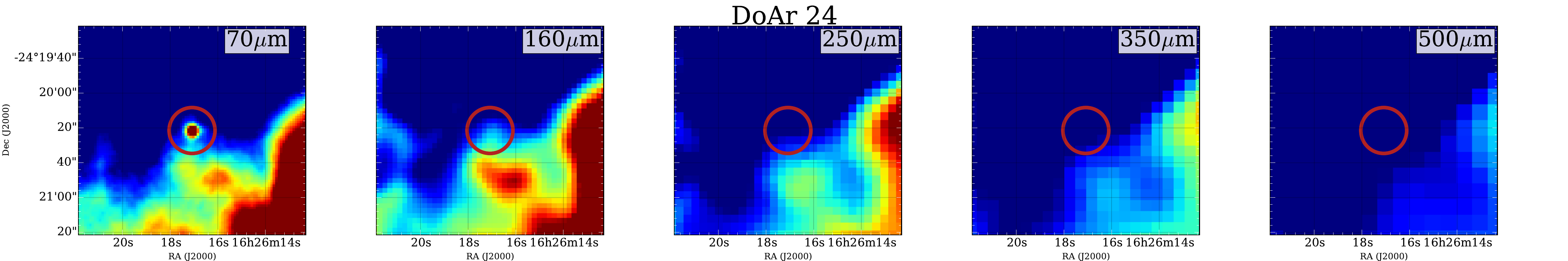}
    \includegraphics[width=1\textwidth,page=9]{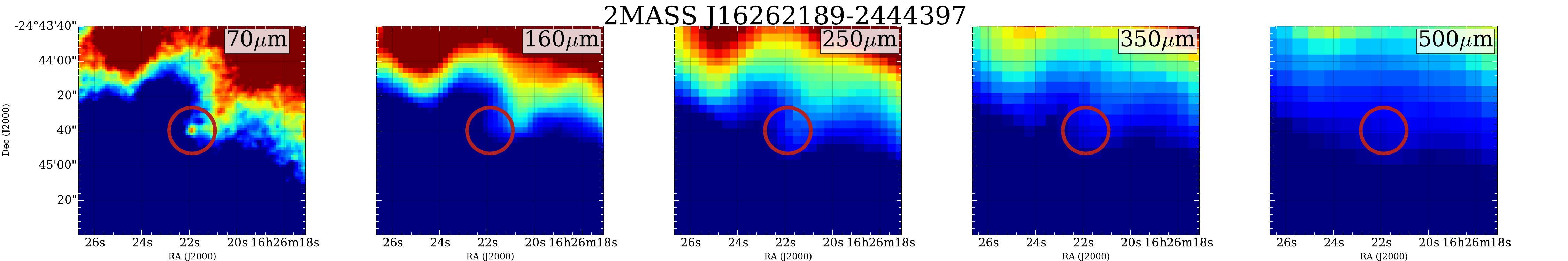}
    \includegraphics[width=1\textwidth,page=10]{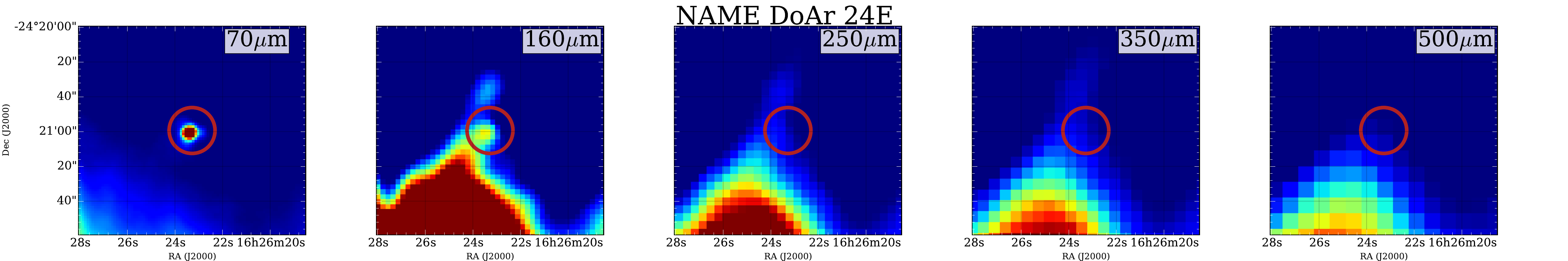}
    \includegraphics[width=1\textwidth,page=11]{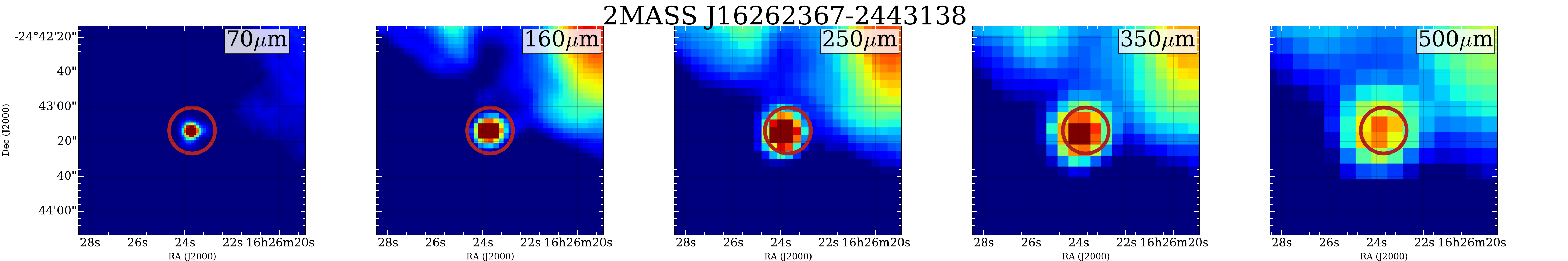}
    \includegraphics[width=1\textwidth,page=12]{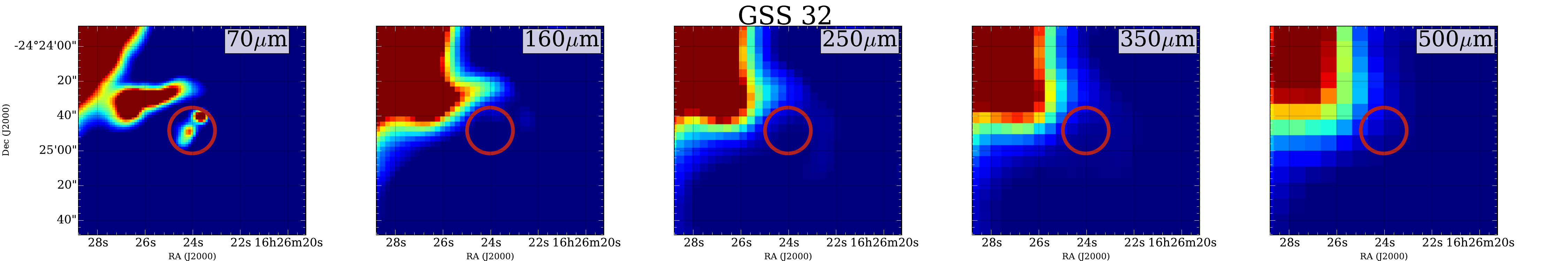}
    \includegraphics[width=1\textwidth,page=13]{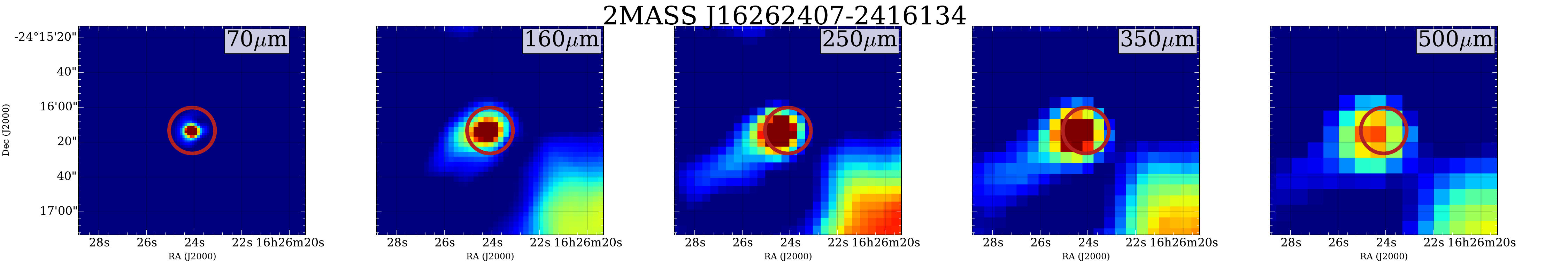}
    \includegraphics[width=1\textwidth,page=14]{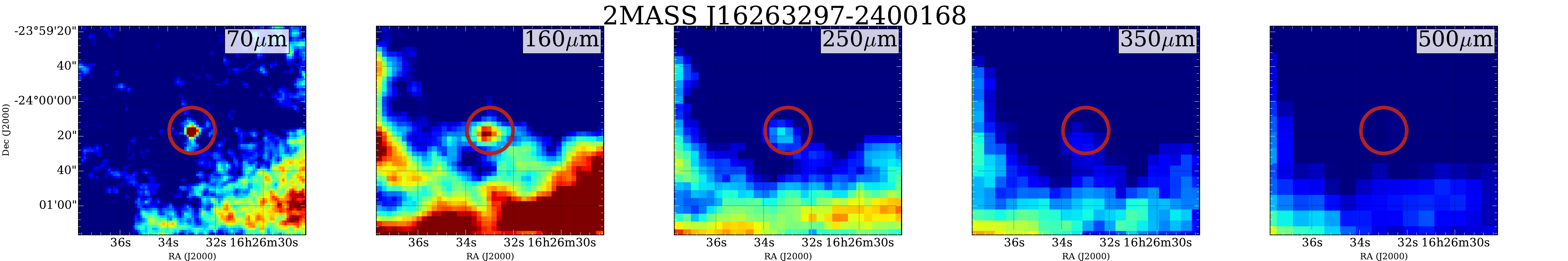}
  \label{fig:images2}
\end{figure*}
\begin{figure*}[HT]
  \centering
\caption{Fig. \ref{fig:images} continued.}
    \includegraphics[width=1\textwidth,page=15]{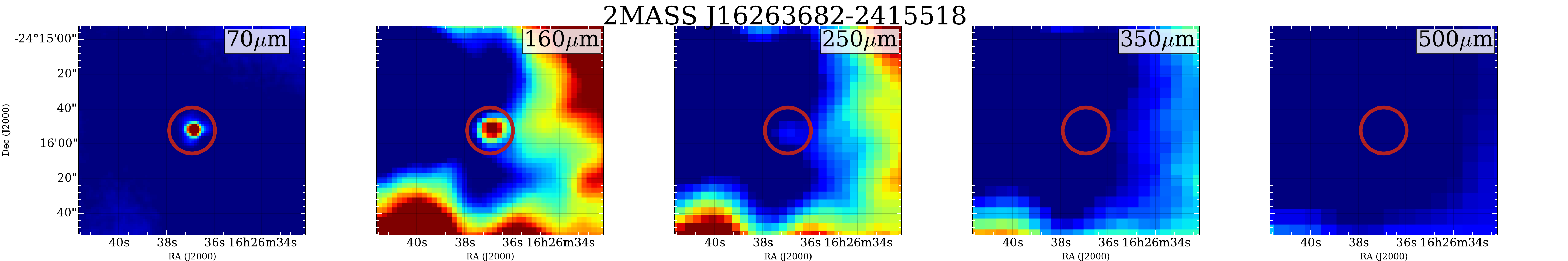}
    \includegraphics[width=1\textwidth,page=16]{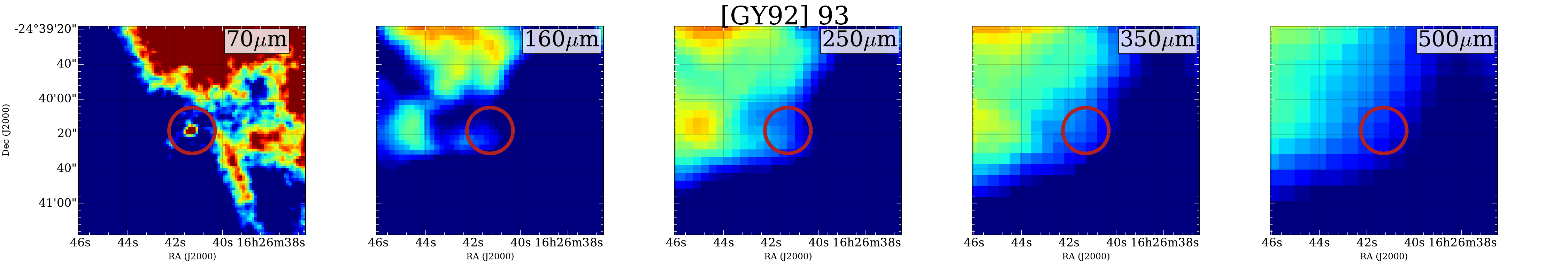}
    \includegraphics[width=1\textwidth,page=17]{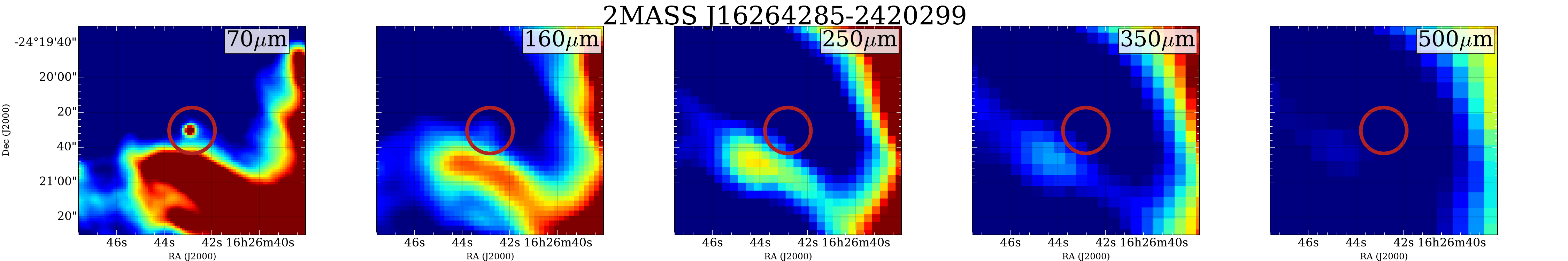}
    \includegraphics[width=1\textwidth,page=18]{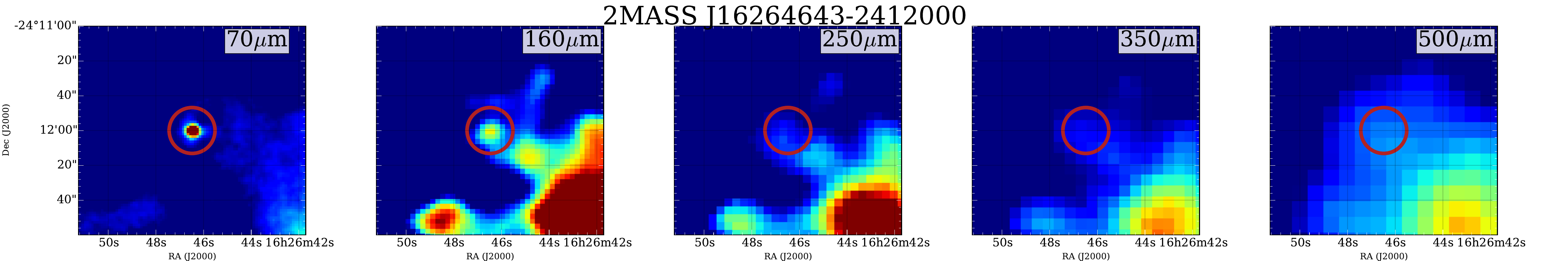}
    \includegraphics[width=1\textwidth,page=19]{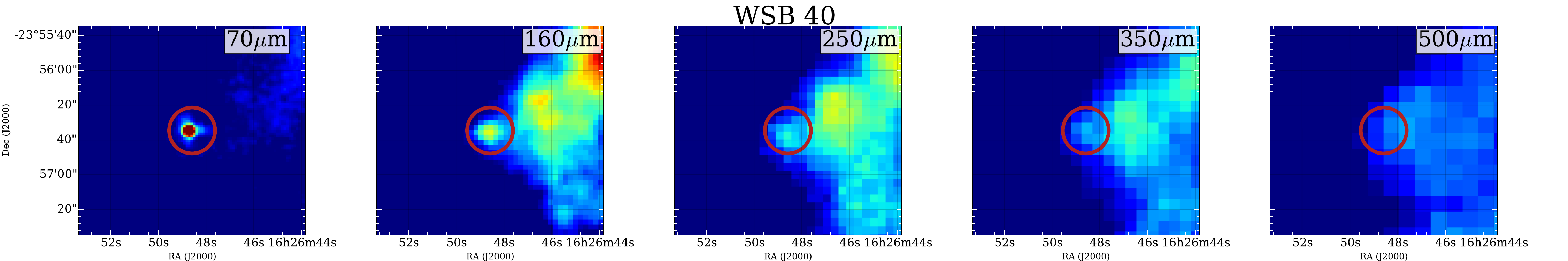}
    \includegraphics[width=1\textwidth,page=20]{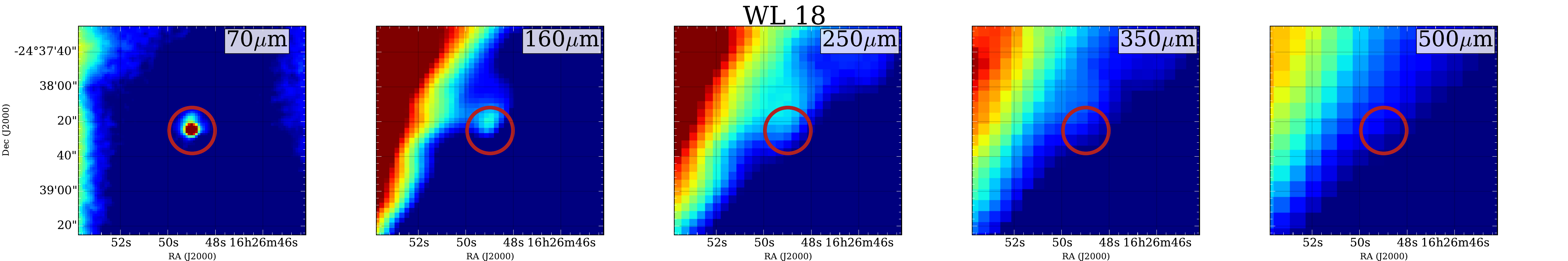}
    \includegraphics[width=1\textwidth,page=21]{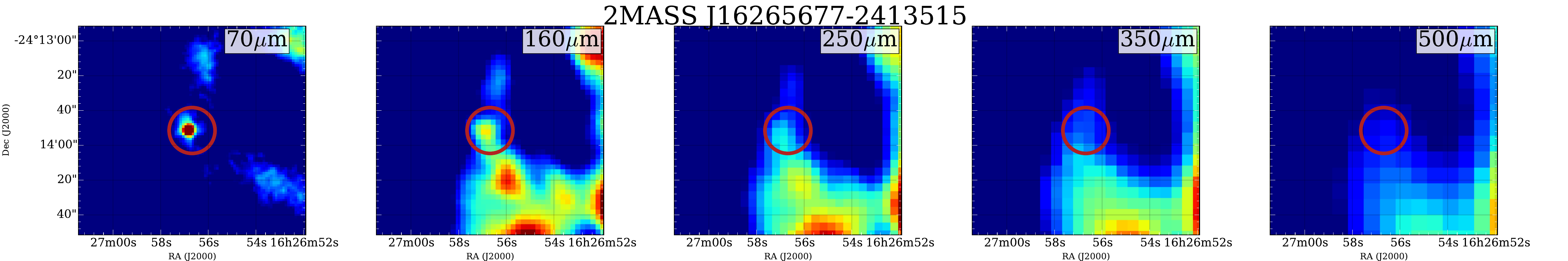}
  \label{fig:images3}
\end{figure*}
\begin{figure*}[HT]
  \centering
\caption{Fig. \ref{fig:images} continued.}
    \includegraphics[width=1\textwidth,page=22]{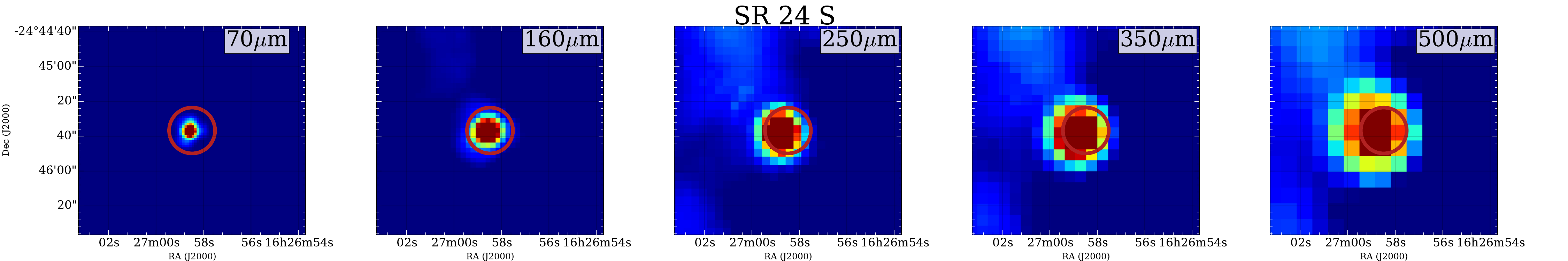}
    \includegraphics[width=1\textwidth,page=23]{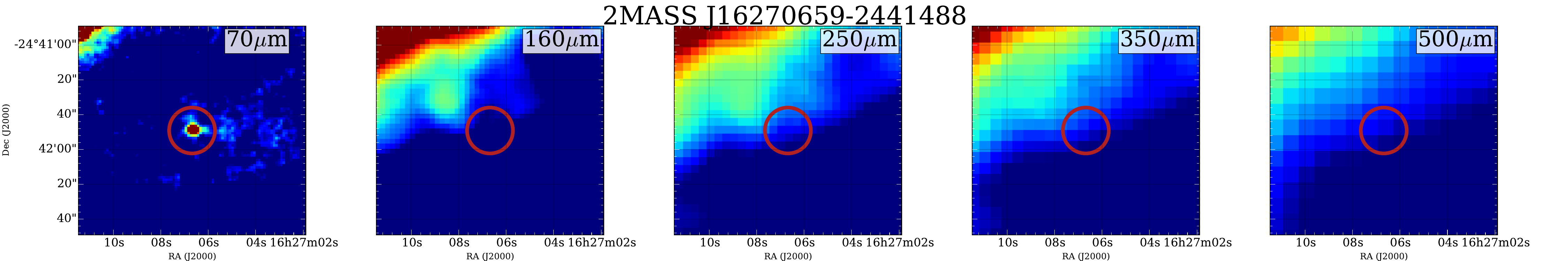}
    \includegraphics[width=1\textwidth,page=24]{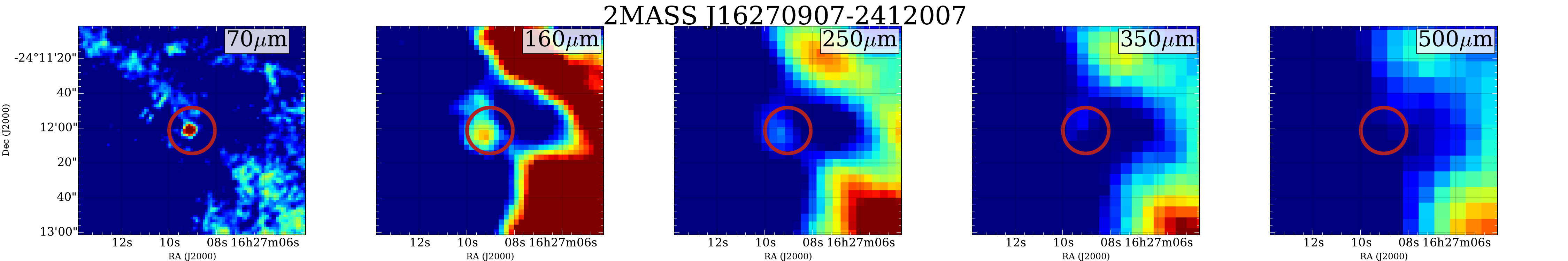}
    \includegraphics[width=1\textwidth,page=25]{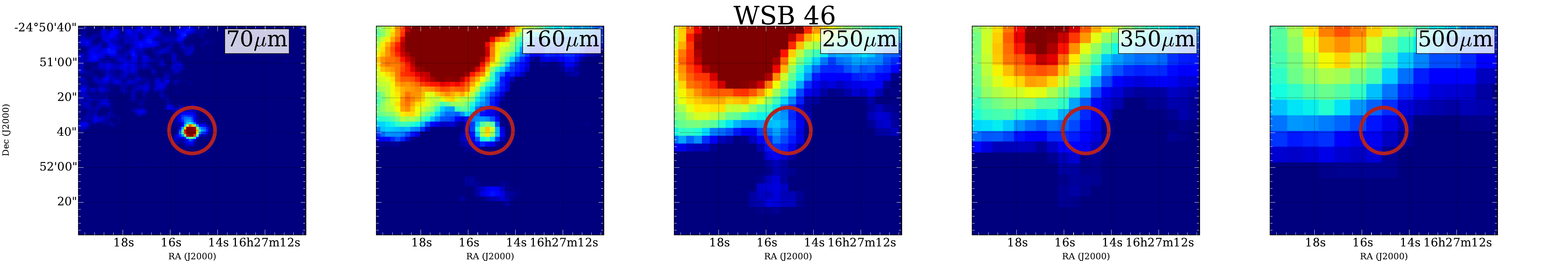}
    \includegraphics[width=1\textwidth,page=26]{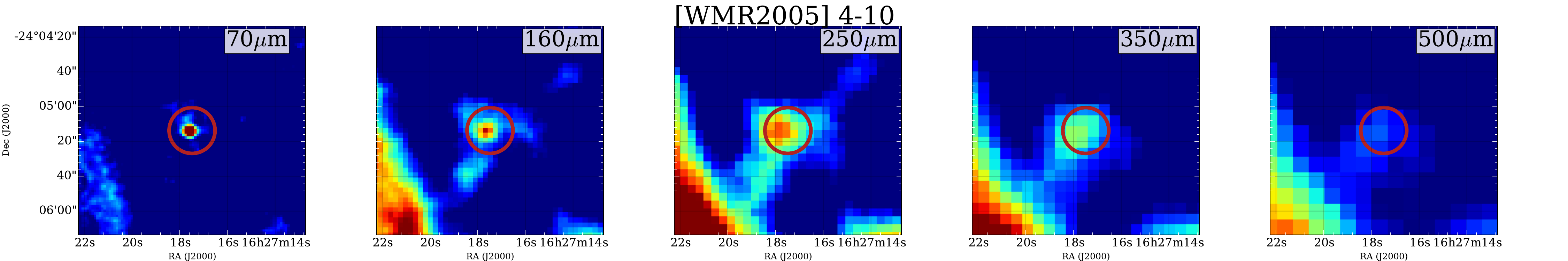}
    \includegraphics[width=1\textwidth,page=27]{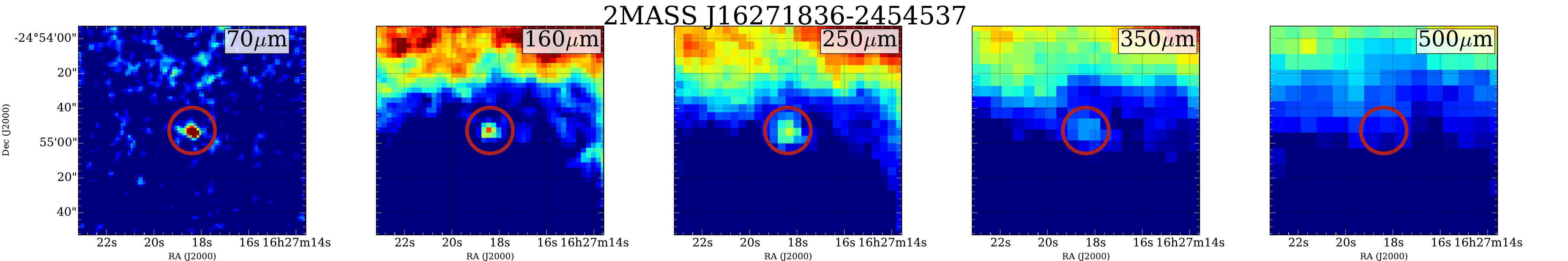}
    \includegraphics[width=1\textwidth,page=28]{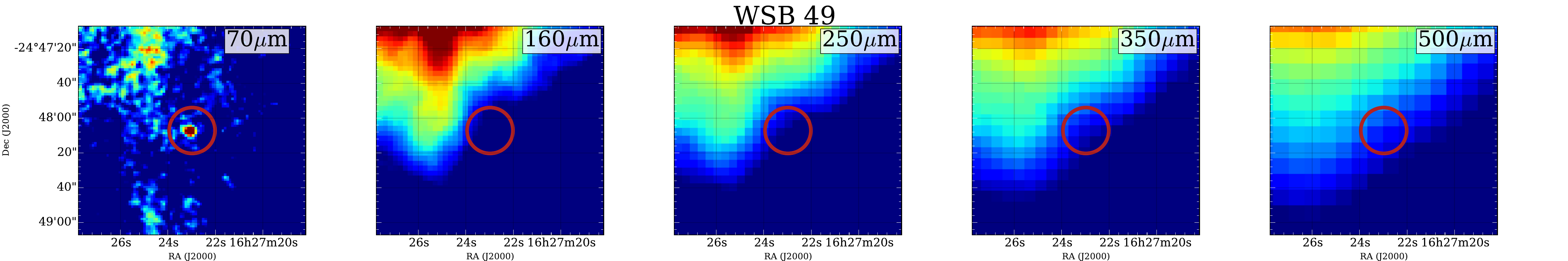}
  \label{fig:images4}
\end{figure*}
\begin{figure*}[HT]
  \centering
\caption{Fig. \ref{fig:images} continued.}
    \includegraphics[width=1\textwidth,page=29]{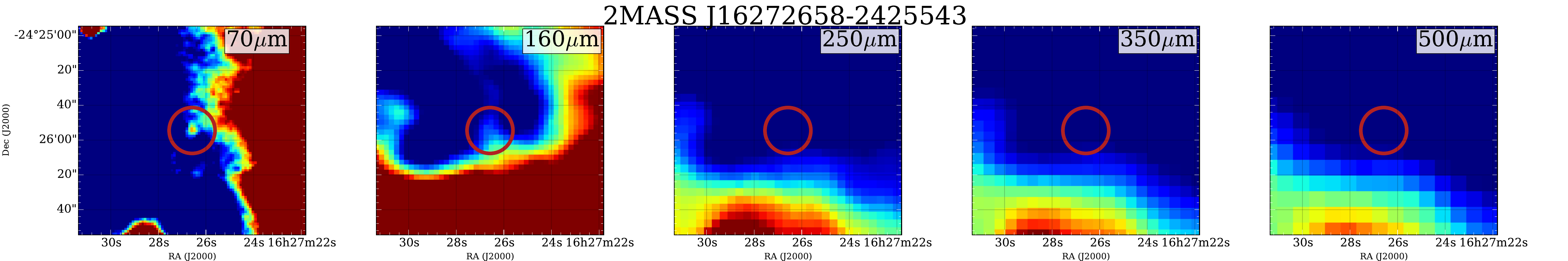}
    \includegraphics[width=1\textwidth,page=30]{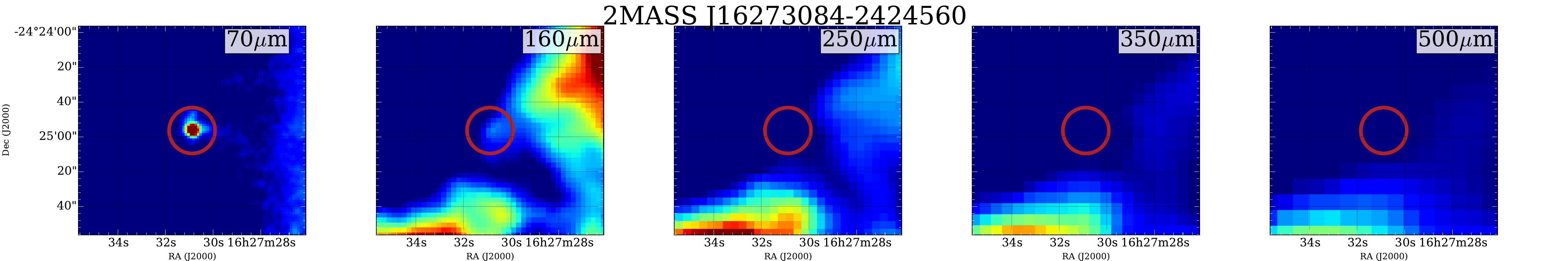}
    \includegraphics[width=1\textwidth,page=31]{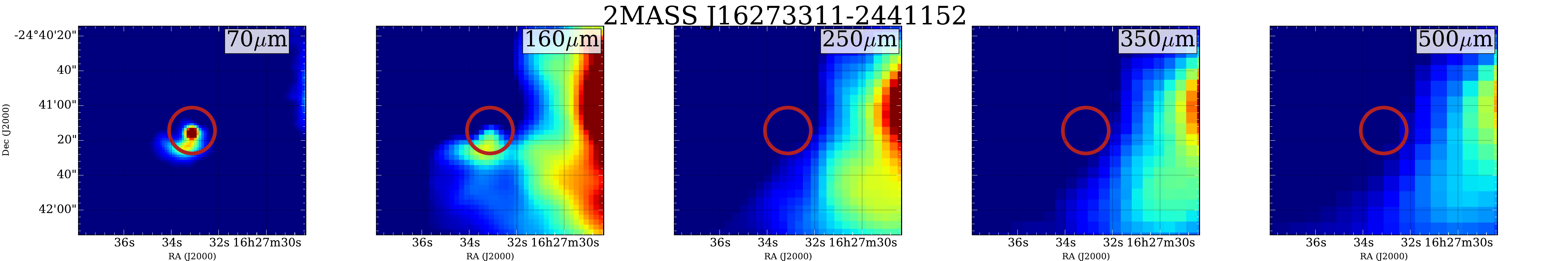}
    \includegraphics[width=1\textwidth,page=32]{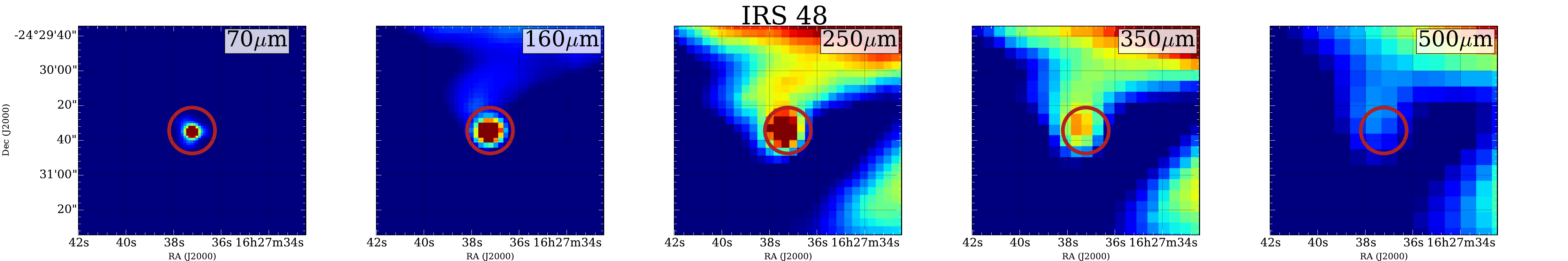}
    \includegraphics[width=1\textwidth,page=33]{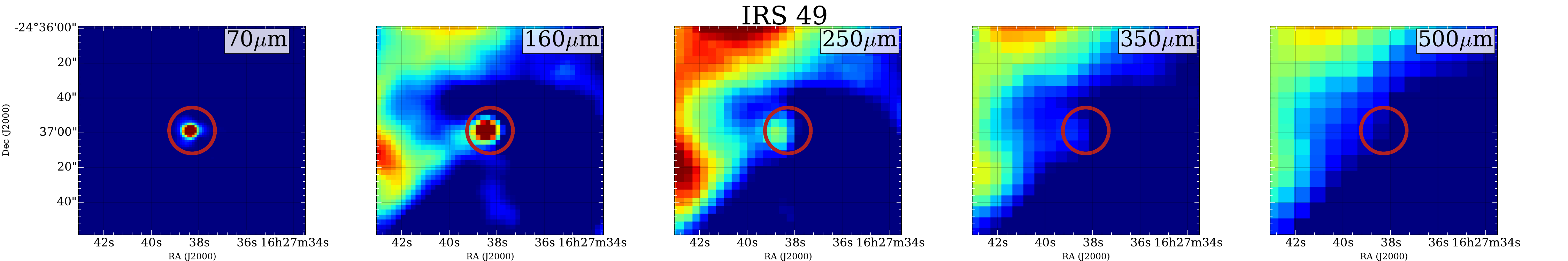}
    \includegraphics[width=1\textwidth,page=34]{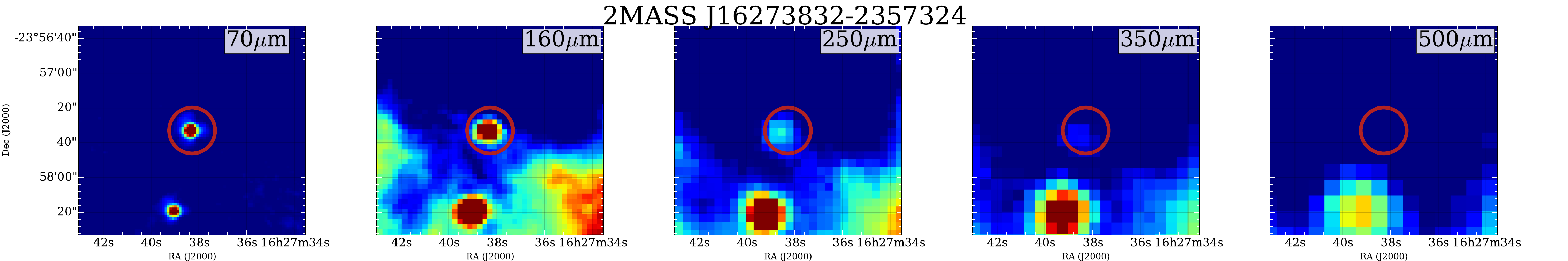}
    \includegraphics[width=1\textwidth,page=35]{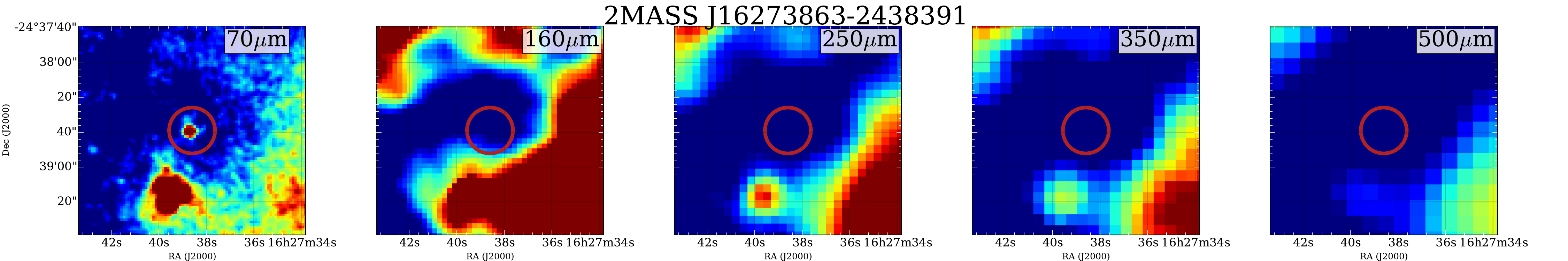}
  \label{fig:images5}
\end{figure*}
\begin{figure*}[HT]
  \centering
\caption{Fig. \ref{fig:images} continued.}
    \includegraphics[width=1\textwidth,page=36]{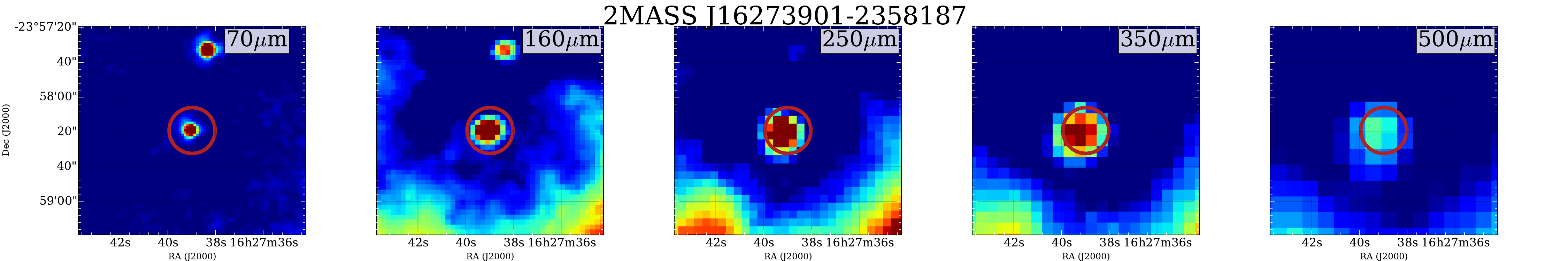}
    \includegraphics[width=1\textwidth,page=37]{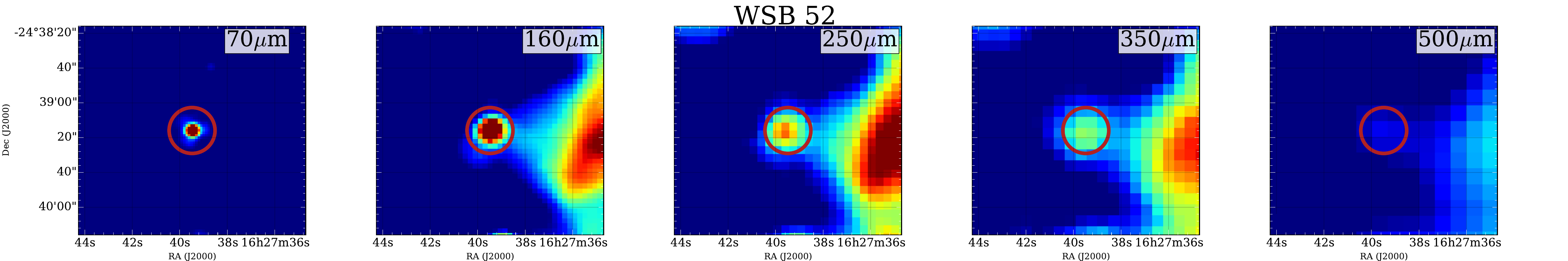}
    \includegraphics[width=1\textwidth,page=38]{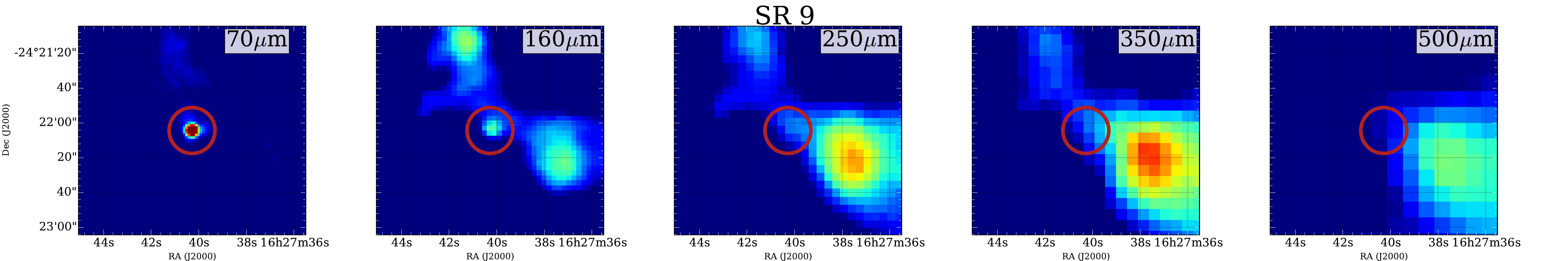}
    \includegraphics[width=1\textwidth,page=39]{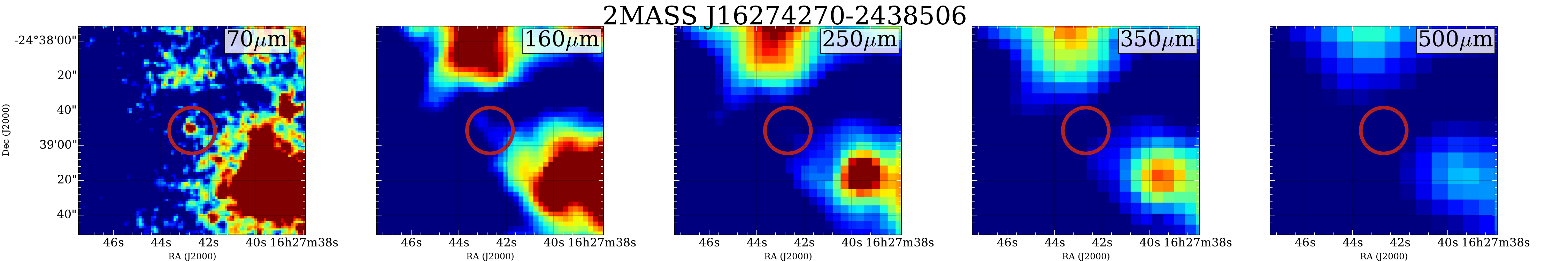}
    \includegraphics[width=1\textwidth,page=40]{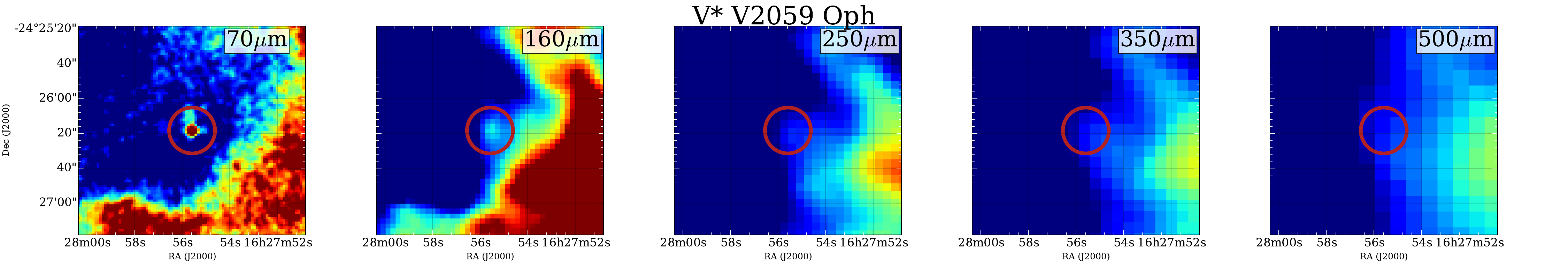}
    \includegraphics[width=1\textwidth,page=41]{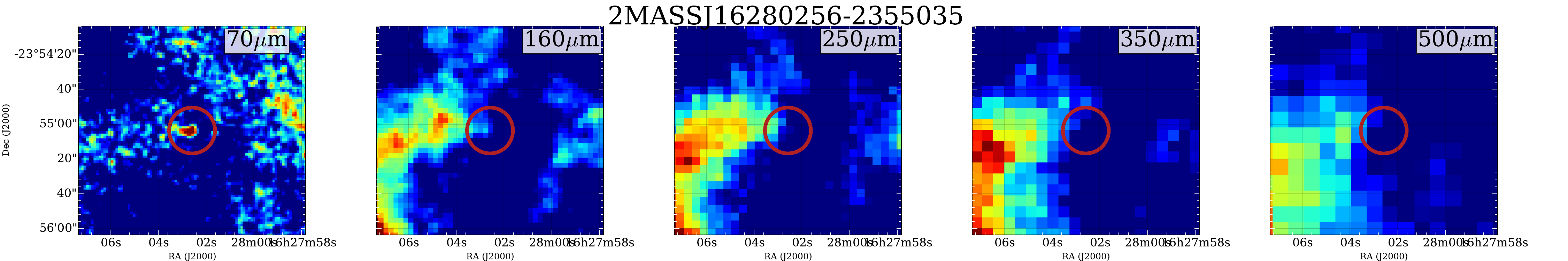}
    \includegraphics[width=1\textwidth,page=42]{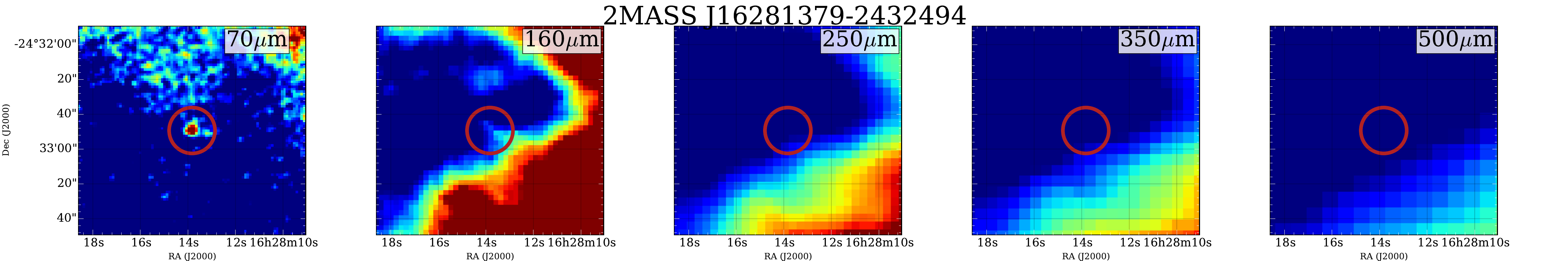}
  \label{fig:images6}
\end{figure*}
\begin{figure*}[HT]
  \centering
\caption{Fig. \ref{fig:images} continued.}
    \includegraphics[width=1\textwidth,page=43]{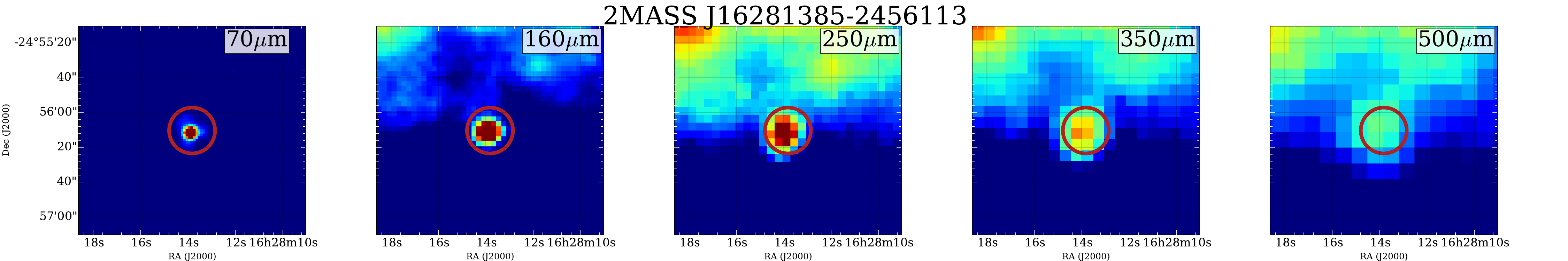}
    \includegraphics[width=1\textwidth,page=44]{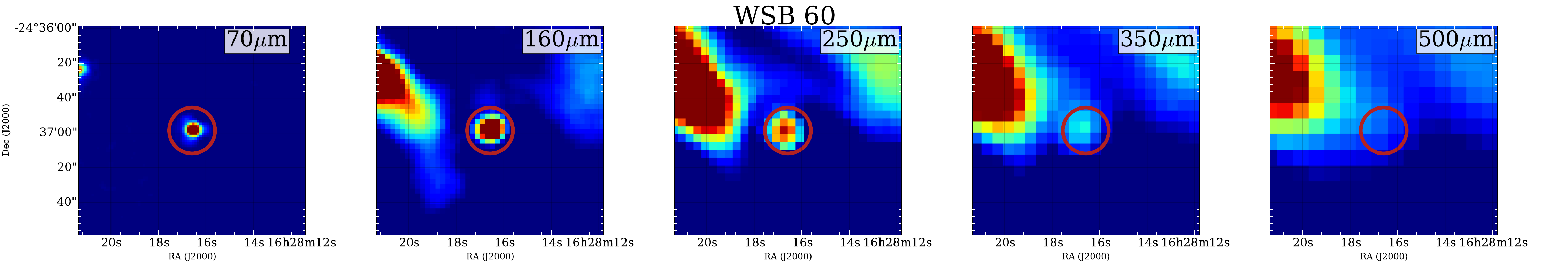}
    \includegraphics[width=1\textwidth,page=45]{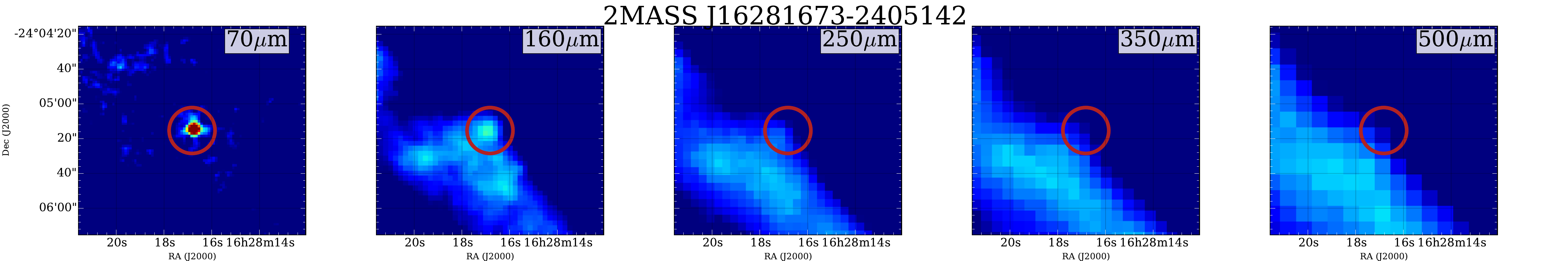}
    \includegraphics[width=1\textwidth,page=46]{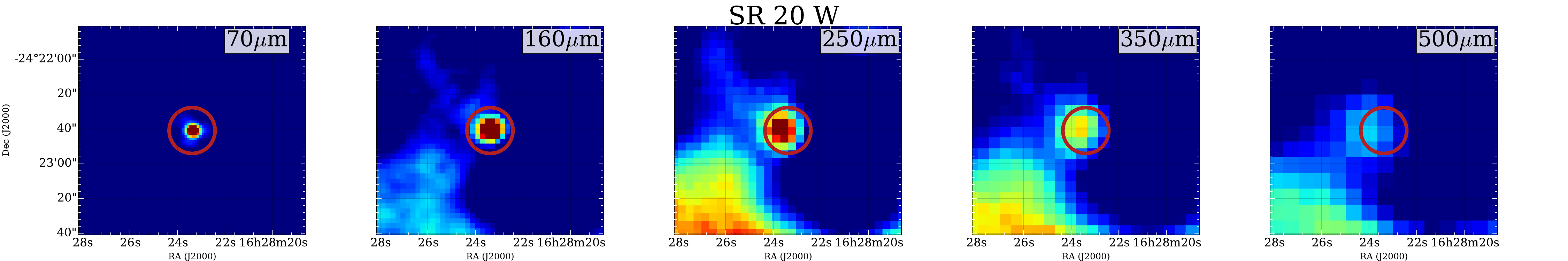}
    \includegraphics[width=1\textwidth,page=47]{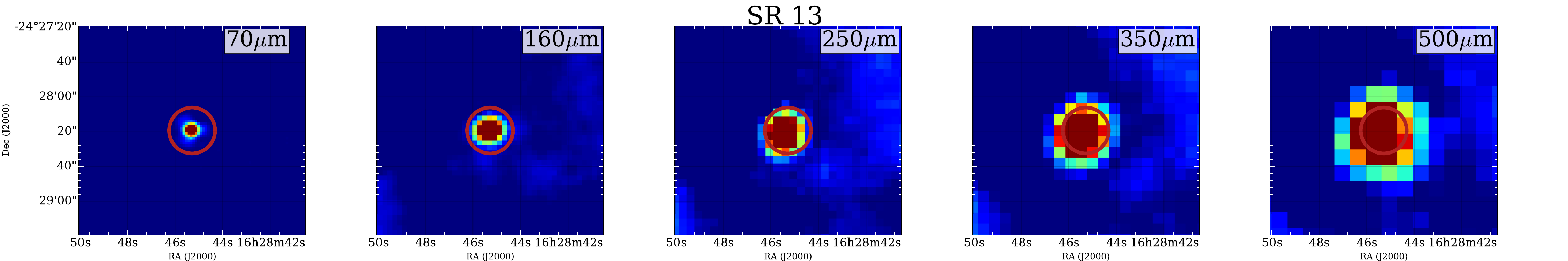}
    \includegraphics[width=1\textwidth,page=48]{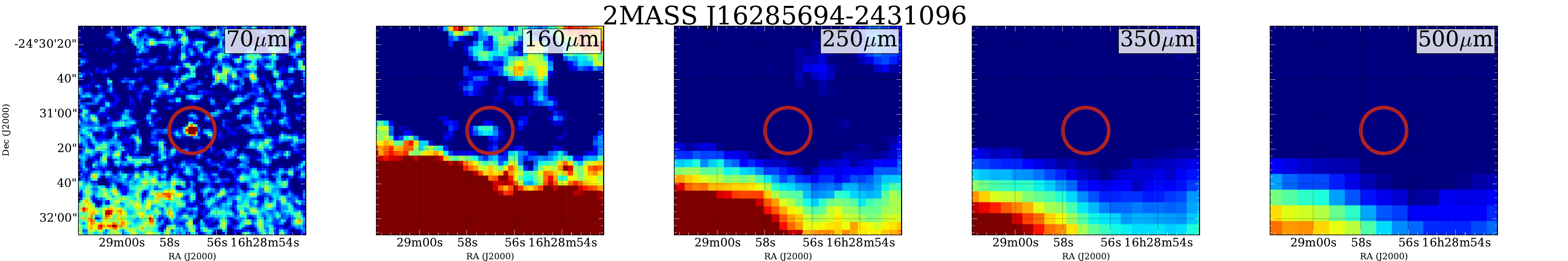}
    \includegraphics[width=1\textwidth,page=49]{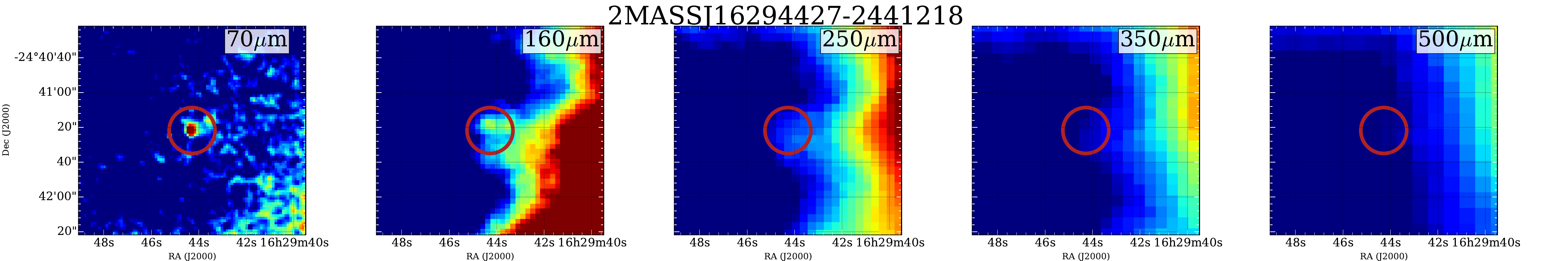}
  \label{fig:images7}
\end{figure*}

\end{appendix}

\end{document}